\documentclass{article}
\usepackage{amsmath}
\usepackage{amssymb}
\usepackage{graphicx}
\usepackage{booktabs}
\usepackage{natbib}
\makeatletter
\renewcommand\@biblabel[1]{#1.}
\makeatother
\setcitestyle{comma,numbers,super,open={},close={}}
\providecommand{\U}[1]{\protect\rule{.1in}{.1in}}

\date{}
\begin{document}
\title{Modelling the Spread of SARS-CoV2 and its variants. Comparison with Real Data. Relations that have to be Satisfied to Achieve the Total Regression of the SARS-CoV2 Infection.}
\author{Giorgio SONNINO, Philippe PEETERS, and Pasquale NARDONE \\Facult{\' e} des Sciences,
Universit\'{e} Libre de Bruxelles (ULB) \\ Bvd du Triomphe, Campus Plaine CP 231 - 1050 Brussels, Belgium\\ Emails: giorgio.sonnino@ulb.be, peeters.philippe@gmail.com\\ and pasquale.nardone@ulb.be\\
{}\\
{}\\
{}Manuscript Accepted for Publication in {\it Medical Research Archives }\\European Society of Medicine (ESMED)}
%Manuscript accepted for publication in \\ {\it Medical Research Archives, European Society of Medicine} (2022)}

\maketitle

\begin{abstract}
A severe acute respiratory syndrome coronavirus 2 (SARS-CoV-2) appeared in the Chinese region of Wuhan at the end of 2019. Since then, the virus spread to other countries, including most of Europe and USA. This work provides an overview on deterministic and stochastic models that have previously been proposed by us to study the transmission dynamics of the Coronavirus Disease 2019 (COVID-19) in Europe and USA. Briefly, we describe realistic deterministic and stochastic models for the evolution of the COVID-19 pandemic, subject to the lockdown and quarantine measures, which take into account the time-delay for recovery or death processes. Realistic dynamic equations for the entire process are derived by adopting the so-called {\it kinetic-type reactions approach}. The lockdown and the quarantine measures are modelled by some kind of inhibitor reactions where susceptible and infected individuals can be {\it trapped} into inactive states. The dynamics for the recovered people is obtained by accounting people who are only traced back to hospitalised infected people. To model the role of the Hospitals we take inspiration from the Michaelis-Menten’s enzyme-substrate reaction model (the so-called {\it MM reaction}) where the {\it enzyme} is associated to the {\it available hospital beds}, the {\it substrate} to the {\it infected people}, and the {\it product} to the {\it recovered people}, respectively. In other words, everything happens as if the hospitals beds act as a {\it catalyser} in the hospital recovery process. The statistical properties of the models, in particular the relevant correlation functions and the probability density functions, have duly been evaluated. We validate our theoretical predictions with a large series of experimental data for Italy, Germany, France, Belgium and United States, and we also compare data for Italy and Belgium with the theoretical predictions of the logistic model. We have found that our predictions are in good agreement with the real world since the onset of COVID 19, contrary to the logistics model that only applies in the first days of the pandemic. In the final part of the work, we can find the (theoretical) relationships that should be satisfied to obtain the disappearance of the virus (corresponding to a value of the {\it effective reproduction number of the infection} lower than 1).


\end{abstract}

\noindent {\bf INTRODUCTION}\label{intro}

\noindent Coronavirus disease 2019 (COVID-19) is caused by a new Coronavirus (SARS-CoV-2) that has spread rapidly around the world. Most infected people have no symptoms or suffer from mild, flu-like symptoms, but some become seriously ill and can die. In recent weeks coronavirus has had too many opportunities to spread again. After successfully tamping down the first surge of infection and death, Europe will most likely be in another coronavirus wave when we shall move in the autumn and winter seasons \cite{cacciapaglia,sonnino2}. In particular, we expect to be subject to the most recent and contagious sub-variants of Omicron. It only took about a month for BA.2.12.1, an Omicron sub-variant, to cause most of the new COVID-19 cases in the U.S. since scientists first spotted it in the country. But even newer iterations of the Omicron variant are spreading rapidly through the USA and are poised to outcompete past versions of the virus, reinfect millions of Americans, and extend the country’s current COVID-19 surge. BA.4 and BA.5 now account for more than 21$\%$ of new cases in the U.S., according to USA Centres for Disease Control and Prevention (CDC) estimates as of 11 June, 2022 \cite{ladyzhets}. These two new sub-variants evolved from the Omicron lineage to become even more contagious and can bypass immunity from a past infection or vaccination. This means people can be reinfected even if they had Omicron earlier this year \cite{ladyzhets}. The newer sub-variants can also bypass monoclonal antibody treatments, which use lab-made immune system proteins developed from earlier strains of SARS-CoV2. So, even though several vaccines for COVID-19 are actually been produced other ways of slowing its spread have to continue to be explored. One way of controlling the disease are the lockdown and the quarantine measures. The lockdown measures are emergency measures or conditions imposed by governmental authorities, as during the outbreak of an epidemic disease, that intervene in situations where the risk of transmitting the virus is greatest. Under these measures, people are required to stay in their homes and to limit travel movements and opportunities for individuals to come into contact with each other such as dining out or attending large gatherings. The lockdown measures are more effective when combined with other measures such as the quarantine. Quarantine means separating healthy people from other healthy people, who may have the virus after being in close contact with an infected person, or because they have returned from an area with high infection rates. Similar recommendations include isolation (like quarantine, but for people who tested positive for COVID-19) and physical distancing (people without symptoms keep a distance from each other). Several governments have then decided that stricter lockdown and quarantine measures are needed to bring down the number of infections. In this work we shall propose interventions which are as targeted as possible. Unfortunately, the greater the number of infections, the more sweeping the measures have to be. Tightening the measures will impact on our society and the economy but this step is needed for getting the coronavirus under control.

\noindent Up to now, while facing the same problem, albeit with different methods, the models proposed in the literature are all united by a single {\it common thread}:  the overall objectives of these works is to obtain the dynamics describing realistic situations of spread of SARS-CoV2 infection by means of macroscopic descriptions. It should immediately be said that that we can consider this task as achieved if we are able to
\begin{enumerate}
\item model the distribution of hospitals in a country;
\item model the distribution of the poles of attraction of susceptible people (e.g., shopping centres, workplaces, etc.);
\item identify a mechanism that allows to establish when a pole of attraction becomes {\it saturated} with infected people by proposing alternative poles of attraction;
\item modelling the Lockdown and the Quarantine measures adopted by the Government of the Country;
\item determine the nature of the intrinsic (i.e., spontaneous) fluctuations to which a macroscopic system is subjected, determining the correlation function by statistical mechanics.
\end{enumerate}
\noindent To our knowledge, the state-of-the-art of the current alternative techniques are unable to resolve the issues listed above. As evident from a comparison between the theoretical results and experimental data, although these models give a trend of the features exhibited by the time-series data, it hardly represents the actual trends. For instance, as the effect of latent time has not been considered, growth in active cases of infections, as predicted by the SIRD model, remains very steep. Further, as quarantine effects have not been considered, the decay predicted by the SIRD model is much slower than reality. The predicted value of total number of deaths is also much higher than actual. Hence, the SIRD model needs proper modifications to corroborate all the three data sets - infected, recovered, and dead - simultaneously. Lockdown and quarantine measures and the role of the time-delay play a significant role in the way the infection spreads over time. Hence, we need to incorporate these factors into the model. When several factors are involved simultaneously in a process, how should we proceed then? A suggestion comes to us from how physicists approached the study of the science of {\it Nonlinear Phenomena and Complex Systems}:
\begin{itemize}
\item First of all, we must realise that it is unrealistic to think to be able to describe a complex phenomenon, in a complete and exhaustive way, by setting up directly {\it macroscopic models}, that in addition are over-simplified, without any microscopic underpinning;
\item Secondly, we must accept the idea that it is not possible to take into consideration, with a single model, all the factors involved in a complex phenomenon;
\item Finally, as physicists currently do to study the dynamics of thermodynamic systems far from equilibrium, the macroscopic model that describes the dynamics of the system must be derived from fundamental processes i.e., by a {\it microscopic description}.
\end{itemize}

\noindent In our works we introduce a {\it kinetic-type reactions} (KTR) approach \cite{sonnino2,sonnino,sonnino3,sonnino4}, calibrated on the COVID-19 outbreak data in Belgium, Italy, France, Germany and USA. Here, by analogy, we are authorised to introduce the following {\it microscopic postulates}:
\begin{enumerate}
\item  \textit{The microscopic detailed balance principle is respected}. The overall  COVID-19 spreading process may be decomposed into elementary processes (contacts among individuals, or steps, or \textit{elementary reactions}). It states that at equilibrium, each elementary process is in equilibrium with its reverse process. It should be noted that this principle has important repercussions at the macroscopic level such as, for example, the validity of the reciprocity relations of the coefficients that appear in the macroscopic model.
\item \textit{The law of mass action is satisfied}. The rate at which an elementary step proceeds is directly proportional to the product of the concentrations of the {\it reactants} (in our case the "populations"). It explains and predicts behaviours of populations in dynamic equilibrium. Specifically, it implies that for a system in equilibrium, the ratio between the "reacting" populations density and the produced populations density is constant.
\item \textit{The Th. De Donder principle is satisfied} \cite{prigogine1,prigogine2}. The Th. De Donder principle establishes that a chemical reaction, however complex, can always be reduced to a finite series of elementary chemical steps. In this principle lies all the real power of the KTR approach. It is easily checked that several current models applied to
a different data set violate the Th. De Donder principle.
\end{enumerate}
\noindent It is worth recalling that, as can be easily understood, the three above axioms provide strict constraints to the coefficients appearing in the macroscopic model which, contrarily to the models described in the works illustrated above, can no longer be chosen arbitrarily. We shall see that the KTR approach is very promising and flexible. Indeed, the KRT approach
\begin{itemize}
\item models each actor by a dedicated “chemical species” that can only be created or destroyed as the result of one, or several, elementary steps, 
\item allows to determine the dynamics of the system starting from this set of elementary steps;
\item  thanks to its flexibility, allows to analyse complex situations where several variables are involved, such as $R$, $Q$, $R_h$, $I_h$ etc;
\end{itemize}
\noindent To the best of our knowledge, this approach, at fundamental level, has never been proposed in the literature.

\noindent The manuscript is organised as follows. In Section~\ref{exp} we derive the law of growth for a Malthusian population. This law foresees that, in absence of lockdown measures, in the initial phase of the pandemic the number of infected people increases exponentially. In Section~\ref{DMSection}, we study the deterministic $(SIS)_L$-model i.e., the dynamics of the compartments {\it Susceptible} $\rightarrow$ {\it Infectious} $\rightarrow$ {\it Susceptible in presence of the lockdown measures}. We shall see that as soon as the lockdown measures are stopped, the spread of the Coronavirus begins to grow back vigorously if a certain "safety threshold" is not reached. The trends of the curves related to the SARS-CoV2 infection are investigated in Subsection~\ref{an}. Section~\ref{comp} shows the comparison between the theoretical predictions of the $(SIS)_L$-model with experimental data for Italy, Belgium, USA, and France. The {\it stochastic} version of the $SIS$-model subjected to the lockdown measures is introduced and analysed in Section~\ref{SM}. In particular, we determine the solutions of the stochastic differential equations governing the dynamics of the infected people for Italy, USA and France and the behaviour of the relevant correlation functions for Italy. The role of the Hospitals is initially investigated in Section~\ref{Hs} by a very simple model referred to as the $(SISI_h)_L$-{\it model}. In Section~\ref{kinetic} we develop a more realistic model to study the spreading of the SARS-CoV-2 that takes into account the role of the Hospitals as well as the lockdown and quarantine measures. This model is obtained by a {\it kinetic-type reactions} (KTR) approach. More specifically, in the KTR model, the lockdown and quarantine measures are modelled by a very simple {\it door function} (see Subsection~\ref{LQM}). The dynamics of the hospitalised individuals (i.e., the infectious, recovered, and deceased people) can be found in Subsection~\ref{H}. The corresponding evolution equations are obtained by considering the {\it Michaelis-Menten’s enzyme-substrate reaction model} (the so-called MM reaction). The equations governing the dynamics of the full process and the related {\it basic reproduction number} are reported in Section~\ref{TODEs} and Section~\ref{BRN}, respectively. Finally, Section~\ref{Applications} shows the good agreement between the theoretical predictions with real data for Belgium, France and Germany. The (theoretic) relations that must be satisfied for achieving the total disappearance of the SARS-CoV2 Infection are shown in Section~\ref{dp}. The perspectives of this work, in particular the applicability of our kinetic type approach for treating the dynamics of the emerging BA.4 and BA.5 sub-variants of Omicron can be found in Section~\ref{C}. Concluding remarks are reported in Section~\ref{D}.

\section{Modelling the onset of the SARS-CoV2 Pandemic }\label{exp}

\begin{figure}
\centering\centering\includegraphics[scale=.35]{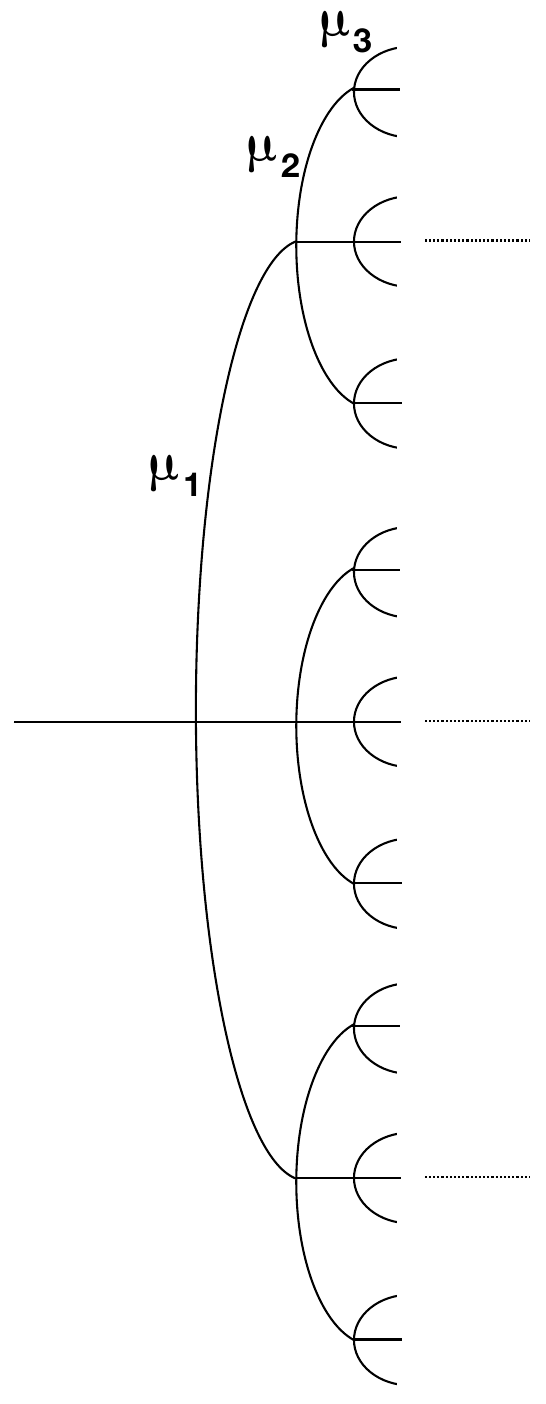}
\caption{
{\it Schematic dynamics of respiratory virus in the absence of the lockdown measures} In this graphics, for illustrative purpose only, we set $R_0=3$. However, for SARS-CoV-2, the value of $R_0$ is $2$ even at the beginning of the outbreak in China and Italy.  After a period of time $\mu_1$, an infected individual can infect $R_0$ other individuals. In turn, after a period $\mu_2$, each of these newly infected individuals can infect other $R_0$ people, and so on. After $n$ steps the elapsed time is $t=\sum_{i=1}^n\mu_i$.
}
\label{Fig.1}
\end{figure}

\begin{figure*}[htb]
  \hfill
  \begin{minipage}[t]{.48\textwidth}
    \centering
    \includegraphics[width=5cm,height=5cm]{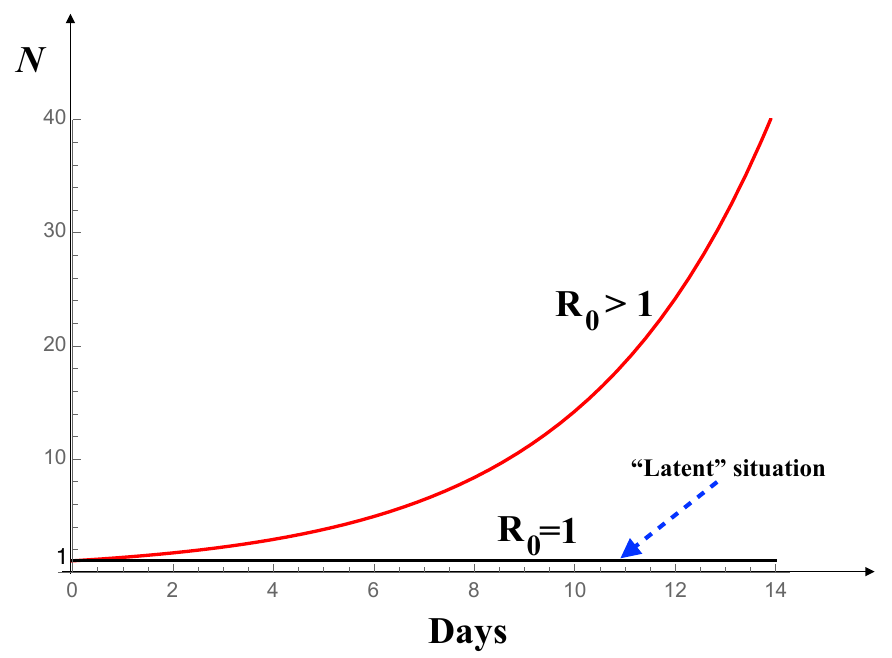}
    \caption{{\it Situation before the lockdown measures}. Number of infected people corresponding to the exponential law. The red line represents the case $R_0>1$, such as the situation before the adoption of lockdown measures. The black line corresponds to the case $R_0=1$, the {\it latent situation} in which the virus is substantially ineffective.}
    \label{Fig.2}
  \end{minipage}
  \hfill
    \begin{minipage}[t]{.48\textwidth}
      \centering
      \includegraphics[width=5cm,height=5cm]{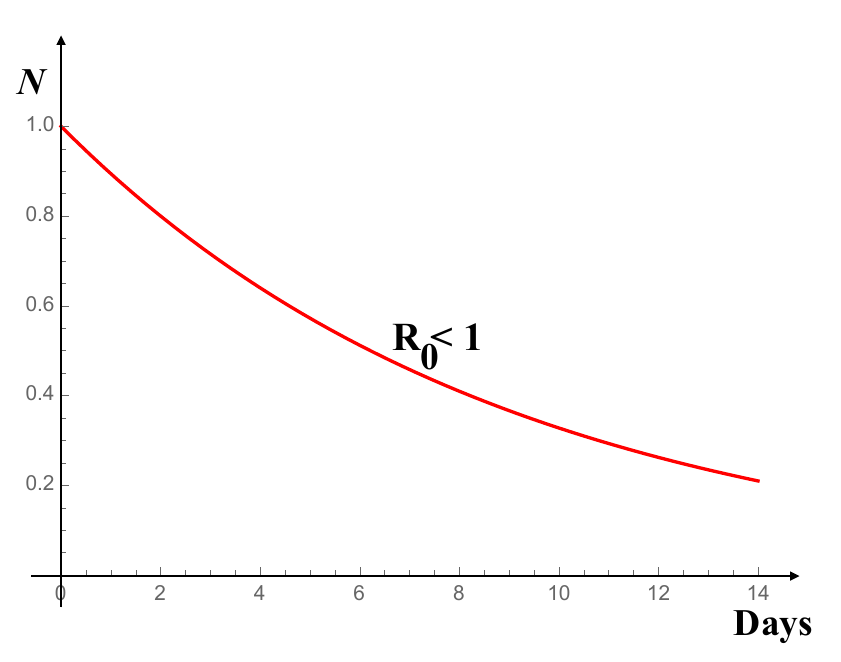}
      \caption{Number of infected people corresponding to the exponential law. The red line represents the case $R_0<1$. In this situation the number of infected people decreases exponentially and the virus disappears after a few weeks.}
      \label{Fig.3}
    \end{minipage}
  \hfill
\end{figure*}

We start by introducing the definition of the {\it basic reproduction number of an infection} $R_0$, defined as {\it the number of infected people derived from a first case in a population where all the others are susceptible}. So, it is not possible to modify $R_0$, in any case, but it is possible to get a different effective $R$\footnote{More rigorously, in epidemiology, {\it the basic reproduction number of an infection, $R_0$, is the expected number of cases directly generated by one case in a population where all individuals are susceptible to infection in absence of any deliberate intervention in disease transmission} (see, for example, \cite{hystory}).}. This parameter is strictly linked to the {\it replication time of a virus}, indicated with $\mu_i$, defined as \textit{the time interval after which the number of infected people has increased by $R_0$ times}. Fig.~\ref{Fig.1} schematically represents the diffusion dynamics of the virus. 
%%%%%%%%%%%%%%%%%%%%%%%%%%%%%%%%%%%%%%%%%%
%%%%%%%%%%%%%%%%%%%%%%%%%%%%%%%%%%%%%%%%%
\noindent By indicating with $I$ the number of infected people, after $n$ steps we get\footnote{In this Section we shall follow the definitions and the expressions reported in standard books such as, for example, \cite{murray}.}:
\begin{equation}\label{In1}
N=R_0^n
\end{equation}
\noindent Of course, after $n$ steps, the elapsed time is $t = \sum_{i=1}^n\mu_i$ and, if there are $M$ outbreaks of infectious viruses, Eq.~(\ref{I1}) can be cast into the form\footnote{Actually, Eq.~(\ref{In2}) applies only if the $M$ outbreaks of the virus are exactly at the same conditions. In general, the correct expression reads $N=\sum_{i=1}^MR_0^{t/{\widetilde \mu}_i}$, with ${\widetilde \mu}_i$ indicating the replication time of the virus for the $i$-$th$ outbreak.}
\begin{equation}\label{In2}
I=M R_0^{t/\mu}
\end{equation}
\noindent with $\mu\equiv 1/n\sum_{i=1}^n\mu_i$. Note that the two parameters $R_0$ and $\mu$ are not independent (see, for example, \cite{anderson,population,gyorgy})\footnote{In ref.~\cite{anderson}, the doubling time is used to calculate $R_0$, by means of the equation $R_0=1+(\gamma+\rho)log(2)/\mu$ where $\gamma$ is the duration of the incubation period, $\rho$ is the duration of the symptomatic period, and $\mu$ is the doubling time (see \cite{anderson}). In this respect, we would also like to mention another excellent work recently produced by G. Steinbrecher~\cite{gyorgy}.}. It is more convenient to work in the Euler base $e$ rather than in base $R_0$; in the Euler base Eq.~(\ref{I2}) provides the {\it law of growth for a Malthusian population} \cite{murray}.
\begin{equation}\label{In3}
I=M\exp(t/\tau)\quad\textrm{ where}\ \ \tau=\frac{\mu}{\log(R_0)}
\end{equation}
\noindent In literature, $\tau$ is referred to as the {\it characteristic time of the exponential trend}.  So, in the absence of containment measures the number of infected people follows the exponential law~(\ref{In3}). Let us now analyse Eq.~(\ref{In3}) more in depth. We have three possible scenarios:
\begin{enumerate}
 \item $R_0>1$ (as is the current world's situation). For Italy, for example, before the adoption of (severe) containment measures, the value of $\tau$ was about $\tau\sim 3.8$ days (and $\mu\sim 2.6$ days). In this case the number of the infected people increases exponentially.
\item $R_0=1$ If the infection-capacity of the virus is of the type {\it one-to-one} (i.e., a person infected by SARS-CoV-2 can in turn infects only another person), we get the stationary situation corresponding to $I = 1$. This situation is referred to as the {\it latent situation} i.e., the virus is still present but does not spread. In this limit case, the SARS-CoV-2 is substantially ineffective. Scenarios (1) and (2) are illustrated in Fig.~\ref{Fig.2}.
\item $0<R_0<1$. We may also imagine that the capacity of infection of SARS-CoV-2 is less than $1$. This means that the virus is no longer able to be spread (e.g., thanks to protective measures, or to the production of vaccines and anti-virals, or because people who overcame the disease became immune. In this case, the value of $\tau$ is negative and the number of infected people decreases ever time. That is, the infection eventually disappears. The rate of decrease of the number of infected people depends on the value of $\tau$. This scenario is depicted in Fig.~\ref{Fig.3}.
\end{enumerate}
%%%%%%%%%%%%%%%%%%%%%%%%%%
%%%%%%%%%%%%%%%%%%%%%%%%%%%%

\subsection{Comparison with the Real Data for COVID-19 before the Lockdown Measures}
It is understood that the main objective of the lockdown measures established by most European governments and health organisations is to reduce the ability of a virus to spread. From a mathematical point of view, we would like to have $R_0=1$ (or, better, $R_0<1$), in Eq.~(\ref{In3}) instead of $R_0>1$. In practical terms, this means reducing the frequency of all involuntary contacts with a large number of people, reducing unnecessary movements to avoid encounters, and to prolong the closure of schools. Although these measures cannot prevent the spread of the infection in the long term, they can reduce the number of new infections daily. This has the benefit of leaving room for seriously-ill patients by avoiding to overload the healthcare system. We can easily realise what are the consequences if the lockdown measures are not set up. To make a comparison between the theoretical predictions and the experimental data in absence of lockdown measures, we have to consider the correct reference period. More specifically, we saw that the number of positive cases grows in the course of time by following the law~{(\ref{In3}). Hence, at the reference time $t_0$, the number of people infected by the virus is
\begin{equation}\label{In4}
N_0=M\exp(t_0/\tau)
\end{equation}
\noindent After a period of time, say $t$, Eq.~(\ref{In3}) reads
\begin{equation}\label{In5}
N=M\exp(t/\tau)
\end{equation}
\noindent Hence,
\begin{equation}\label{In6}
N=N_0\exp((t-t_0)/\tau)
\end{equation}
\noindent Eq.~({\ref{In6}) is the equation that we use for comparing the mathematical predictions with experimental data during the initial phase where the spread of SARS-CoV-2, causing the COVID-19, follows the exponential law, and $(t-t_0)$ is our \textit{reference period}. For the case of COVID-19 we get (see, for example,~\cite{murray})
\begin{itemize}
\item All infectious outbreaks are exactly at the same conditions. So, Eq.~(\ref{In2}) applies;
\item $R_0= 2$;
\item All the $\mu_i$ are equal with each other: $\mu_i=const=\mu$ (see also \cite{murray,anderson}).
\end{itemize}
\noindent In this case, $\mu$ is referred to as the {\it doubling time}. So, the {\it doubling time is the amount of time it takes for a given quantity to double in size or value at a constant growth rate} \cite{population}. If we do not apply the locking measures, the evolution in the course of time of the number of infected people is best approximated by an exponential curve with $R=2$, even though we have to stress that $R_0$ is only associated with the beginning of the epidemic and, with certain approximations, with the early stages, but not beyond. Fig.~\ref{Fig.6} and Fig.~\ref{Fig.7} respectively show the comparison between the theoretical predictions and the experimental data for Italy and Belgium before the lockdown measures. We get $\tau\simeq 3.8$ days and $\mu\simeq 2.6$ days for Italy, and $\tau\simeq 5.2$ days and $\mu\simeq 3.7$ days for Belgium. We conclude this Introduction by mentioning that there are several methods currently proposed in Literature to derive by mathematical models, the value of $R_0$. For example, in ref.~\cite{gyorgy}, we can find a short numerical code, written in $R$-programming language for statistical computing and graphics, able to compute the estimated $R_0$ values for the following 17 infectious diseases: {\it Chickenpox (varicella) (Transmission: Aerosol), Common cold (Transmission: Respiratory Droplets), COVID-19 (Transmission: Respiratory Droplets), Diphtheria (Transmission: Saliva), Ebola - 2014 Ebola outbreak (Transmission:: Body fluids, HIV/AIDS (Transmission: Body fluids), Influenza - 1918 pandemic strain (Transmission:: Respiratory Droplets), Influenza - 2009 pandemic strain (Transmission: Respiratory Droplets, Influenza - seasonal strains (Transmission: Respiratory Droplets), Measles (Transmission: Aerosol), MERS (Transmission: Respiratory Droplets), Mumps (Transmission: Respiratory Droplets), Pertussis (Transmission: Respiratory Droplets), Polio (Transmission: Fecal oral route), Rubella (Transmission: Respiratory Droplets), SARS (Transmission: Respiratory Droplets), Smallpox (Transmission: Respiratory Droplets)}. However, this task is particularly problematic if there are intermediate vectors between hosts, such as malaria. 
%%%%%%%%%%%%%%%%%%%%%%%%%%
\begin{figure*}[htb]
  \hfill
  \begin{minipage}[t]{.45\textwidth}
    \centering
    \includegraphics[width=5cm,height=5cm]{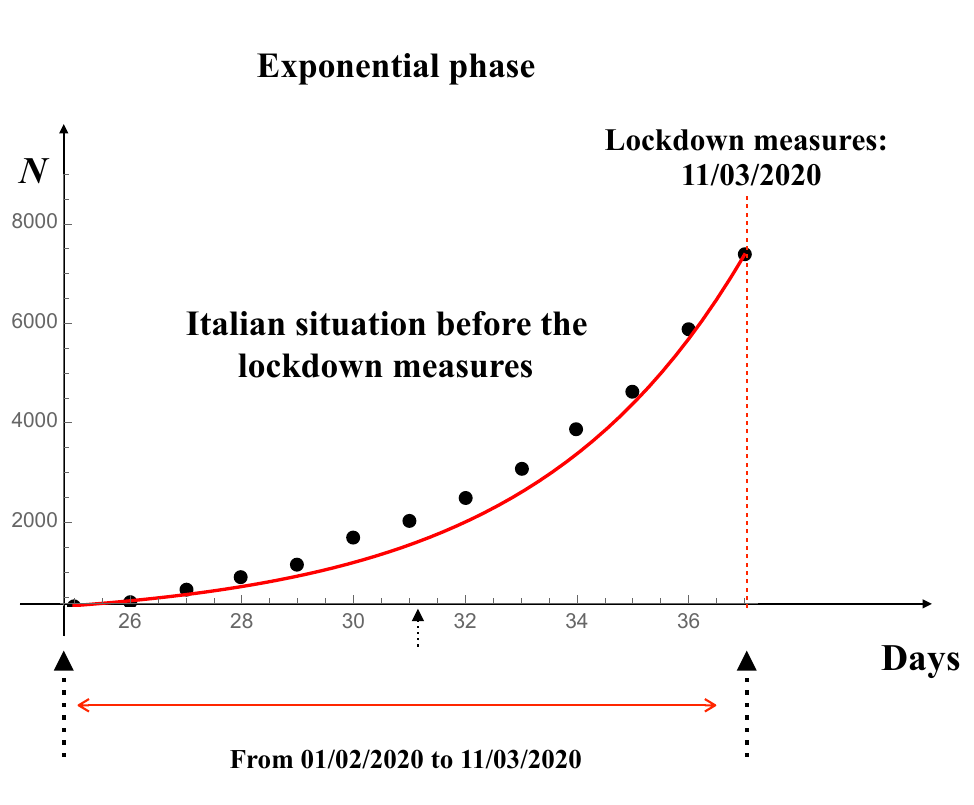}
    \caption{\textit{Number of infected people in Italy on the 10th of March 2020 (before the adoption of lockdown measures)}. The blue line corresponds to the theoretical predictions and the black dots correspond to experimental data. The values of the parameters $\tau_{IT}$ and $\mu_{IT}$ are $\tau_{IT}\simeq 3.8$ days and $\mu_{IT}\simeq 2.6$ days, respectively.}
    \label{Fig.6}
  \end{minipage}
  \hfill
  \begin{minipage}[t]{.45\textwidth}
    \centering
    \includegraphics[width=5cm,height=5cm]{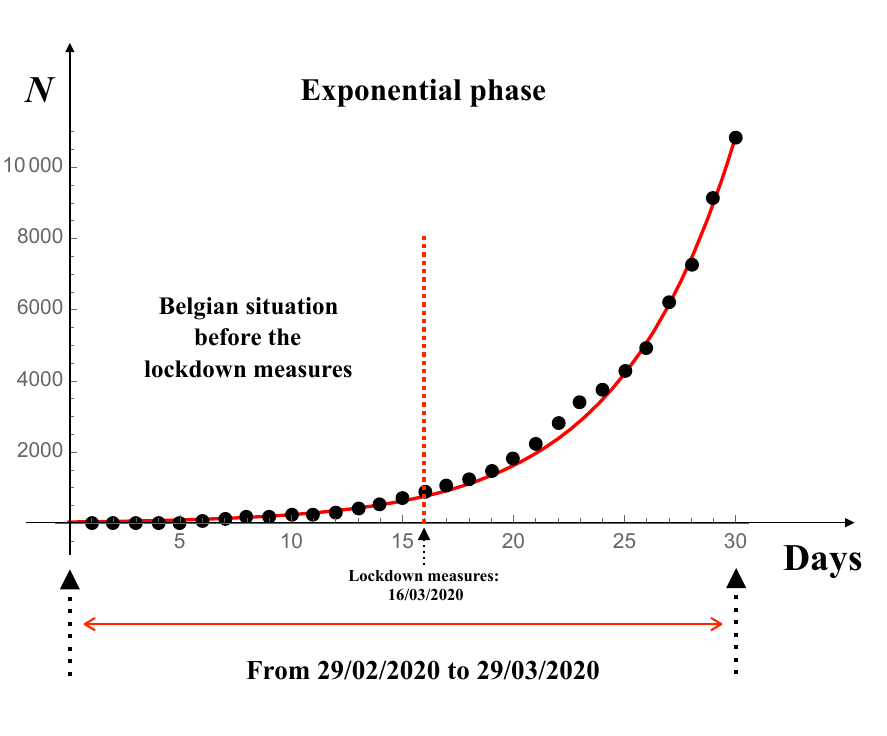}
    \caption{\textit{Exponential phase in Belgium}. The lockdown measure have been adopted on the 16the of March 2020 (however, initially not so strict as in Italy).The red line corresponds to the theoretical predictions and the black dots correspond to experimental data. The values of the parameters $\tau_{BE}$ and $\mu_{BE}$ are $\tau_{BE}\simeq 5.3$ days and $\mu_{BE}\simeq 3.7$ days, respectively.}
    \label{Fig.7}
  \end{minipage}
  \hfill
\end{figure*}
%%%%%%%%%%%%%%%%%%%%%%%%%%%%

\section{A Simple Model in Presence of the Lockdown Measures - The $(SIS)_L$-Model}\label{DMSection}
\subsection{A Simple Deterministic $(SIS)_L$-Model}\label{DMs}
\noindent The current work starts from the following hypothesis commonly supported by the most accredited virologists: \textit{the SARS-CoV-2 behaves like other viruses which cause respiratory diseases} (see, for instance,~\cite{ronchetti}). The common cold and influenza, do not confer any long-lasting immunity. Such infections do not give immunity upon recovery from infection, and individuals become susceptible again. Hence, according to the above-cited hypothesis we propose the following simplest compartmental model:
\begin{equation}\label{DM0}
S +I \xrightarrow{\mu} 2I
\end{equation}
\noindent In our model the SARS-CoV-2 infection does not leave any immunity, thus individuals return back into the $S$ compartment. Hence, infectious people, after recovery, return back to the compartmental $S$. This added detail can be shown by including an $R$ class in the middle of the model \cite{sonnino}
\begin{align}\label{DM1}
&S +I \xrightarrow{\mu} 2I\\
&I \xrightarrow{\gamma} R\xrightarrow{\gamma_1} S\nonumber
\end{align}
\noindent In this work, scheme (\ref{DM1}) has to be interpreted as follows. The entire process is described by adopting a \textit{kinetic-type reactions} approach where the \textit{lockdown measures} are modelled by some kind of \textit{inhibitor reactions} where \textit{susceptible individuals} can be \textit{trapped} into \textit{inactive states}. In addition, the \textit{substrate} is associated to the \textit{infected people} and the \textit{product} to the \textit{recovered people}, respectively. From scheme~(\ref{DM1}), we get O.D.E.s for $S$, $R$, and $I$: 

\begin{align}\label{DM2}
&\frac{dS}{dt}=-\sigma\frac{S}{N_{Tot.}} I+\gamma_ 1R\\
&\frac{dI}{dt}=\sigma\frac{S}{N_{Tot.}} I-\gamma I\nonumber\\
&\frac{dR}{dt}=\gamma I-\gamma_1 R\nonumber
\end{align}
\noindent with $N_{Tot.}$ denoting the total population and $\sigma\equiv\mu N_{Tot.}$. By assuming that the dynamics of $R$ is much faster that those of $S$ and $I$, we may set $d R/dt\simeq 0$ and system~(\ref{DM2}) reduces to\footnote{This assumption is not strictly necessary, but we adopt it just to show that it is possible to obtain relevant results in an analytical way, without having to solve the equations numerically. We may easily convince ourselves that it is not difficult to perform (numerically) the same calculations below relaxing this hypothesis.}
\begin{align}\label{DM3}
&\frac{dS}{dt}\simeq-\sigma\frac{S}{N_{Tot.}} I+\gamma I\\
&\frac{dI}{dt}=\sigma\frac{S}{N_{Tot.}} I-\gamma I\nonumber
\end{align} 
\noindent which corresponds to the model
\begin{align}\label{DM4}
&S +I \xrightarrow{\mu} 2I\\
&I \xrightarrow{\gamma} S\nonumber
\end{align}
\noindent In literature, the model~(\ref{DM4}) is referred to as the $SIS$-{\it model} (see, for example, \cite{SIS}). From Eq.~(\ref{DM3}) we get the conservation relation
\begin{equation}\label{DM5}
\frac{dS}{dt}+\frac{dI}{dt}=0\qquad{\rm or}\quad S+I=N_{Tot.}
\end{equation} 
\noindent Hence, the dynamics of infectious is governed by the logistic model
\begin{equation}\label{DM6}
\frac{dI}{dt}=(\sigma-\gamma)I\left(1-\frac{\sigma}{N_{Tot.}(\sigma-\gamma)}I\right)
\end{equation} 
\noindent or
\begin{align}\label{DM7}
&\frac{dI}{dt}={\tilde\alpha} I\left(1-\frac{I}{K}\right)\qquad{\rm with}\\
&{\tilde\alpha}\equiv\sigma\left(1-\frac{\gamma}{\sigma}\right)\quad;\quad K\equiv N_{Tot.}\left(1-\frac{\gamma}{\sigma}\right)\nonumber
\end{align} 
\noindent where $\tilde\alpha$ and $K$ denote the {\it linear growing rate of the COVID-19} and the {\it carrying capacity}, respectively. 

\noindent The lockdown measures are mainly based on the isolation of the susceptible individuals, eventually with the removal of infected people by hospitalisation. In our model, the effect of the lockdown measures are taken into account by introducing in the $SIS$-model the {\it lockdown-induced decrease rate} $c(t)$
\begin{align}\label{DM8}
&S +I \xrightarrow{\mu} 2I\\
&I \xrightarrow{\gamma+c(t)} S \nonumber
\end{align}
\noindent where
\begin{align}\label{DM9}
&c(t_L)=0\qquad {\rm for} \quad t=t_L\\
&c(t)>0\qquad\ \ {\rm for} \quad t>t_L\nonumber
\end{align}
\noindent in which $t_L$ denotes the time when the lockdown measures are applied. The meaning of "{\it kinetic reactions}"~(\ref{DM8}) is the following. Clearly, the lockdown measures act on the susceptible people but they are unable to affect the "{\it chemical capacity}" (i.e. to decrease the infection capacity) of the Coronavirus (therefore the value of the "{\it kinetic reaction}" $\mu$ must remain constant). However, the final effect of the lockdown measures is to increase the number of susceptible people "at the expense" of the infected people (since $S+I= const.$). According to the model, this can be done only by increasing the value of the "{\it kinetic constant}" $\gamma$\footnote{Anyhow, by writing the corresponding O.D.E.s, it is easy checked that, if we assume that the effect of the lockdown is to decrease the value of the "{\it kinetic constant}" $\mu$, we would get the no-sense result that the carrying capacity of the infected people increases in time.}. The corresponding deterministic differential equations for the COVID-19 model in presence of the lockdown measures reads then:
\begin{align}\label{DM10}
&\frac{dS}{dt}=-\sigma\frac{S}{N_{Tot.}} I+\gamma I+c(t)I\\
&\frac{dI}{dt}=\sigma\frac{S}{N_{Tot.}} I-\gamma I-c(t)I\nonumber
\end{align} 
\noindent Scheme~(\ref{DM8}) may be referred to as the $(SIS)_L$-\textit{model} where $L$ stands for $Lockdown$. The general expression for the lockdown-contribution, $c(t)$, may be obtained as follow.

\noindent {\bf 1)} At $t = t_L$, $c (t_L) = 0$. This corresponds to the requirement that the lockdown measures start at time $t = t_L$;

\noindent {\bf 2)} $c (t)$ must always be a positive function. Notice that this requirement must be verified for all $t > t_L$ and for {\it all kind of scenarios}. This imposes several restrictions on the coefficients;

\noindent {\bf 3)} $ c (t)$ must be a function able to bend, downwardly, the trend of the curve of infected people. Even this requirement must be verified for all $t > t_L$ and for {\it all kind of scenarios}.

\noindent Afer that, we may proceed as usual: we look for a class of functions satisfying the three requirements above under the form of polynomials (or {\it fractions of polynomials}) by bearing in mind that the degrees of the polynomials must chosen such that they are able to bend downwardly the curve of the infected people. This requirement must be satisfied by {\it each term of the series}. 

\noindent {\bf 4)} Ultimately, we have to prove that the obtained series constitutes a complete base of functions\footnote{Here, we shall not discuss on the completeness of the basis functions $\{ {\hat t}^{i+1}-{\hat t}^{-j}\}_{i,j=0,1,\dots}$.}. This will be subject of a future work. 

\noindent Finally, the general expression for the lockdown-contribution, $c(t)$, may be cast into the form \cite{sonnino}
\begin{equation}\label{DM11}
c({\hat t})=\sum_{i,j=0}^\infty\beta_{ij}\left({\hat t}^{i+1}-{\hat t}^{-j}\right)\qquad{\rm with}\quad {\hat t}\geq 1
\end{equation} 
\noindent with ${\hat t}$ denoting the \textit{normalised time} ${\hat t}\equiv t/t_L$ and $\beta_{ij}$ are real numbers, subject to the condition $c({\hat t})>0$ for ${\hat t}>1$, respectively. Note that expression~(\ref{DM11}) satisfies the three requirements above. We are looking for a lockdown expression that takes into account only the most relevant terms. So, the first two terms of expression~(\ref{DM11}) are
\begin{equation}\label{DM12}
c({\hat t})\simeq \beta_{00}({\hat t}-1) +\beta_{01} \left(\frac{{\hat t}^2-1}{\hat t}\right)\qquad{\rm with}\quad {\hat t}\geq 1
\end{equation} 
\noindent In refs~ \cite{sonnino1,sonnino2,mathur}) we can find huge works of fittings with the experimental data carried out for several Countries and for different SARS-type pathologies. The results of the fittings have shown that the most relevant contribution is expressed by the term $(t^2-t^2_L / t)$ while the term $(t-t_L)$ provides an irrelevant contribution in comparison with the previous one\footnote{Of course, there are situations where the term $(t-t_L)$ turns out to be significant e.g., in {\it chemotherapy} that is started with a {\it log-kill effect}. In this case c(t) is the therapy-induced death rate, which is modelled by the term $(t-t_C)$ since it is imposed, and {\it finely calibrated}, from the outside}. Finally, we get
\begin{equation}\label{DM13}
c({\hat t})\simeq \beta\left(\frac{{\hat t}^2-1}{\hat t}\right)\qquad{\rm with}\quad\beta>0\quad {\rm and}\quad {\hat t}\geq 1
\end{equation}
\noindent with $\beta\equiv \beta_{01}=const.$ denoting the {\it intensity of the lockdown measures}. Eq.~(\ref{DM13}) is the expression for the lockdown measures that will be considered in this work. Notice that Eq.~(\ref{DM13}) generalises the lockdown term introduced in ref.~\cite{sonnino2}. By taking into account Eq.~(\ref{DM7}), the deterministic differential equation for the infectious people, in presence of the lockdown measures, reads
\begin{equation}\label{DM14}
\frac{dI}{d{\hat t}}=\alpha I\left(1-\frac{I}{K}\right)-\beta\left(\frac{{\hat t}^2-1}{\hat t}\right) I
\end{equation}
\noindent with $\alpha\equiv{\tilde\alpha} t_L$. The exact solution of Eq.~(\ref{DM14})can be brought into the form \cite{sonnino}
\begin{equation}\label{DM16}
I=\frac{I_0{\hat t}^\beta\exp((1-\alpha/\beta)^2/\sigma)\exp(-({\hat t}-\alpha/\beta)^2/\sigma)}{1+(I_0\alpha/K)\exp((1-\alpha/\beta)^2/\sigma)\int_1^{\hat{t}}x^\beta\exp(-(x-\alpha/\beta)^2/\sigma)dx}
\end{equation} 
\noindent with $I_0$ denoting the value of the total cases at the time when the lockdown measures are applied, i.e. $I_{{\hat t}=1}=I_0$. Notice that for large values of the carrying capacity, Eq.~(\ref{DM16}) tends to the expression
\begin{equation}\label{DM17}
I\simeq I_0{\hat t}^\beta\exp((1-\alpha/\beta)^2/\sigma)\exp(-({\hat t}-\alpha/\beta)^2/\sigma)
\end{equation}
\noindent Fig.~\ref{N_IT}. shows two solutions of the model~(\ref{DM14}) for Italy, during the first wave of infection by SARS-CoV-2. The red dotted line refers to the solution without the lockdown measures and the dark dotted line to the solution when the lockdown measures were applied.
%%%%%%%%%%%%%%%%%%%%%%%%%%%%%%%%%%%%%%%%%%
\begin{figure}[hbt!]
\centering\centering\includegraphics[scale=.40]{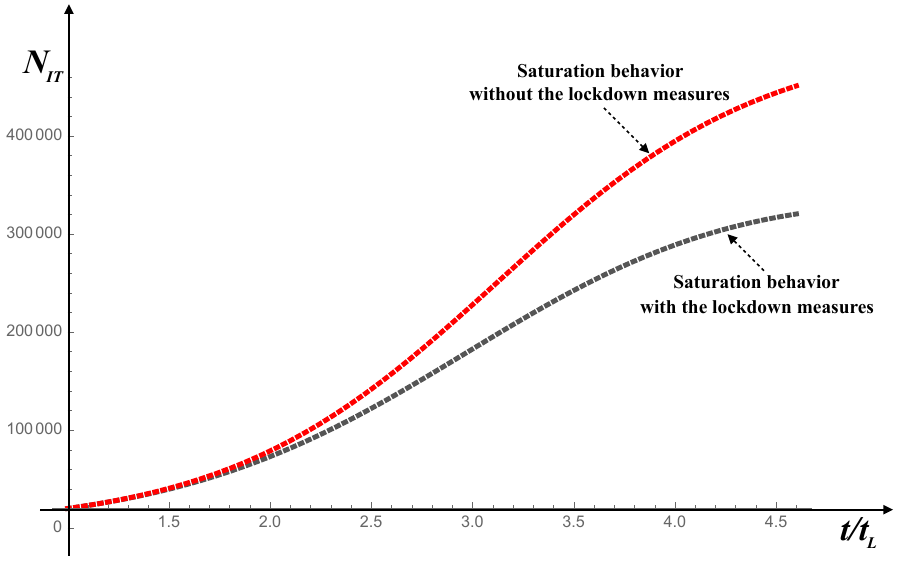}
\caption{
\textit{{\bf Solution of Eq.~(\ref{DM14}) for Italy, first wave of infection by SARS-CoV2.} The black and the red dotted lines refer to the solutions of Eq.~(\ref{DM14}) with and without the application of the lockdown measures, respectively. The values of the parameters are $I_0$=5000, $K=150000$, $\alpha=1.5$, and $\beta=0.1$, respectively.} 
}
\label{N_IT}
\end{figure}
%%%%%%%%%%%%%%%%%%%%%%%%%%%%%%%%%%%%%%%%%%

\subsection{Analysis of SARS-CoV2 Infection Curves}\label{an}
The aim of this Section is to report a summary of the main characteristics observed by analysing the trends of the curves for the (total) infectious capacity of SARS-CoV2. To achieve this goal, it is necessary to consider all categories affected by the virus, i.e., currently infected individuals, people who were previously infected and subsequently recovered, and people who died of SARS-CoV2 infection. Let $N$ denote the sum of the individuals belonging to the aforementioned compartments. We can find \cite{sonnino2} a preliminary study of the dynamics of $N$. Here, we report the main conclusions and the comparison between theoretical predictions and experimental data. For Italy and Belgium one observes two distinct phases related to the dynamics of the COVID-19, which we may classify as {\it period before the adoption of the lockdown measures} and {\it period corresponding to a few days after the adoption of the lockdown measures}. The question therefore arises whether these two periods are separated by a well-defined transition. We will see that this is the case. Indeed, we can identify three different periods, which may be classified as follows:
\vskip0.1cm
%%%%%%%%%%%%%%%%%%%%%%%%%%
\begin{figure*}[t]
  \hfill
    \begin{minipage}[t]{.45\textwidth}
    \begin{center}
      \includegraphics[width=5cm,height=5cm]{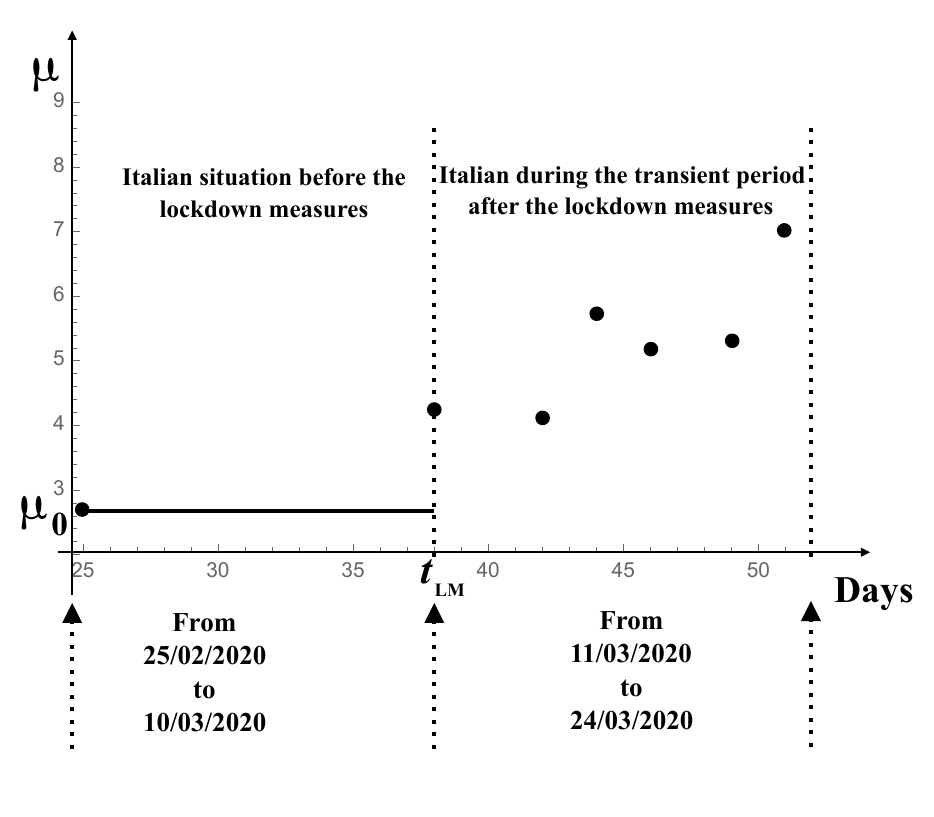}
      \caption{\textit{Italian transient period (from the 10th of March 2020 to the 24th of March 2020)}. During this period, the doubling time $\mu$ oscillates over time. $\mu_0$ indicates the (constant) doubling time during the exponential period (for Italy $\mu_0\simeq 2.6$ days).}
      \label{Fig.8}
    \end{center}
  \end{minipage}
  \hfill
  \begin{minipage}[t]{.45\textwidth}
    \begin{center}
      \includegraphics[width=5cm,height=5cm]{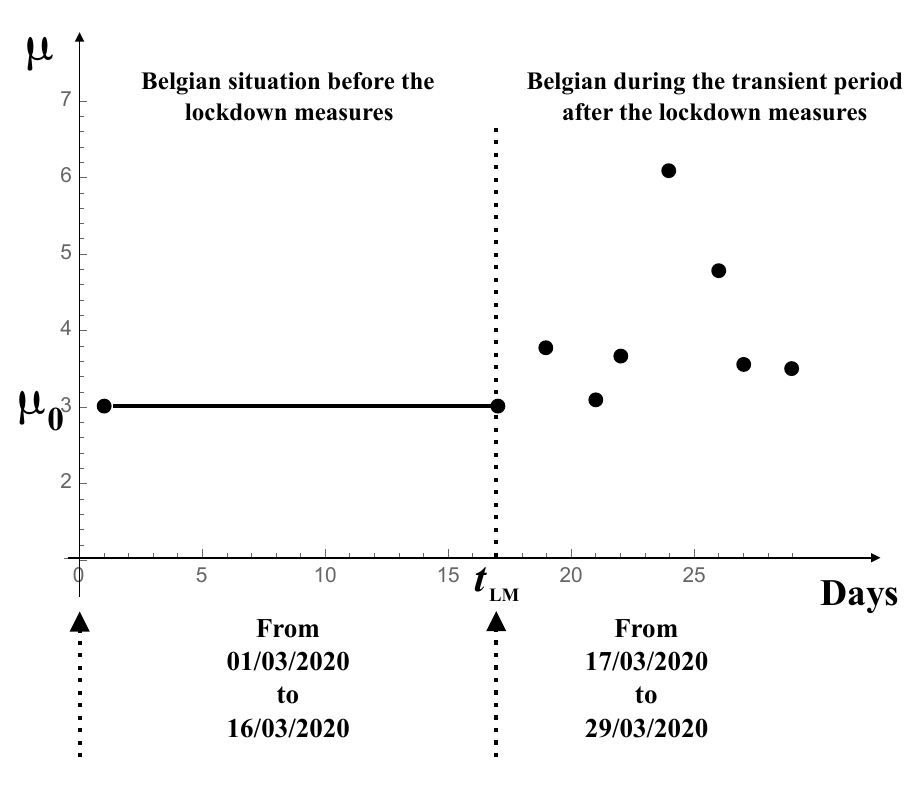}
      \caption{\textit{Belgian transient period (from the 17th of March 2020 to the 29th of March 2020)}. During this period, the doubling time $\mu$ oscillates over time. $\mu_0$ indicates the (constant) doubling time during the exponential period (for Belgium $\mu_0\simeq 3.7$ days).}
      \label{Fig.9}
    \end{center}
  \end{minipage}
  \hfill
\end{figure*}
%%%%%%%%%%%%%%%%%%%%%%%%%%%%
\noindent
\begin{enumerate}

\item The {\it exponential period}. As seen in Section~\ref{exp}, before the adoption of lockdown measures, the \textit{exponential trend} is the intrinsic behaviour of the grow rate of the COVID-19. In this period the doubling time $\mu$ is a {\it constant parameter} versus time.\noindent
\item The \textit{transient period}. The transient period {\it starts} after immediately having applied the severe lockdown measures. Fig.~\ref{Fig.8} and Fig.~\ref{Fig.9} show the behaviour of the parameter $\mu$ versus time for Italy and Belgium, respectively. 
\item The {\it bell-shaped period} (or the {\it post-transient period}). In the bell-shaped period parameter $\mu$ is a (typical) function of time. Several theoretical models can be used to investigate the post-transient period (e.g., by using the logistic model~(see, the logistic model \cite{wiki}) or Gompertz's law~\cite{gompertz}). 
\end{enumerate}
\noindent In this Section, we shall compare the experimental data with two theoretic solutions: the solution of the differential equation for the compartment $N$ \cite{sonnino2} and the {\it logistic function} \cite{hilbe}. Notice that the number of free parameters of these two models are exactly the same. More specifically,
\begin{description}
\item{{\bf a)}} The logistic model possesses two free parameters : {\it the carrying capacity} $K$ and the {\it time} $t_{0L}$ {\it when the lockdown measures were applied}. Notice that the {\it grow rate} is not free since it is linked to the doubling time $\mu$;
\item{{\bf b)}} Also our model possesses two free parameter: {\it the carrying capacity} $K_N$ and $t_0$, the {\it time value of the sigmoid's point}. 
\end{description}
\noindent  Figures~(\ref{Fig.10}) and (\ref{Fig.11}) compare the predictions of our model (blue lines) and the {\it logistic model} (red lines) with experimental data for Italy and Belgium, respectively. We recall the expression of the {\it logistic function}:
\begin{equation}\label{log1}
N_L=\frac{K}{1+\exp(-\alpha(t-t_{0L})}
\end{equation}
\noindent Experimental data for Italy have been found in \cite{sole24ore} and for Belgium in \cite{sciensano} and \cite{covid-19wiki}, respectively. They are updated to the 15th of May 2020. The values of the parameters $\tau$, $K_N$, and $t_0$ for the solution and the parameters $\tau$, $t_{0L}$ and $K$ for the logistic function are shown in the figure captions \cite{sonnino2}. 
%%%%%%%%%%%%%%%%%%%%%%%%%%%%%%%%%%%%%%%%%%
\begin{figure}[th!]
\centering\centering\includegraphics[scale=.40]{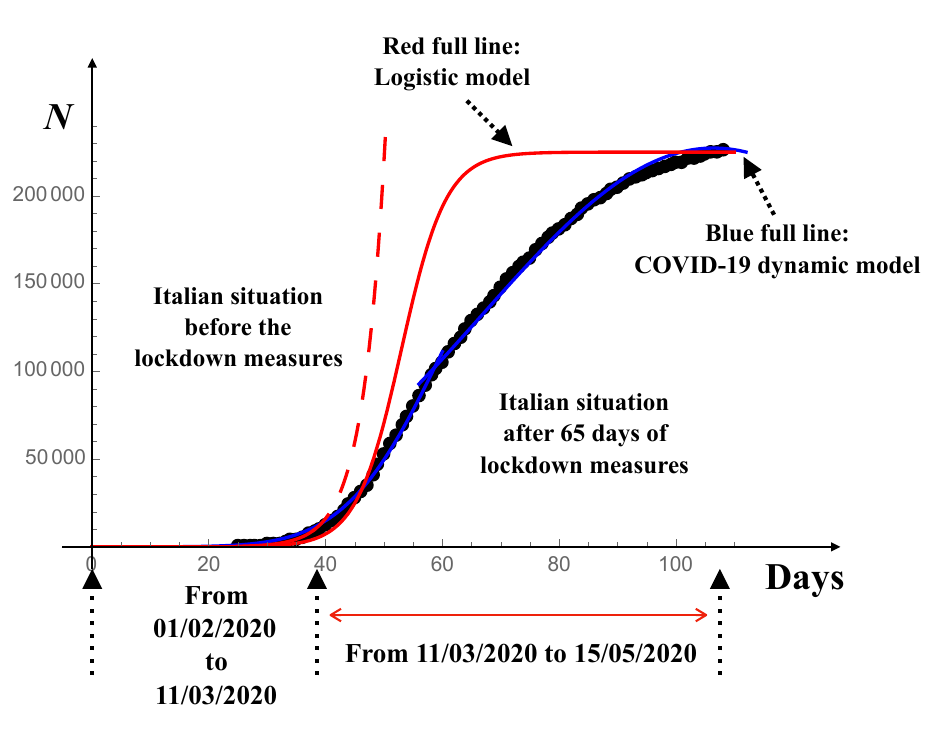}
\caption{\textit{Situation in Italy on 15 May 2020---before, and 65 days after, the adoption of lockdown measures}. The black dots correspond to experimental data. The red dotted line corresponds to the situation in Italy before the adoption of the lockdown measures. The blue and the red solid lines correspond to the theoretical predictions for Italy according to the solution of the differential equation for the compartment $N$ \cite{sonnino2} and the logistic model, respectively. Solution of the differential equation for the compartment $N$ fits well all the experimental data from the initial days (i.e., from the 1st of February 2020) \cite{sonnino2}, while the logistic model applies only to the first days. The values of the parameters of solution in Sonnino {\it et al.} \cite{sonnino2} and the logistic function are: $\tau_{IT}\simeq 3.8$ days ($\mu_{IT}=2.6$ days), $K_N^{IT}\simeq 355250$, and $t_{0IT}\simeq 72.5$ days for the solution in Sonnino {\it et al.} \cite{sonnino2}, and $\tau_{IT}\simeq 3.8$ days ($\mu_{IT}=2.6$ days), $K_{IT}=225000$, $t_{0LIT}=53$ days for the Logistic function.}
  \label{Fig.10}
\end{figure}
%%%%%%%%%%%%%%%%%%%%%%%%%%%%%%%%%%%%%%%%%%
%%%%%%%%%%%%%%%%%%%%%%%%%%%%%%%%%%%%%%%%%%
\begin{figure}[th!]
\centering\centering\includegraphics[scale=.40]{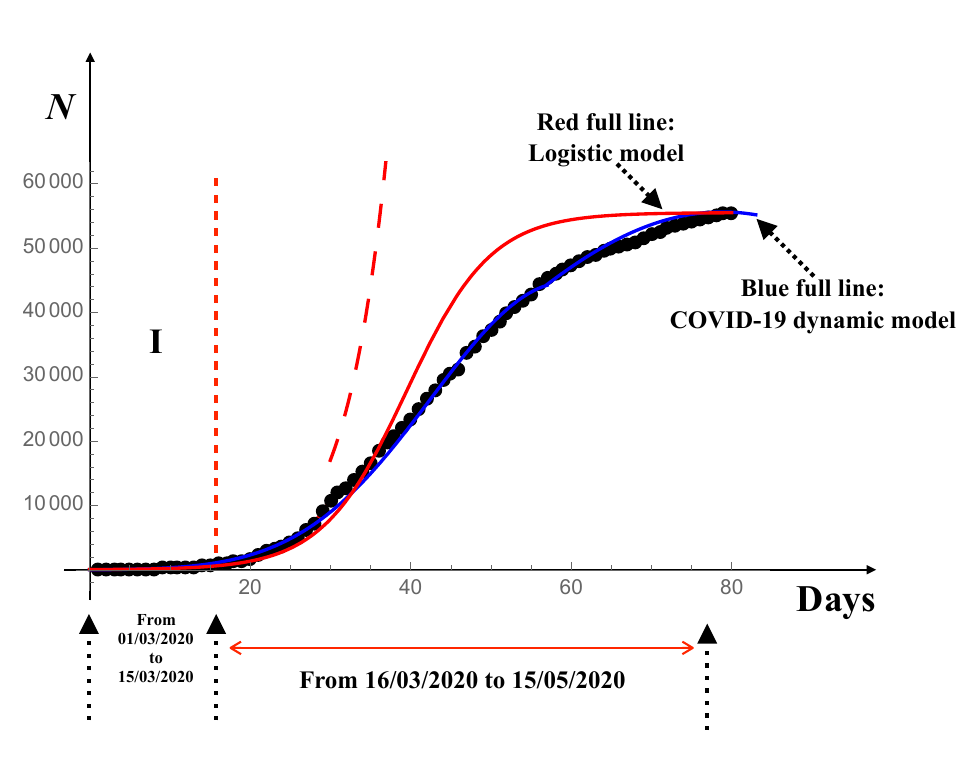}
\caption{{\it Situation in Belgium on 15 May 2020---before, and 60 days after, the adoption of lockdown measures}. The black dots correspond to real data. The blue dotted line corresponds to the situation in Belgium before the adoption of the lockdown measures. The blue and the red solid lines correspond to the theoretical predictions for Belgium according to the solution of in \cite{sonnino2} and the logistic model, respectively. solution of the differential equation for the compartment $N$ shown in \cite{sonnino2} fits well all the experimental data from the initial days (i.e., from the 29th of February 2020), while the logistic model applies only to the first data. The values of the parameters of the differential equation for the compartment $N$ shown in Sonnino {\it et al.} \cite{sonnino2} and the logistic function are: $\tau_{BE}\simeq 5.3$ days ($\mu_{BE}=3.7$ days), $K_N^{BE}\simeq 42626$, and $t_{0BE}\simeq 53.4$ days for solution in Sonnino {\it et al.} \cite{sonnino2}, and $\tau_{BE}\simeq 5.3$ days ($\mu_{BE}=3.7$ days), $K_{BE}=111000$, $t_{0LBE}=39.5$ days for the Logistic function, respectively. The zone $I$ corresponds to the period before the adoption of the lockdown measures.}
\label{Fig.11}
\end{figure}
%%%%%%%%%%%%%%%%%%%%%%%%%%%%%%%%%%%%%%%%%%
\noindent  As we can see, for both Countries our predictions are in excellent agreement with the real world since the onset of COVID 19, contrary to the the logistics model that only applies in the first days of the pandemic. These curves tend to reach the plateau at the time $t_{Max}$ given by 
\begin{equation}\label{c2}
t_{MaxIT}\simeq 80\ \textrm{ days}\quad\textrm{ and}\quad t_{MaxBE}\simeq 60\ \textrm{ days}
\end{equation}
\noindent corresponding to $t_{MaxIT}\simeq 21\ \textrm{ April}\ 2020$ and $t_{MaxBE}\simeq 2\ \textrm{ May}\ 2020$ for Italy and Belgium, respectively.

\section {\bf Comparison between the Theoretical Predictions of Eq.~(\ref{DM14}) and Real Data for USA and France}\label{comp}

Refs~\cite{usa,france1}, report the links to the official sites of the number of infected individuals in US and France, respectively. As the data show, France is currently subjected to the second wave of Coronavirus while the USA data may induce to think that they are in a full second (or third) wave. However, looking at the behaviour of the infectious curve we may also argue that, in agreement with~\cite{usasituation2}, {\it USA as a whole is not in a second (or third) wave because the first wave never really stopped. The virus is simply spreading into new populations or resurgent in places that let down their guard too soon}. So, we are interested in analysing both of these scenarios. Fig.~\ref{USA3w} shows the comparison between the theoretical predictions (blue line) and real data (black dots) for USA. This prediction has been obtained by assuming that US population is currently subjected to the third wave of Coronavirus, respectively. Fig.~\ref{Francefw}. and Fig.~\ref{Francesw}. show the comparison between the theoretical predictions (blue line) and real data (black dots) for France. The values of the parameters for both cases, USA and France, are reported in the figure captions.

%%%%%%%%%%%%%%%%%%%%%%%%%%%%%%%%%%%%%%%%%%
\begin{figure}[hbt!]
\hskip 0.5truecm
\centering\centering\includegraphics[scale=.40]{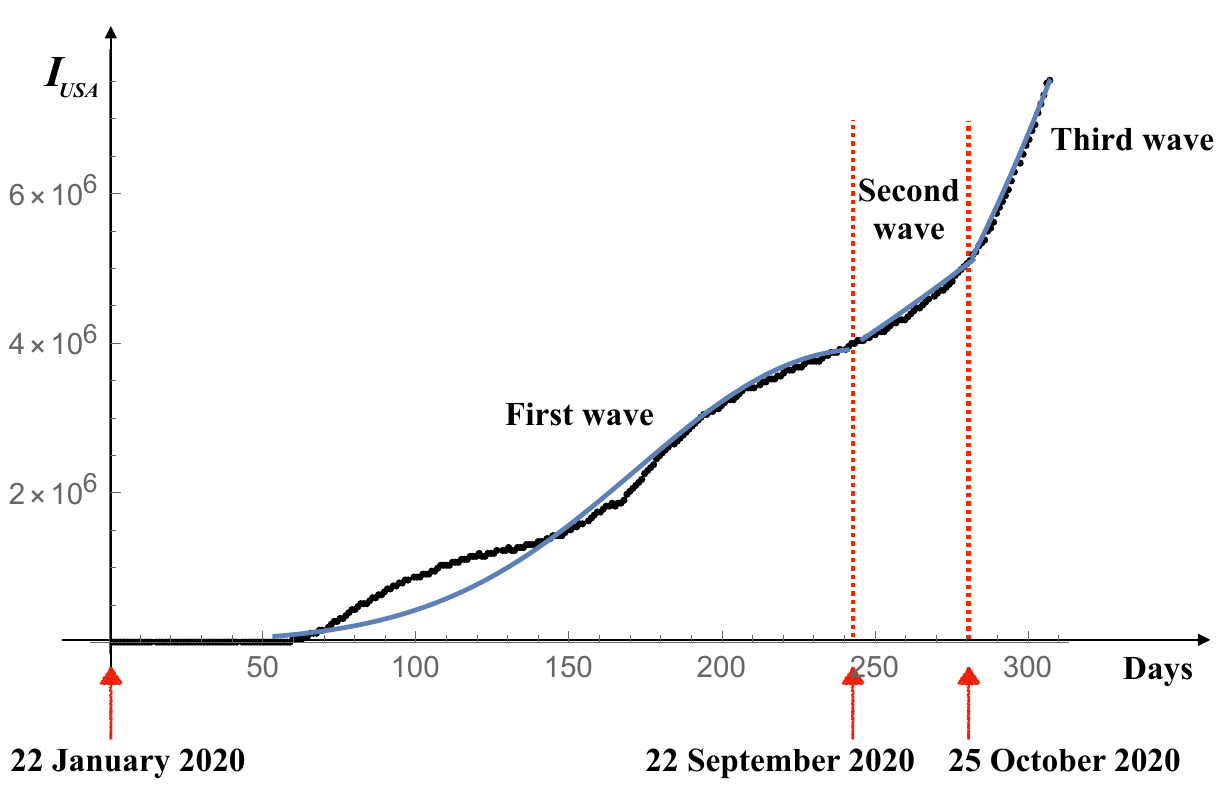}
\caption{
\textit{Comparison between the theoretical predictions (blue lines) and real data (black dots) for USA by assuming that US is in the third wave of Coronavirus. The three series of values of the parameters, corresponding to the three waves, are, respectively: (${\tilde\alpha}=0.043\ {\rm day}^{-1}$, $\beta=0.45\ {\rm day}^{-2}$) ; (${\tilde\alpha}=0.021\ {\rm day}^{-1}$, $\beta=0.008\ {\rm day}^{-2}$), and (${\tilde\alpha}=0.01\ {\rm day}^{-1}$, $\beta=0.02\ {\rm day}^{-2}$). $K=25000000$ and $t_L=54\ {\rm days}$.}
}
\label{USA3w}
\end{figure}
%%%%%%%%%%%%%%%%%%%%%%%%%%%%%%%%%%%%%%%%%%
%%%%%%%%%%%%%%%%%%%%%%%%%%%%%%%%%%%%%%%%%%
\begin{figure}[hbt!]
\hskip 0.5truecm
\centering\centering\includegraphics[scale=.40]{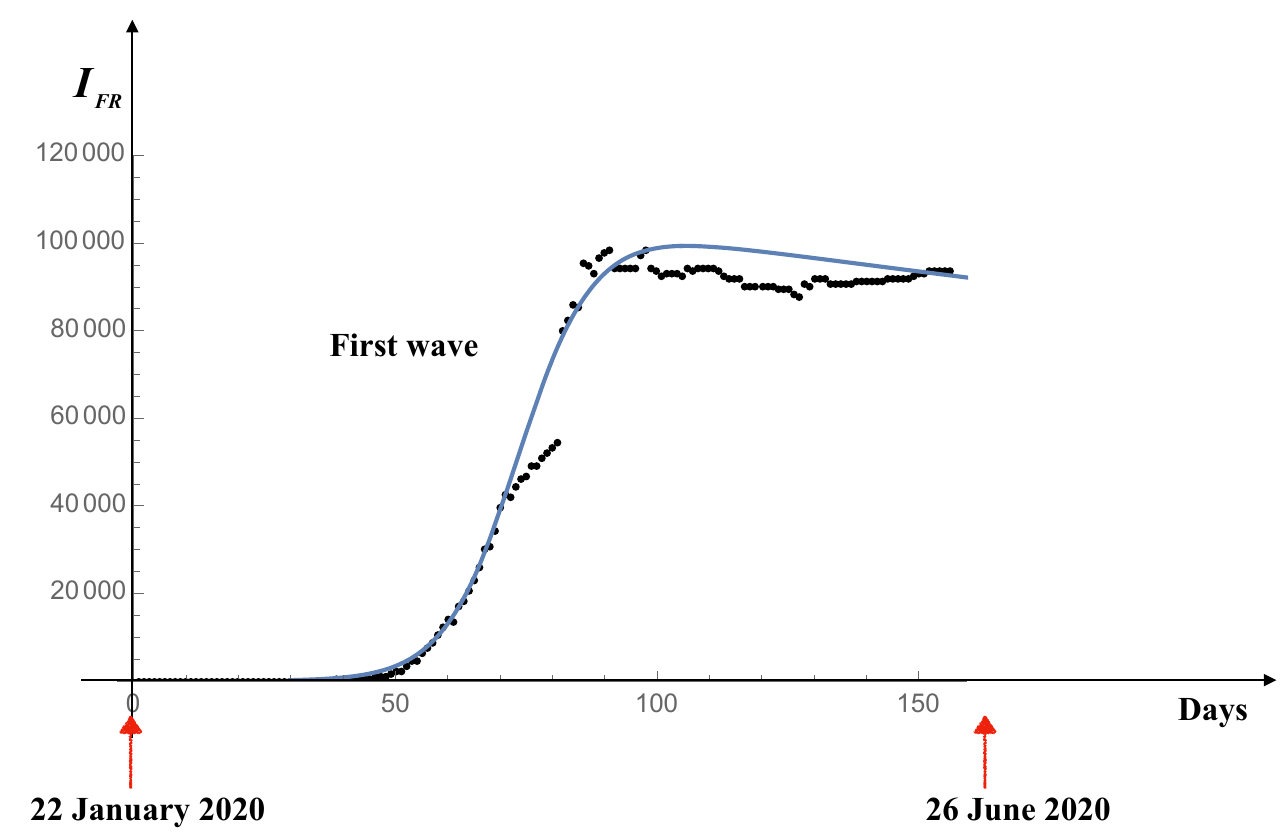}
\caption{
\textit{Comparison between the theoretical predictions (blue line) and real data (black dots) for France during the first wave of Coronavirus. The values of the parameters are; ${\tilde\alpha}=0.145\ {\rm day}^{-1}$, $\beta=0.5\ {\rm day}^{-2}$, $K=110000$, and $t_L=53\ {\rm days}$, respectively.}
}
\label{Francefw}
\end{figure}
%%%%%%%%%%%%%%%%%%%%%%%%%%%%%%%%%%%%%%%%%%
%%%%%%%%%%%%%%%%%%%%%%%%%%%%%%%%%%%%%%%%%%
\begin{figure}[hbt!]
\hskip 0.5truecm
\centering\centering\includegraphics[scale=.40]{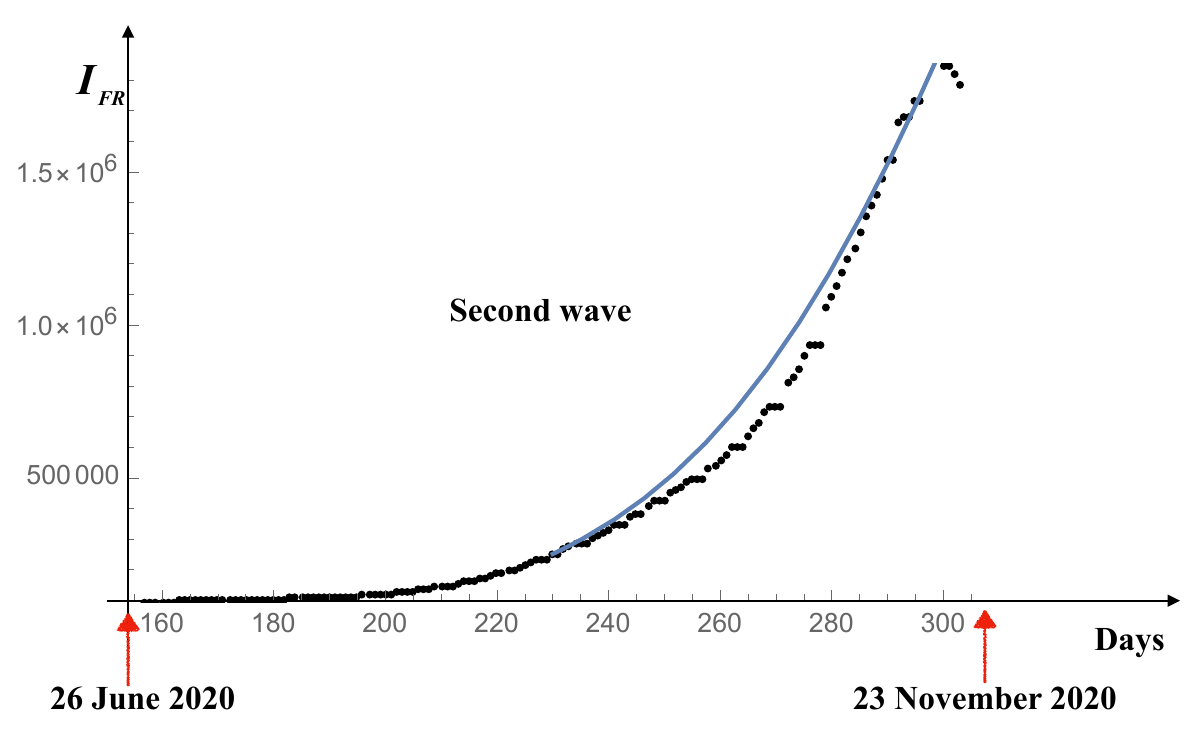}
\caption{
\textit{Comparison between the theoretical predictions (blue line) and real data (black dots) for France during the second wave of Coronavirus. The values of the parameters are; ${\tilde\alpha}=0.035\ {\rm day}^{-1}$, $\beta=0.5\ {\rm day}^{-2}$, $K=5000000$, and $t_L=278\ {\rm days}$, respectively.}
}
\label{Francesw}
\end{figure}
%%%%%%%%%%%%%%%%%%%%%%%%%%%%%%%%%%%%%%%%%%
\section{The Stochastic $(SIS)_L$-Model}\label{SM}
\noindent The Stochastic version of Eq.~(\ref{DM14}) reads
\begin{equation}\label{S1}
\frac{dI}{d{\hat t}}=\alpha I\left(1-\frac{I}{K}\right)-\beta\left(\frac{{\hat t}^2-1}{{\hat t}}\right)I+\ \xi({\hat t})
\end{equation}
\noindent with $\xi({\hat t})$ denoting a white noise
\begin{align}\label{S2}
&<\xi({\hat t})>=0\\
&<\xi({\hat t})\xi({\hat t}')>=\eta\delta({\hat t}-{\hat t}')\nonumber
\end{align}
\noindent and $\sqrt\eta$ is the intensity of the noise. $\delta$ denotes the Dirac delta function (distribution). It is shown \cite{sonnino} that under a very general assumption, we may consider that the system is governed by the {\it Grand Ensemble} statistical mechanics. So, according to this statistics, the relative fluctuations of the  number of the infectious people behaves as $K_{Country}^{-1/2}$ \cite{reif}, with $K_{Country}$ denoting the capacity  of the Country's population. This implies that the intensity of the noise in Eq.~(\ref{S2}) is of the order of 
\begin{equation}\label{S2a}
\eta\sim t_L K_{Country}^{-1/2}
\end{equation}
\noindent with $t_L$ denoting the time when the lockdown measures have been applied. We may object that this reasoning is based on {\it equilibrium conditions}. However, we would like to point out that the aim of this Section is only to provide a (rough) estimate of the order of magnitude of the noise. Let us now consider the system at the reference state $I_{RS}$, which is the solution of the deterministic equation 
\begin{equation}\label{S3}
\frac{dI_{RS}}{d{\hat t}}=\alpha I_{RS}\left(1-\frac{I_{RS}}{K}\right)-\beta\left(\frac{{\hat t}^2-1}{\hat t}\right)I_{RS}
\end{equation}
\noindent subject to a perturbation of small amplitude $\delta I$, i.e.
\begin{equation}\label{S4}
I({\hat t})=I_{RS}({\hat t})+\delta I({\hat t})
\end{equation}
\noindent Our aim is to compute the relevant statistical correlation functions of this processes i.e. $<\delta I({\hat t})\xi({\hat t})>$ and $<\delta I({\hat t})\delta I({\hat t}')>$. We can find \cite{sonnino} the expression for the second moment and the correlation function for $\delta I $:
\begin{align}\label{S11a}
&<\left(\delta I({\hat t}\right))^2>=\left({\hat t}^{2\beta}\exp(2\alpha({\hat t}-1)-\beta({\hat t}^2-1)-4\alpha G({\hat t})/K\right)\!\!\Big(<\!\left(\delta I(1)\right)^2\!>\nonumber\\
&\!+\eta \int_1^{\hat t}x^{-2\beta}\exp(-2\alpha (x-1)+\beta (x^2-1)+4\alpha G(x)/K)\!dx\Bigl)
\end{align}
\begin{equation}\label{S11b}
<\delta I(1)\delta I({\hat t})> ={\hat t}^\beta\exp(\alpha({\hat t}\!-\!1)\!-\!\beta/2({\hat t}^2\!-\!1)\!-\!2\alpha G({\hat t})/K)\!<(\delta I(1))^2>
\end{equation}
\noindent Fig.~\ref{I_Stoch_IT1} shows the simulation of 200 trajectories of Eq.~(\ref{S1}) for Italy using the Order-2 Stochastic Runge-Kutta integration method. The thick black curve is the numerical solution of the deterministic equation~(\ref{DM14}). The noise intensity is $\eta=0.05$ and the values of other parameters are reported in the caption of Fig.~\ref{I_Stoch_IT1}.
%%%%%%%%%%%%%%%%%%%%%%%%%%%%%%%%%%%%%%%%%%
\begin{figure}[hbt!]
\hskip 0.5truecm
\centering\centering\includegraphics[scale=.25]{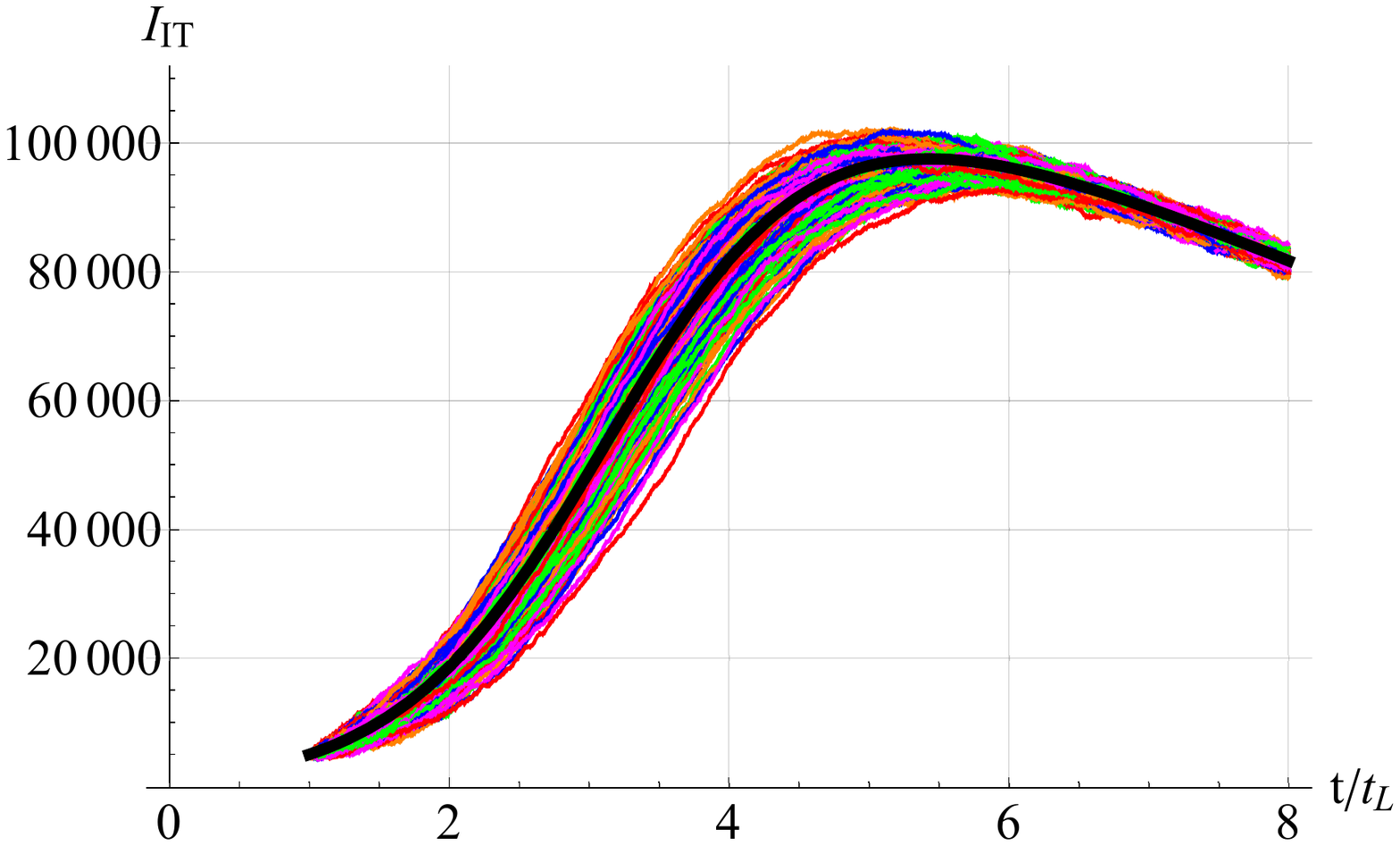}
\caption{
\textit{{\bf Solution of the stochastic equation for Italy.} Solutions of Eqs~(\ref{S1}) and (\ref{S2}) for Italy - first wave of infection by SARS-CoV-2 - with $\eta=0.05$ and for 200 realisations.  The values of the other parameters are $I_0$=5000, $K=150000$, $\alpha=1.5$, and $\beta=0.1$, respectively. The black thick curve is the numerical solution of the deterministic equation~(\ref{DM14}).} 
}
\label{I_Stoch_IT1}
\end{figure}
%%%%%%%%%%%%%%%%%%%%%%%%%%%%%%%%%%%%%%%%%%
\noindent Figs.~(\ref{CFII}) and (\ref{CFI0I}) show the correlation functions $<(\delta I({\hat t}))^2>$ and $<\delta I(1)\delta({\hat t})>$ for Italy, first wave of infection by SARS-CoV-2. The values of the parameters are reported in the figure captions.
%%%%%%%%%%%%%%%%%%%%%%%%%%%%%%%%%%%%%%%%%%
\begin{figure}[hbt!]
\hskip 0.5truecm
\centering\centering\includegraphics[scale=.40]{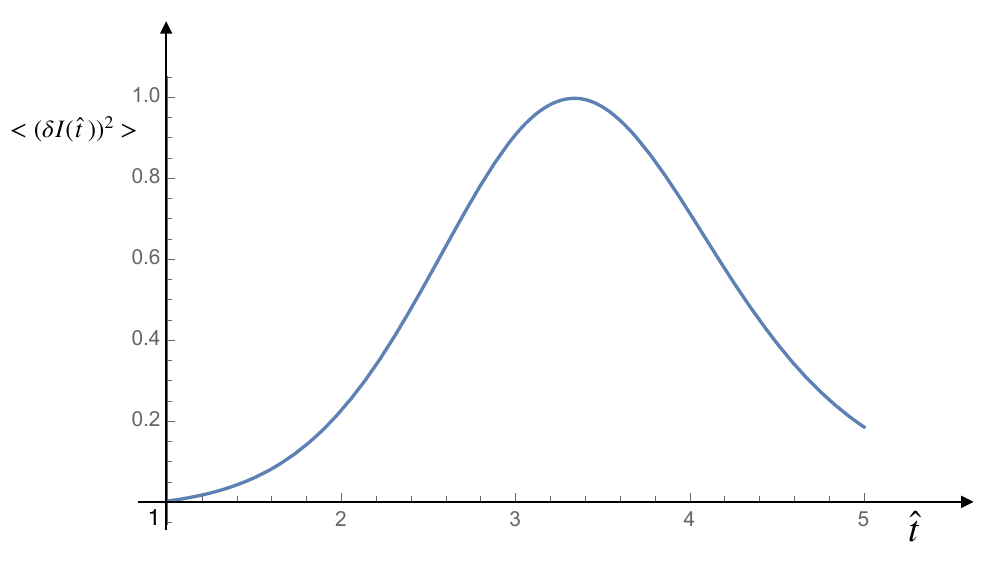}
\caption{
\textit{{\bf Correlation Function} $<(\delta I({\hat t}))^2>$ {\bf for Italy, first wave of infection by SARS-CoV-2.} This correlation function corresponds to Eq.~(\ref{S11a}). The values of the parameters are $I_0$=5000, $K=150000$, $\alpha=1.5$, and $\beta=0.1$, respectively.} 
}
\label{CFII}
\end{figure}
%%%%%%%%%%%%%%%%%%%%%%%%%%%%%%%%%%%%%%%%%%

%%%%%%%%%%%%%%%%%%%%%%%%%%%%%%%%%%%%%%%%%%
\begin{figure}[hbt!]
\hskip 0.5truecm
\centering\centering\includegraphics[scale=.40]{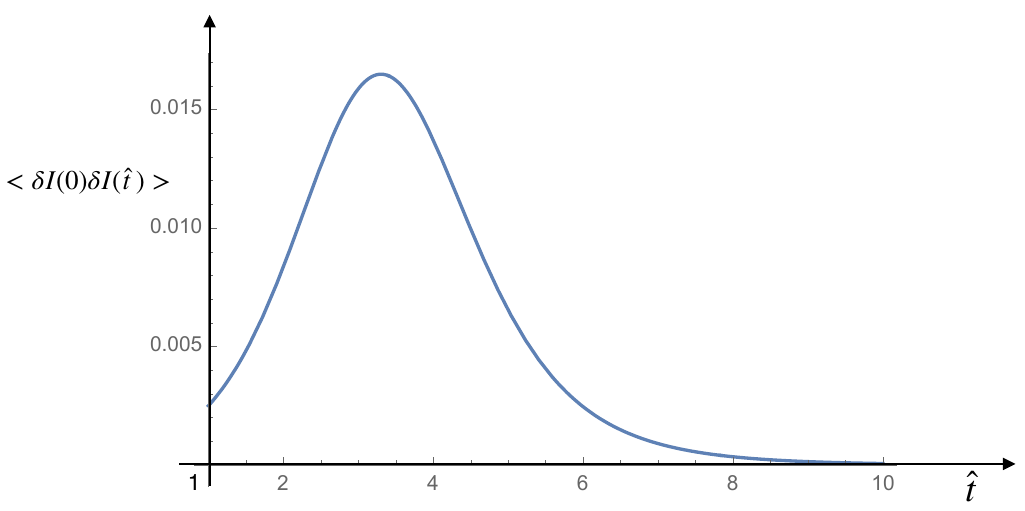}
\caption{
\textit{{\bf Correlation Function for the Infectious people} $<\delta I(1)\delta({\hat t})>$. Plot of Eq.~(\ref{S11b}) with $I_0$=5000, $K=150000$, $\alpha=1.5$, and $\beta=0.1$, respectively.}
}
\label{CFI0I}
\end{figure}
%%%%%%%%%%%%%%%%%%%%%%%%%%%%%%%%%%%%%%%%%%
\noindent Fig.~\ref{FRfwStoc}. and Fig.~\ref{FRswStoc}. illustrate the comparison between the theoretical predictions (blue lines) and real data (black dots) for France concerning the first and the second waves of SARS-CoV-2, respectively. The intensity of the noise has been estimated by using Eq.~(\ref{S2a}). As we can see, the predictions of our model are in a fairly good agreement with real data. Fig.~\ref{US3w1Stoc}. refers to the stochastic differential equation for USA. In this case the intensity of the noise corresponds to the value estimated by Eq.~(\ref{S2a}) (we have ${\tilde\eta}=0.0002$). Here, we get a fairly good agreement with real data. 
%%%%%%%%%%%%%%%%%%%%%%%%%%%%%%%%%%%%%%%%%%
\begin{figure}[hbt!]
\hskip 0.5truecm
\centering\centering\includegraphics[scale=.20]{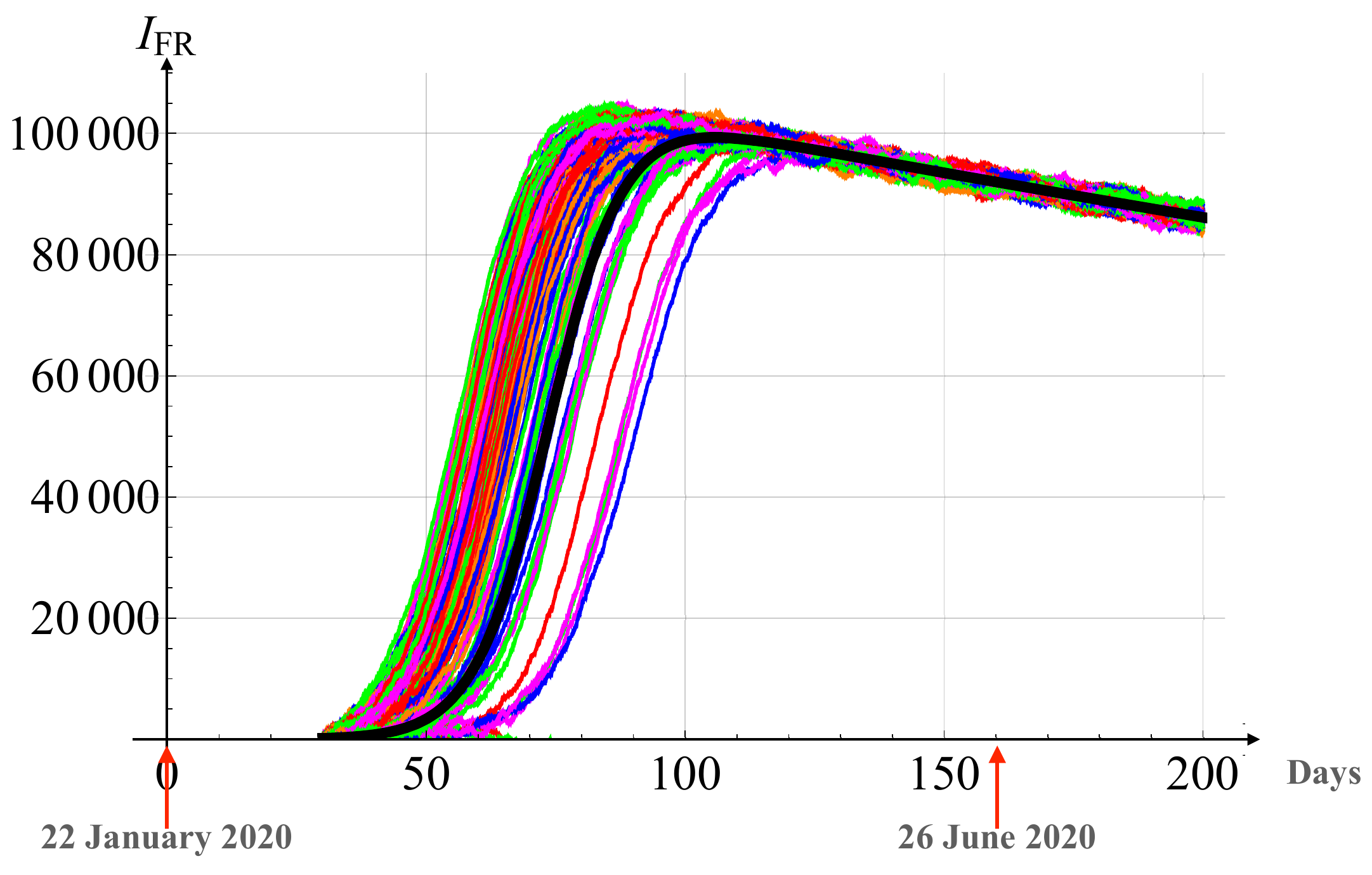}
\caption{
{\textit{{\bf Solutions (200 realisations) of the stochastic differential equation for France - First wave}. The values of the parameters are those reported in Fig.~\ref{Francefw}. ${\tilde\eta}=0.002$, which has been estimated according to Eq.~(\ref{S2a}). The black curve is the solution of the deterministic equation.}}
}
\label{FRfwStoc}
\end{figure}
%%%%%%%%%%%%%%%%%%%%%%%%%%%%%%%%%%%%%%%%%%
%%%%%%%%%%%%%%%%%%%%%%%%%%%%%%%%%%%%%%%%%%
\begin{figure}[hbt!]
\hskip 0.5truecm
\centering\centering\includegraphics[scale=.40]{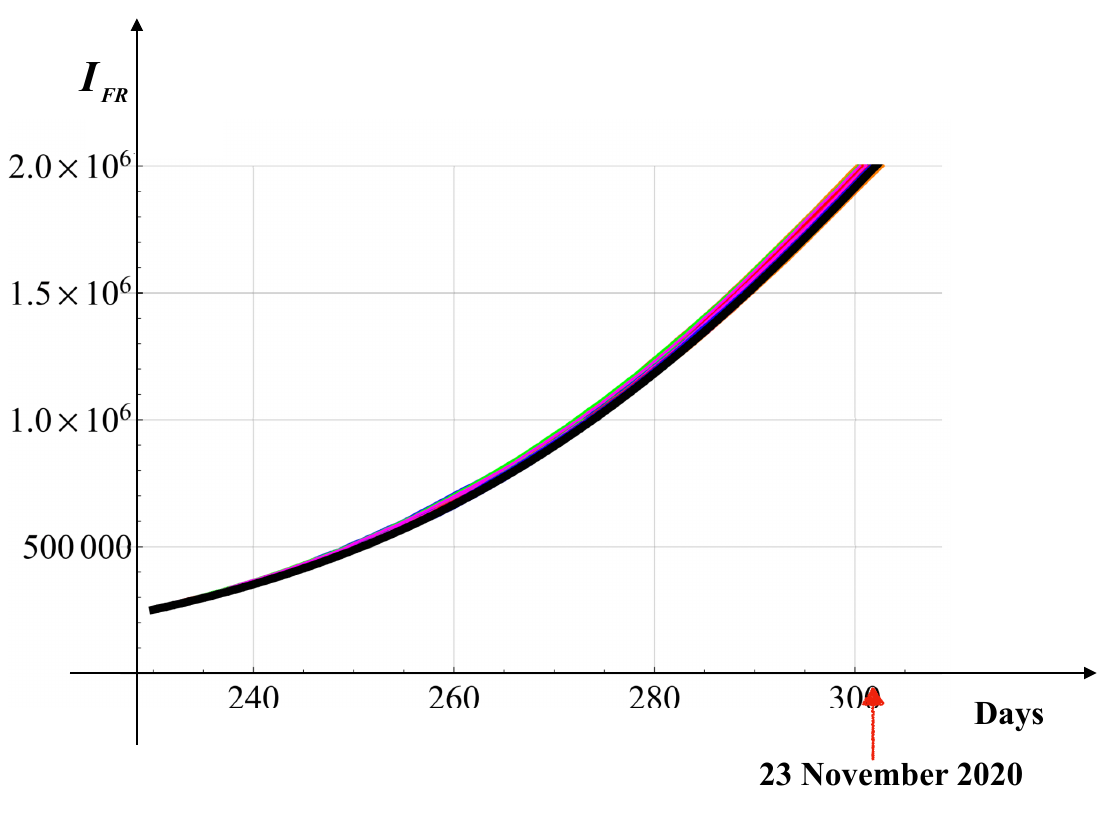}
\caption{
{\textit{{\bf Solutions (200 realisations) of the stochastic differential equation for France - Second wave}. The values of the parameters are those reported in Fig.~\ref{Francesw}. ${\tilde\eta}=0.001$, which has been estimated according to Eq.~(\ref{S2a}).}}
}
\label{FRswStoc}
\end{figure}
%%%%%%%%%%%%%%%%%%%%%%%%%%%%%%%%%%%%%%%%%%
%%%%%%%%%%%%%%%%%%%%%%%%%%%%%%%%%%%%%%%%%%
\begin{figure}[hbt!]
\hskip 0.5truecm
\centering\centering\includegraphics[scale=.30]{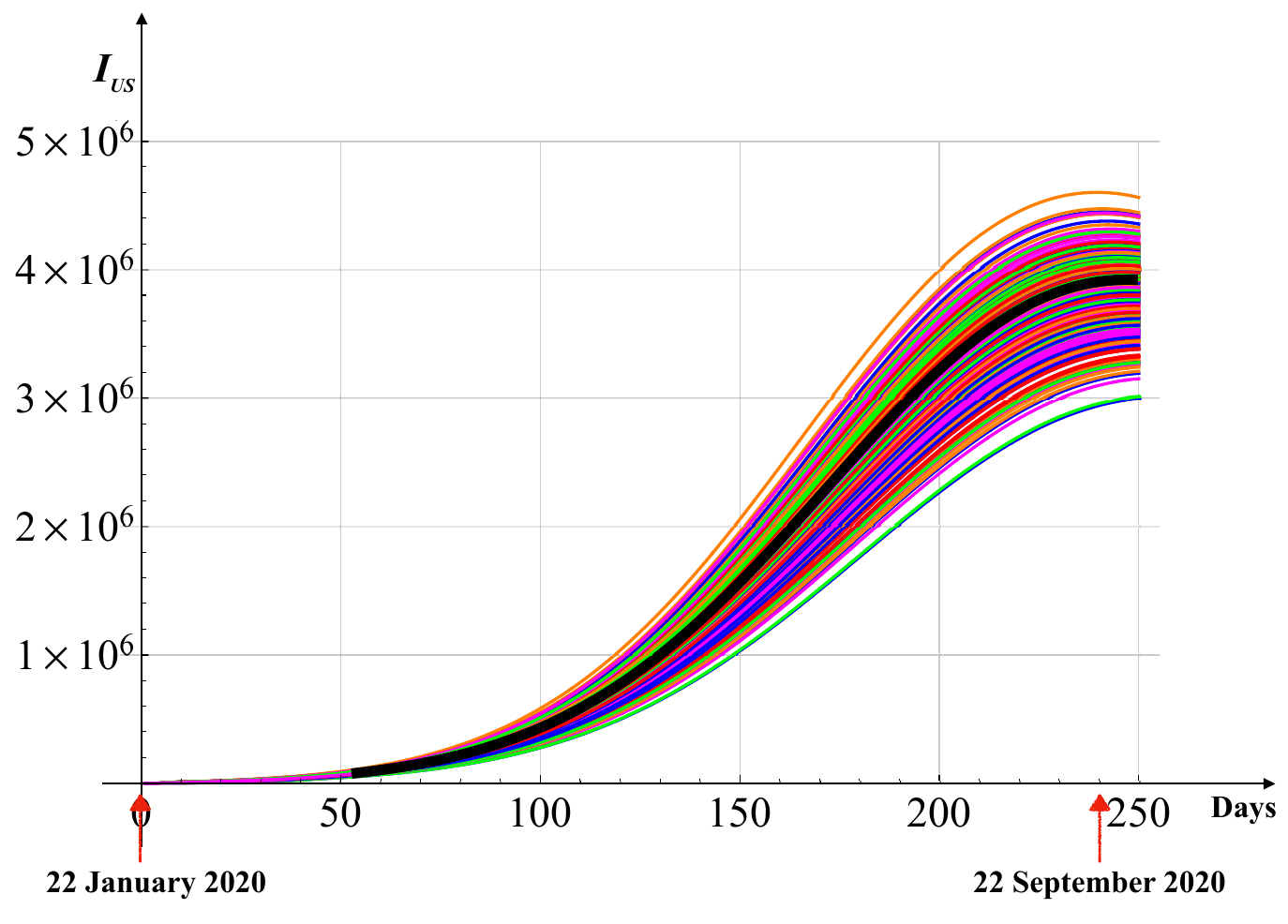}
\caption{
{\textit{{\bf Solutions (200 realisations) of the stochastic differential equation for USA}. The values of the parameters are those reported in Fig.~\ref{USA3w}. The intensity of the noise is ${\tilde\eta}=0.0002$, which has been obtained by using Eq.~(\ref{S2a}). The black curve is the solution of the deterministic equation.}}
}
\label{US3w1Stoc}
\end{figure}
%%%%%%%%%%%%%%%%%%%%%%%%%%%%%%%%%%%%%%%%%%

\section{\bf Modelling the Role of the Hospitals}$\label{Hs}$
\noindent As can be seen, according to the $SIS$-model, after the lockdown measures the number of infectious people starts to increase again. Hospitals and health institutions play a crucial role in hindering the spread of the Coronavirus. In this Section, we propose a model which accounts for people who are only traced back to hospitalised infectious individuals. In our approach, the dynamics of the Health Institutes is obtained by taking inspiration from the {\it Michaelis- Menten’s enzyme-substrate reaction model} (the so-called {\it MM reaction} \cite{MM1,MM3}) where the {\it enzyme} is associated to the {\it available hospital beds}, the {\it substrate} to the {\it infected people}, and the {\it product} to the {\it recovered people}, respectively. In other words, everything happens as if the hospitals beds act as a {\it catalyser} in the hospital recovery process \cite{sonnino2} and \cite{sonnino3}. We propose the following model:
\begin{align}\label{Hs1}
&S+I \xrightarrow{\sigma} 2I\\
&I\xrightarrow{\gamma+c(t)}S\nonumber\\
&I + b \xrightarrow{k_1} I_h \xrightarrow{k_r} r_h+b\nonumber\\
&\qquad\qquad\ \! I_h \xrightarrow{k_d} d_h+b\nonumber\\
&\qquad\qquad\ \! r_h \xrightarrow{\gamma_1} S\nonumber
\end{align}
\noindent where the hypothesis that an individual acquires immunity, after having contracted the Coronavirus and being recovered, is not adopted. In scheme~(\ref{H1}), $b$ denotes the number of available {\it hospital beds}, $I_h$ the number of {\it infected people blocking an hospital bed}, $r_h$ the number of {\it recovered people previously hospitalised}, and $d_h$ the number of {\it people deceased in the hospital}, respectively. According to scheme~(\ref{Hs1}), people, once recovered, are subjected to the same existing lockdown measures as any other people. We refer the simple model, based on scheme~\ref{Hs1} to as the  $(SISI_h)_L$-model}. 

\noindent Concerning the {\it recovered people}, we would like to make clear the following. $R$ stands for the {\it total number of the recovered people} (i.e., the number of recovered people previously hospitalised, plus the number of the asymptomatic people, plus the infected people who have been recovered without being previously hospitalised). However, the natural question is: \textit{how can we count $R$ and compare this variable with real data ?}. The current statistics, produced by the Ministries of Health of various Countries, concern the people released from hospitals. Apart from Luxembourg (where almost the entire population has been subjected to the COVID-19 test), no other Countries are in a condition to provide statistics regarding the total people recovered by COVID-19. Hence, it is our opinion that the equation for $R$, is not useful since it is practically impossible to compare the theoretical predictions for $R$ with real data. We then proceed by adopting approximations and by establishing the differential equations where the solution may realistically be subjected to experimental verification. More specifically, firstly, we assume that $R$ is given by three contributions:
\begin{equation}\label{Rs1}
R=r_h+r_{A}+r_{I}
\end{equation}
\noindent with $r_h$, $r_{A}$, and $r_{I}$ denoting the {\it total number of the recovered people previously hospitalised}, {\it the total number of asymptomatic people}, and the {\it total number of people immune to SARS-CoV-2}, respectively. Secondly, we assume that the two contributions $r_{A}$  and $r_{I}$ are negligible, i.e., we set $r_A\approx 0$ and $r_{I}\approx 0$ \footnote{We consider that the SARS-CoV-12  has just appeared for the first time. So, we do not consider the asymptomatic people who are immune to the virus without any medical treatment.}. Finally, due to lack of reliable statistics, we are forced to limit ourselves to consider the (very) simplified case 
\begin{equation}\label{Rs2}
R\simeq r_h
\end{equation}
\noindent Of course, we have
\begin{equation}\label{Hs2}
I_h+b=C_h=const.\qquad {\rm where}\quad{C_h={\rm Total\ hospital's\ capacity}}
\end{equation}
\noindent The dynamical equations for the entire process are then:
\begin{align}\label{Hs3}
&\frac{d}{d{\hat t}}S=-\sigma\frac{S}{N_{Tot.}} I+\gamma I+\gamma_1 r_h+\beta\left(\frac{{\hat t}^2-1}{\hat t}\right) (I+r_h)\\
&\frac{d}{d{\hat t}}I=\sigma\frac{S}{N_{Tot.}} I-\gamma I-k_1I(C_h-I_h)-\beta\left(\frac{{\hat t}^2-1}{\hat t}\right) I\nonumber\\
&\frac{d}{d{\hat t}}I_h= k_1I(C_h-I_h)-k_r{I_h}-k_d{I_h}\nonumber\\
&\frac{d}{d{\hat t}}r_h=k_r{I_h}-\gamma_1 r_h\nonumber\\
&\frac{d}{d{\hat t}}d_h=k_d {I_h}\nonumber
\end{align}
\noindent where, at this stage, for simplicity, the {\it average recovery time delay} and the {\it average death time delay} have been neglected \cite{sonnino}. In this case $I$ stands for the {\it infectious individuals not hospitalised}. From system~(\ref{Hs3}) we get the following conservation law
\begin{equation}\label{Hs4}
S+I+I_h+r_h+d_h=N_{Tot.}=const.
\end{equation}
\subsection{\bf The Deterministic $(SISI_h)_L$-model}\label{ASM}
\noindent To simplify as much as possible the set of O.D.E.s~(\ref{Hs3}), we adopt several hypotheses that will not compromise the validity of our model. First, we assume that that $S+I+I_h\simeq N_{Tot.}=const.$ Secondly, let $\gamma\simeq\gamma_1$. Finally, we take into account the current Belgian hospital-protocol: "{\it Only the seriously sick people are hospitalised, the remaining infectious individuals have to be sent home and they must be subjected to quarantine measures}". Hence, $I\gg I_h $ and the total number of recovered people, $R$, is much larger than the total number of recovered people, previously hospitalised (i.e. $R\gg r_h$). Under these assumptions, the model simplifies to
\begin{align}\label{AH1}
&S+I \xrightarrow{\sigma} 2I\\
&I\xrightarrow{\gamma+c(t)}S\nonumber\\
&I + b \xrightarrow{k_1} I_h
\end{align}
\noindent with $I_h+b=C_h$. Hence, under these assumptions, after hospitalisation, individuals will be removed from the disease, either due to immunisation (e.g. due to vaccination or special health care received) or due to death. The governing O.D.E.s, associated to the model~(\ref{AH1}), read
\begin{align}\label{AH2}
&\frac{d}{d{\hat t}}S\simeq-\sigma\frac{S}{N_{Tot.}} I+\gamma I+\beta\left(\frac{{\hat t}^2-1}{\hat t}\right) I\\
&\frac{d}{d{\hat t}}I=\sigma\frac{S}{N_{Tot.}} I-\gamma I-k_1I(C_h-I_h)-\beta\left(\frac{{\hat t}^2-1}{\hat t}\right) I\nonumber\\
&\frac{d}{d{\hat t}}I_h\simeq k_1I(C_h-I_h)\nonumber
\end{align}
\noindent The recovered people $r_h$ and the deceased people $d_h$ may be obtained by solving, respectively, the following O.D.E.s
\begin{equation}\label{AH3}
\frac{d}{d{\hat t}}r_h\simeq k_r{I_h}\qquad ;\qquad \frac{d}{d{\hat t}}d_h=k_d {I_h}
\end{equation}
\noindent From system~(\ref{AH2}), we obtain 
\begin{align}\label{AH4}
&\frac{d}{d{\hat t}}I=\alpha_1I\left(1-\frac{I}{K_1}\right)-\rho I I_h-\beta\left(\frac{{\hat t}^2-1}{\hat t}\right) I\\
&\frac{d}{d{\hat t}}I_h= k_1C_h I-k_1I I_h\qquad{\rm with}\nonumber\\
&\alpha_1=\sigma\!\left(\!1-\frac{\gamma}{\sigma}-\frac{k_1}{\sigma}C_h\!\right)\ ;\ K_1=N_{Tot.}\left(\!1-\frac{\gamma}{\sigma}-\frac{k_1}{\sigma}C_h\!\right)\ ;\ \rho=\frac{\sigma}{N_{Tot.}}\!-\!k_1\nonumber
\end{align}
\noindent In absence of the lockdown measures ($\beta=0$), we have the following scenarios: $\forall I_0>0$
\begin{align}\label{AH5}
&{\bf i)}\  {\rm if}\quad C_h<C_{hCrit.}\equiv\frac{\sigma}{k_1}\left(1-\frac{\gamma}{\sigma}\right)\  {\rm the\ equilibrium\ with}\ I=K_1\ {\rm is\ stable}\\
&{\bf ii)}\ {\rm if}\quad C_h>C_{hCrit.}\equiv\frac{\sigma}{k_1}\left(1-\frac{\gamma}{\sigma}\right)\  {\rm the\ equilibrium\ with}\ I=0\ {\rm is\ stable}
\nonumber
\end{align}
\noindent In words: 

\noindent $\bullet$ for the case ${\bf i}$), there will be a proper epidemic outbreak with an increase of the number of the infectious people;

\noindent $\bullet$ for the case ${\bf ii}$), independently of the initial size of the susceptible population, the disease can never cause a proper epidemic outbreak.

\noindent This result highlights the crucial role of the Hospitals and the Health Care Institutes: 

\noindent \textit{If the threshold of the hospital capacities exceeds a lower limit, the spread of the Coronavirus tends to decrease over time, and the stable solution corresponds to zero infectious individuals}.

\noindent Fig.~\ref{Hs1}. and \ref{Hs0}. illustrate the Italian situation. Notice that, with the values of parameters reported in the corresponding figure captions, $C_{hCrit.}=18434$.

%%%%%%%%%%%%%%%%%%%%%%%%%%%%%%%%%%%%%%%%%%
\begin{figure}[hbt!]
\hskip 0.5truecm
\includegraphics[width=11cm, height=7cm]{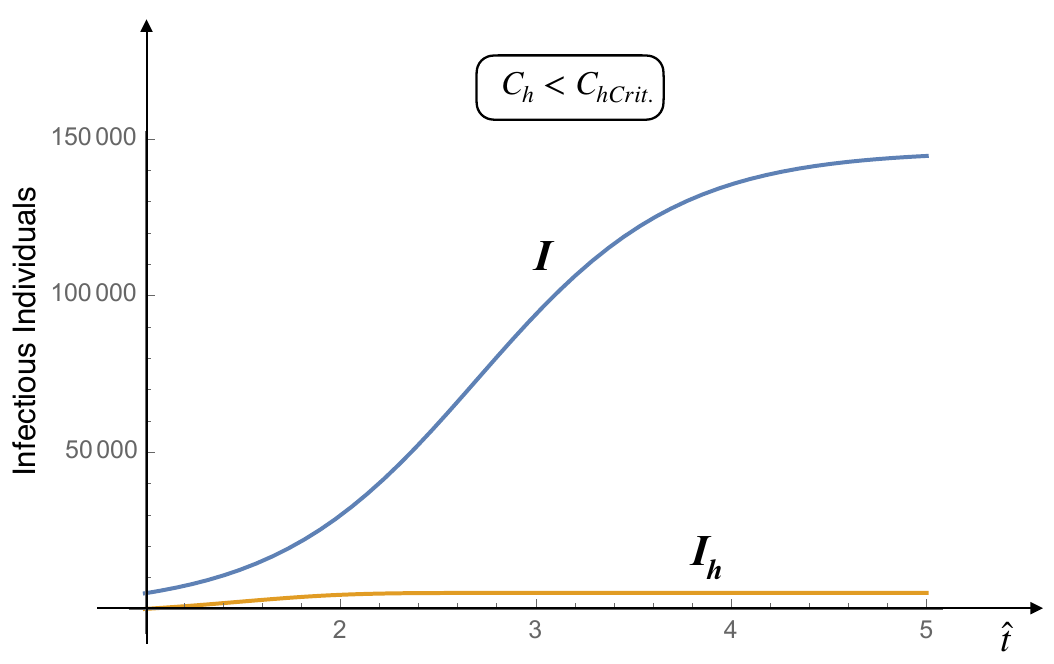}
\caption{
\textit{{\bf Infectious people and Infectious hospitalised people vs time - $C_h>C_{HCrit.}$}. In this case, there is an epidemic outbreak with an increase of the number of the infectious people. The values of the parameters are: $\beta=0$, $k_1=0.00001$, $\rho=0.0001$, $K_1=150000$, $\gamma=0.0001$, $\sigma=2.7651$, and $C_h=5100$.}
}
\label{Hs1}
\end{figure}
%%%%%%%%%%%%%%%%%%%%%%%%%%%%%%%%%%%%%%%%%%

%%%%%%%%%%%%%%%%%%%%%%%%%%%%%%%%%%%%%%%%%%
\begin{figure}[hbt!]
\hskip 0.5truecm
\includegraphics[width=11cm, height=7cm]{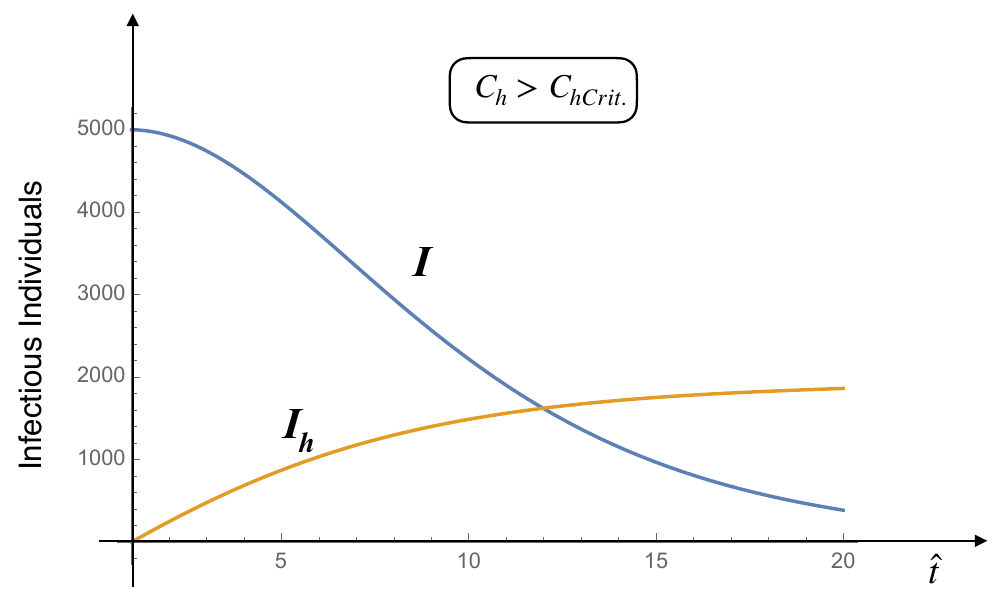}
\caption{
\textit{{\bf Infectious people and Infectious hospitalised people vs time - $C_h>C_{HCrit.}$}. Independently of the initial size of the susceptible population, the disease can never cause a proper epidemic outbreak. The values of the parameters are: $\beta=0$, $k_1=0.00001$, $\rho=0.0001$, $K_1=150000$, $\gamma=0.0001$, $\sigma=2.7651$, and $C_h=18436$.}
}
\label{Hs0}
\end{figure}
%%%%%%%%%%%%%%%%%%%%%%%%%%%%%%%%%%%%%%%%%%
\noindent Similar analysis leading to Eq.~(\ref{AH5}), allowing the calculation of the critical threshold for France and US hospital capacities, may also be performed. By summarising, the $(SISI_h)_L$-model shows first the crucial role of hospital capacity, second highlights its limits. For instance, for the Italian situation, to obtain a relevant {\it dampening effect} of the COVID-19 infection, the capacity of hospitals in Italy would have to increase by about 4 times its current value (which is enormous). Hence the need to combine and coordinate the two actions at the same time: to increase the hospitals' capacity  as much as possible and to distribute effective vaccines. The current analysis is mainly addressed to Countries that do not have the possibility to buy and distribute vaccines on a mass level (such as, for example, some African Countries). In this case, the role of the hospitals becomes crucial and basically it represents the only real remedy to stop the spread of the pandemic.

\subsection {\bf The Stochastic $(SISI_h)_L$ Model}\label{SMS}

\noindent If the dynamics is subjected to white noise, the related stochastic equations read
\begin{align}\label{SM1}
&\frac{d}{d{\hat t}}I=\alpha_1I\left(1-\frac{I}{K_1}\right)-\rho I I_h-\beta\left(\frac{{\hat t}^2-1}{\hat t}\right) I+\xi_1({\hat t})\\
&\frac{d}{d{\hat t}}I_h= k_1C_h I-k_1I I_h+\xi_2({\hat t})\nonumber
\end{align}
\noindent where
\begin{align}\label{SM2}
&<\xi_i(t)>=0\qquad (i=1,2)\\
&<\xi_i(t)\xi_j(t')>=\eta_{ij}\delta_{ij}\delta(t-t')\qquad{\rm with}\quad \eta_{12}=\eta_{21}\nonumber
\end{align}
\noindent with $\delta_{ij}$ denoting Kronecker's delta. The statistical properties of this processes, i.e. $<\delta I({\hat t})\delta I({\hat t})>$, $<\delta I({\hat t})\delta I_h({\hat t})>$ and $<\delta I_h({\hat t})\delta I_h({\hat t})>$, may be obtained, firstly, by determining the reference state. This state satisfies the following O.D.E.s \cite{sonnino}
\begin{align}\label{SM3}
&\frac{d}{d{\hat t}}I_{RS}=\alpha_1I_{RS}\left(1-\frac{I_{RS}}{K_1}\right)-\rho I_{RS} I_{hRS}-\beta\left(\frac{{\hat t}^2-1}{\hat t}\right) I_{RS}\\
&\frac{d}{d{\hat t}}I_{hRS}= k_1C_h I_{RS}-k_1I_{RS} I_{hRS}\nonumber
\end{align}
\noindent Let us choose, for example, $I_{hRS}=C_h$, then \cite{sonnino}
\begin{align}\label{SM4}
&I_{RS}({\hat t})=\frac{I_{0RS}\exp((1-\alpha_2/\beta)^2/\sigma){\hat t}^\beta\exp(-({\hat t}-\alpha_2/\beta)^2/\sigma)}{1+(I_{0RS}\alpha_2/K_2)\exp((1-\alpha_2/\beta)^2/\sigma)\int_1^{\hat{t}}x^\beta\exp(-(x-\alpha_2/\beta)^2/\sigma)dx}\nonumber\\
&\alpha_2\equiv\sigma\left(1-\frac{\gamma}{\sigma}-\frac{C_h}{N_{Tot.}}\right)\quad ;\quad K_2=N_{Tot.}\left(1-\frac{\gamma}{\sigma}-\frac{C_h}{N_{Tot.}}\right)
\end{align}
\noindent We have \cite{sonnino}
\begin{align}\label{SM8}
&\frac{d}{d{\hat t}}\delta I=\left(\alpha_1-\rho C_h-2\frac{\alpha_1}{K_1}I_{RS}-\beta\left(\frac{{\hat t}^2-1}{\hat t}\right)\right)\delta I-\rho I_{RS}\delta I_h+\zeta_1({\hat t})\nonumber\\
&\frac{d}{d{\hat t}}\delta I_h=-k_1I_{RS}(t)\delta I_h+\zeta_2({\hat t})
\end{align} 
\noindent For Italy, the value of $C_h$ (the total Italian hospitals' capacity) may be obtained by making reference to the data published in \cite{hospitalsIT}. More specifically, in 2017, when there were $518$ public hospitals and $482$ accredited private ones, in Italy there were $151646$ beds for ordinary hospitalisation in public hospitals ($2.5$ per $1000$ inhabitants) and $40458$ in private ones ($0.7$ per $1000$ inhabitants), for a total of over $192$ thousand beds ($3.2$ per $1000$ inhabitants). The number of public and private beds destined for intensive care was 5.090 (a number very close to the $5100$ cited by the newspapers these days), about $8.42$ per $100000$ inhabitants \cite{hospitalsIT}. Fig.~\ref{CFIhIh}. shows the correlation function $<\delta I_h(t)\delta I_h(t)>$. As se can see, this is a typical correlation function at equilibrium for system subjected to random fluctuations. This is not surprising as our reference state correspond to the maximum capacity of the hospitals (i.e. $I_ {hRS}=C_h$) so, fluctuations at equilibrium are the only possible ones. The three correlation functions $<\delta I(t)\delta I(t)>$, $<\delta I_h(t)\delta I_h(t)>$, and $<\delta I(t)\delta I_h(t)>$ are shown in Fig.~\ref{CFs} \cite{sonnino}.
%%%%%%%%%%%%%%%%%%%%%%%%%%%%%%%%%%%%%%%%
\begin{figure}[hbt!]
\hskip 2truecm
\includegraphics[width=8cm, height=6cm]{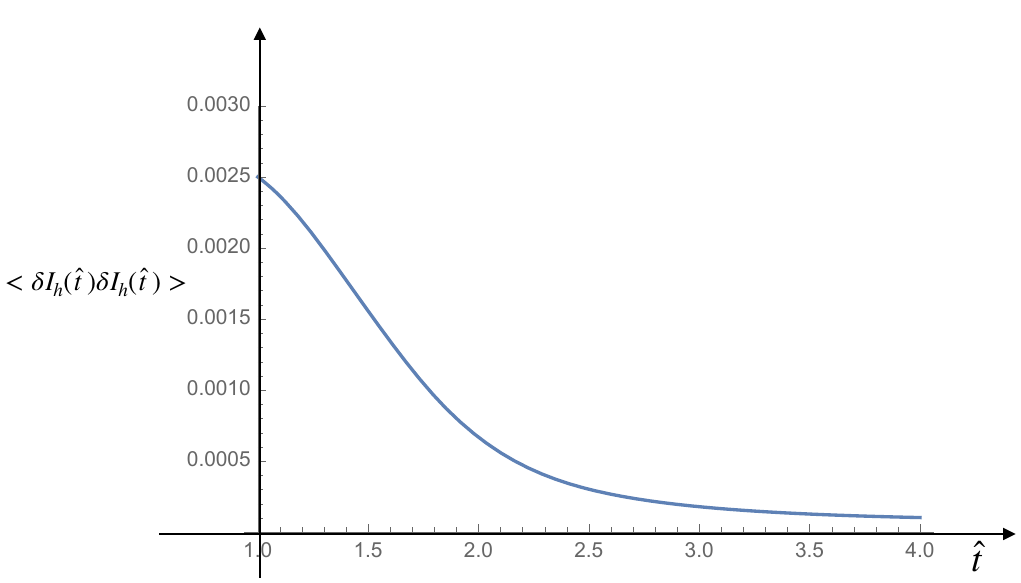}
\caption{
\textit{{\bf Correlation function $<(\delta I_h(t))^2>$ for Italy, first wave of infection by SARS-CoV-2}. Having chosen our reference state $I_{hRS}=C_h$,  we get a typical correlation function for random fluctuations \textit{at equilibrium}.} 
}
\label{CFIhIh}
\end{figure}
%%%%%%%%%%%%%%%%%%%%%%%%%%%%%%%%%%%%%%%%
%%%%%%%%%%%%%%%%%%%%%%%%%%%%%%%%%%%%%%%%
\begin{figure}[hbt!]
\hskip 2truecm
\includegraphics[width=8cm, height=6cm]{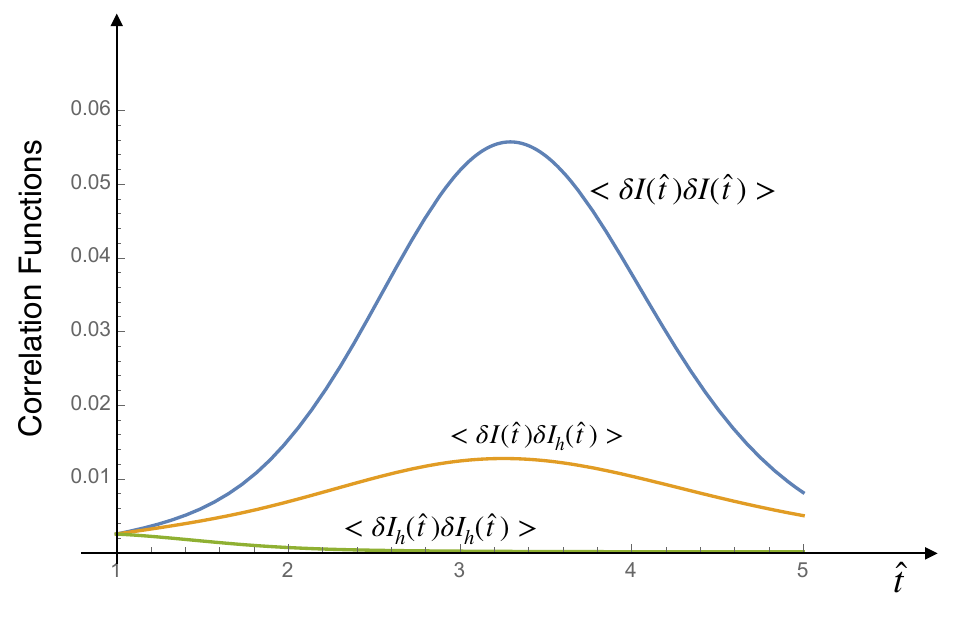}
\caption{
\textit{{\bf Correlation functions} $<\delta I(t)\delta I(t)>$, $<\delta I_h(t)\delta I_h(t)>$, and $<\delta I(t)\delta I_h(t)>$.} 
}
\label{CFs}
\end{figure}
%%%%%%%%%%%%%%%%%%%%%%%%%%%%%%%%%%%%%%%%

\section{Modelling the Spreading of the SARS-CoV-2 in Presence of the Lockdown and Quarantine Measures through the Kinetic-Type Reactions}\label{kinetic}
We are now in a position to propose a more realistic model governing the dynamics of the infectious, recovered, and deceased people when population is subject to lockdown and quarantine measures imposed by governments. We shall see that the combined effect of the restrictions measures with the action of the Hospitals and Health Institutes is able to contain and even dampen the spread of the SARS-CoV-2 epidemic. The dynamics of the entire process will be obtained by taking into account the theoretical results summarised in the previous Sections and in particular \cite{sonnino1,sonnino2,sonnino}) and by adopting a {\it kinetic-type reactions} approach \cite{sonnino3,sonnino4}. As we did in Section~\ref{DM2}, also here, the dynamics of the Health Institutes is obtained by taking inspiration from the Michaelis- Menten’s enzyme-substrate reaction model (the so-called \textit{MM reaction} \cite{MM1,MM3}). We recall that in this framework, the {\it enzyme} is associated to the {\it available hospital beds}, the {\it substrate} to the \textit{infected people}, and the \textit{product} to the {\it recovered people}, respectively. In other words, everything happens as if the hospitals beds act as a {\it catalyser} in the hospital recovery process \cite{sonnino,sonnino4}. In addition, the time-delay for recovery or death processes will duly be taken into account. More specifically, in this, more sophisticated, model the entire dynamics is governed by eleven compartments which, for easy reference, we list below:

\noindent $S$ = Number of susceptible people. This number concerns individuals not yet infected with the disease at time $t$, but they are susceptible to the disease of the population;

\noindent $S_L$ = Number of susceptible people subject to the lockdown measures;

\noindent $I_h$ = Number of hospitalised infected people;

\noindent $I_Q$ = Number of people in quarantine. This number concerns individuals who may have the virus after being in close contact with an infected person;

\noindent $I$ = Number of people who have been infected and are able of spreading the disease to those in the susceptible category (in this compartment, $I_h$ and $I_Q$ are not accounted);

\noindent $r_h$ = Cumulative  recovered people previously hospitalised;

\noindent $R$ = Cumulative number of recovered people (by excluding people previously hospitalised) meaning specifically individuals having survived the disease and now immune. Those in this category are not able to be infected again or to transmit the infection to others;

\noindent $d_h$ = Cumulative number of people previously hospitalised dead for COVID-19;

\noindent $D$ = Cumulative number of dead people (by excluding the compartment $d_h$), for COVID-19;

\noindent $L$ = Number of \textit{inhibitor sites} mimicking lockdown measures:

\noindent $Q$ = Number of \textit{inhibitor sites} mimicking quarantine measures.

\noindent In addition, $N$, defined in Eq.~(\ref{6.4}), denotes the number of total cases.

\noindent We shall proceed as follows. In Section~\ref{ODE} we derive the deterministic Ordinary Differential Equations (ODSs) governing the dynamics of the infectious, recovered, and deceased people. The lockdown and quarantine measures are modelled in Subsection~\ref{LQM}. The dynamics of the hospitalised individuals (i.e., the infectious, recovered, and deceased people) can be found in Subsection~\ref{H}. As previously mentioned, the corresponding ODEs are obtained by considering the {\it MM reaction model}. The equations governing the dynamics of the full process and the related {\it basic reproduction number} are reported in Section~\ref{TODEs} and Section~\ref{BRN}, respectively. It is worth mentioning that in   it is shown \cite{sonnino4} that, in absence of the restrictive measures and neglecting the role of the Hospitals and the delay in the reactions steps, our model reduces to the classical {\it Susceptible-Infectious-Recovered-Deceased-Model} (SIRD-model \cite{sird}). Finally, Section~\ref{Applications} shows the good agreement between the theoretical predictions with real data for Belgium, France and Germany.

\subsection{\bf Model for COVID-19 in Presence of the Lockdown and Quarantine Measures}\label{ODE}

\noindent As said, the population is assigned to compartments with labels $S$, $I$, $R$ $D$ etc. The dynamics of these compartments is generally governed by deterministic ODEs, even though stochastic differential equations should be used to describe more realistic situations \cite{sonnino3,sonnino4}. In this Section, we shall derive the deterministic ordinary differential equations obeyed by compartments. This task will be carried out by taking into account the theoretical results recently appeared in literature \cite{sonnino1,sonnino2,sonnino} without neglecting the delay in the reactions steps.

\subsection{Modelling the Susceptible People}

\noindent If a susceptible person encounters an infected person, the susceptible person will be infected as well. So, the scheme simply reads
\begin{equation}\label{S1a}
S + I \xrightarrow{\mu} 2I
\end{equation}
\subsubsection{Modelling the Lockdown and Quarantine Measures}

\noindent The lockdown measures are mainly based on the isolation of the susceptible people, (eventually with the removal of infected people by hospitalisation), but above all on the removal of susceptible people. 
\vskip0.2cm

\subsection{Modelling the Lockdown and Quarantine Measures with Chemical Interpretation}\label{LQM}

\noindent It is assumed the lockdown and quarantine measures are modelled by some kind of inhibitor reaction where the susceptible people and the infected can be \textit{trapped} into inactive states $S_L$ and $I_Q$, respectively. Indicating with $L$ and $Q$ the Inhibitor sites mimicking the lockdown and the quarantine measures respectively, we get
\begin{align}\label{LQM1}
&S + L \iff[k_{LMax}-k_L]{k_L} S_L\\
&I  \xrightarrow{k_Q} I_Q\Longrightarrow{k_{QR}, \ t_{QR}} R\nonumber
\end{align}
\noindent In the scheme~(\ref{LQM1}), symbol $\implies$ stands for a \textit{delayed reaction} just like \textit{enzyme degradation processes} for instance. Here, $L_{max}=S_L+L$ hence, if $L\simeq L_{Max}$, an almost perfect lockdown measures would totally inhibit virus propagation by inhibiting all the susceptible people $S$ and the infected people $I$. A not so perfect lockdown measures would leave a fraction of $I$ free to spread the virus. The number of inhibitor sites maybe a fraction of the number of the infected people. Fig.~\ref{LEP}. shows the behaviour of the lockdown efficiency parameter adopted in our model. For simplicity, we have chosen a parameter which is constant $k_{LMax}\neq0$ inside the time-interval $t_1\leq t\leq t_2$ and vanishes outside it. The \textit{inverse Lockdown efficiency parameter} is $k^{-1}_L=k_{LMax}-k_L$, which is equal to $k_{LMax}$ outside the door and vanishes inside the the interval $t_1\leq t\leq t_2$.
%%%%%%%%%%%%%%%%%%%%%%%%%%%%%%%%%%%%%%%%%%
\begin{figure}[hbt!]
\hskip 0.5truecm
\includegraphics[width=10cm, height=6cm]{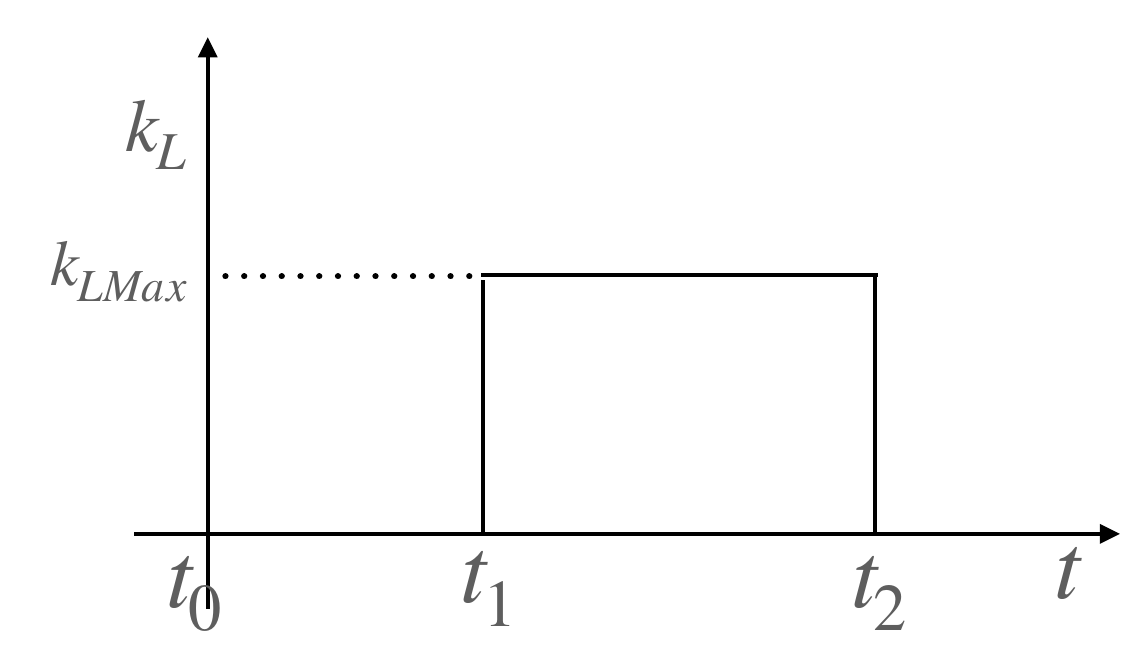}
\caption{
\textit{{\bf Lockdown Efficiency Parameter.} For simplicity, in our model the lockdown efficiency parameter $k_L$ is a \textit{door-step function}. This function is constant, $K_{LMax}\neq 0$,within the range $t_1\leq t\leq t_2$ and zero outside it.} 
}
\label{LEP}
\end{figure}
%%%%%%%%%%%%%%%%%%%%%%%%%%%%%%%%%%%%%%%%%%
Finally, from Schemes~(\ref{S1}) and (\ref{LQM1}), we get the O.D.E.s for $S$, $L$, $Q$, and $I_Q$:
\begin{align}\label{LQM2}
&{\dot S}=-\mu SI - k_L S(L_{Max}-S_L)+(1-k_L)(L_{Max}-L)\\
&{\dot S}_L=k_L SL-k^{-1}_LS_L\nonumber\\
&{\dot I_Q}=k_Q I-\chi{ I_Q}_{(t-t_R)}\nonumber
\end{align}
\noindent with the \textit{dot} above the variables denoting the \textit{time derivative}.
\noindent 

\subsection{ O.D.E. for the Total Recovered People}

\noindent At the first approximation, the O.D.E. for the \textit{total recovered people} $R$ (i.e. the total individuals having survived the disease) is trivially obtained by considering the following \textit{kinetic scheme}:
\begin{align}\label{R1}
&I \Longrightarrow{\chi ,\ t_R} R\\
&I_Q \Longrightarrow{k_{QR} ,\ t_{QR}} R\nonumber 
\end{align}
\noindent That is, the rate of $R_t$ is approximatively proportional to the number of the infected people $I$ at time $t$ i.e.\footnote{Notice that the first \textit{reaction} in the scheme Eq.~(\ref{R1}) is the dynamic equation for the total recovered people adopted in the SIRD-model \cite{sird}.}.
\begin{equation}\label{R}
{\dot R}=\chi I_{(t-t_R)}+\chi R_{(t-t_R)} 
\end{equation}
\noindent where we have introduced the time-delay $t_R$ (the number of the recovered people at time time $t$ is proportional to the infected people at time $t-t_R$). However, it is useful to clarify the following. In Eqs~(\ref{R1}), $R$ stands for the {\it total number of the recovered people} (i.e. the number of the recovered people previously hospitalised, plus the number of the asymptomatic people, plus the infected people who have been recovered without being previously hospitalised). The natural question is: {\it how can we count $R$ and compare this variable with the real data ?}. The current statistics, produced by the Ministries of Health of various Countries, concern the people released from the hospitals. Apart from Luxembourg (where the entire population has been subject to the COVID-19-test), no other Countries are in a condition to provide statistics regarding the total people recovered by COVID-19. Hence, it is our opinion that the equation for $R$, is not useful since it is practically impossible to compare $R$ with the experimental data. We then proceed by adopting approximations and to establish the differential equation whose solution can realistically be subject to experimental verification. More specifically:

\noindent Firstly, we assume that $R$ is given by three contributions:
\begin{equation}\label{R2}
R=r_h+r_{A}+r_{I}
\end{equation}
\noindent with $r_h$, $r_{A}$, and $r_{I}$ denoting the \textit{total number of the recovered people previously hospitalised}, \it the total number of asymptomatic people}, and the \textit{total number of people immune to SARS-CoV-2}, respectively. 

\noindent Secondly, we assume that the two contributions $r_{A}$  and $r_{I}$ are negligible i.e. we set $r_A\approx 0$ and $r_{I}\approx 0$ \footnote{We consider that the SARS-CoV-12  has just appeared for the first time. So, we do not consider the asymptomatic people who are immune to the virus without any medical treatment.}. 

\subsection{O.D.E. for the Recovered People in the Hospitals}\label{H}

\noindent Now, let us determine the dynamics for the recovered people in the hospitals. So, we account people who are only traced back to hospitalised infected people. We propose the following model\footnote{Our model is inspired by \text{Michaelis-Menten's enzyme-substrate reaction}. Of course, the reverse \textit{MM reaction} has no sense in our case and, consequently, the \textit{kinetic constant} is equal to zero.}:
\begin{align}\label{H1}
&I + b_h \xrightarrow{k_1} I_h\Longrightarrow{k_r, \ t_r} r_h+b_h\\
&\qquad\qquad\ \! I_h \Longrightarrow{k_d,\ t_d} d_h+b_h\nonumber
\end{align}
\noindent with $b_h$ denoting the number of available \textit{hospital beds}, $I$ the number of {\it infected people}, $I_h$ the number of \textit{infected people blocking an hospital bed}, $r_h$ the number of \textit{recovered people previously hospitalised}, and $d_h$ the number of {\it people deceased in the hospital}. Of course, 
\begin{equation}\label{H2}
I_h+b_h=C_h=const.\qquad {\rm where}\quad{C_h={\rm Total\ hospital's\ capacity}}
\end{equation}
\noindent The dynamic equations for the processes are then:
\begin{align}\label{H3}
&{\dot I}_h=k_1I(C_h-I_h)-k_r{I_h}_{(t-t_r)}-k_d{I_h}_{(t-t_d)}\\
&{\dot r}_h=k_r{I_h}_{(t-t_r)}\nonumber\\
&{\dot d}_h=k_d {I_h}_{(t-t_d)}\nonumber
\end{align}
\noindent where $t_r$ and $t_d$ are the \textit{average recovery time delay} and the {\it average death time delay}, respectively, and we have taken into account Eq.~(\ref{H2}) i.e., $b_h=C_h-I_h$. In general $t_r\neq t_d\neq 0$. Of course, the variation of $r(t)$ over a period $\Delta t$ is:
\begin{equation}\label{H4}
\Delta {r_h}_t={r_h}_t-{r_h}_{(t-\Delta t)}
\end{equation}

\subsection{O.D.E. for People Tested Positive to COVID-19}

\noindent The number of the infected people may be modelled by the following \textit{kinetic scheme}
\begin{align}\label{I1}
&S + I \xrightarrow{\mu} 2I\\
&I \Longrightarrow{\chi ,\ t_R} R\nonumber\\
&I \Longrightarrow{\alpha ,\ t_D} D\nonumber\\
&I + b \rightarrow{k_1} I_h\nonumber\\
&I \rightarrow{k_Q} I_Q\nonumber
\end{align}
\noindent The scheme~(\ref{I1}) stems from the following considerations
\begin{description}
\item{{\bf a)}} If a susceptible person encounters an infected person, the susceptible person will be infected ;
\item{{\bf b)}} The infected people can either survive and, therefore, be recovered after an average time-delay $t_R$, or die after an average time-delay $t_D$;
\item{{\bf c)}} The schemes~(\ref{LQM1}) and (\ref{H1}), respectively, have been taken into account.
\end{description}
\noindent The differential equation for the infected people is reads then
\begin{equation}\label{I2}
{\dot I}=\mu SI-k_Q IQ-k_1I(C_h-I_h)-\chi I_{(t-t_R)}-\alpha  I_{(t-t_D)}
\end{equation}

\subsection{ O.D.E. for Deaths}

\noindent In this model, we assume that the rate of death is proportional to the infected people, according to the scheme~(\ref{I1}). By also taking into account the scheme~(\ref{LQM1}), we get 
\begin{equation}\label{D1}
I \Longrightarrow{\alpha ,\ t_D} D
\end{equation}
\noindent and the corresponding O.D.E. for deaths reads
\begin{equation}\label{D2}
{\dot D}=\alpha I_{(t-t_D)}
\end{equation}

\subsection{Set of O.D.E.s for the Spread of SARS-CoV-2 when the Lockdown and the Quarantine Measures are Adopted}\label{TODEs}

\noindent By collecting the above O.D.E.s, we get the full system of differential equations governing the dynamics of the number of the infected people, the total number of the recovered people previously hospitalised and the total number of deceased peopled, when the lockdown and the quarantine measures are adopted
\begin{align}\label{6.1}
&{\dot S}=-\mu SI - k_L S(L_{Max}-S_L)+k^{-1}_LS_L\qquad{\rm with}\quad k^{-1}_L=k_{Max}-k_L\\
&{\dot S}_L=-k_L S(L_{Max}-S_L)+k^{-1}_LS_L\nonumber\\
&{\dot I}=\mu SI-k_Q I-k_1I(C_h-I_h)-\chi I_{(t-t_R)}-\alpha  I_{(t-t_D)}\nonumber\\
&{\dot I}_{h}=k_1I(C_h-I_h)-k_r{I_h}_{(t-t_r)}-k_d{I_h}_{(t-t_d)}\nonumber\\
&{\dot I}_{Q}={k_{Q}}I_t-\chi {I_Q}_{(t-t_{R})}\nonumber\\
&{\dot r}_h=k_r{I_h}_{(t-t_r)}\nonumber\\
&{\dot R}=\chi I_{(t-t_R)}+\chi {I_Q}_{(t-t_{R})}\nonumber\\
&{\dot d}_h=k_d {I_h}_{(t-t_d)}\nonumber\\
&{\dot D}=\alpha I_{(t-t_D)}\nonumber
\end{align}
\noindent From Eqs~(\ref{6.1}) we get 
\begin{equation}\label{6.2}
S+S_L+I+I_Q+I_h+R+r_h+D+d_h=const.
\end{equation}
\noindent or, by taking into account that $S+S_L=S_{Tot.}$, $R+r_h=R_{Tot.}$, $D+d_h=D_{Tot.}$, and $I+I_Q+I_h=I_{Tot.}$ we get
\begin{equation}\label{6.3}
S_{Tot.}+I_{Tot.}+R_{Tot.}+D_{Tot.}=const.
\end{equation}
\noindent The number of total cases $N$ is defined as
\begin{equation}\label{6.4}
N=I_{Tot.}+r_h+D_{Tot.}
\end{equation}

\subsection{The Basic Reproduction Number}\label{BRN}

\noindent We note that, in absence of the lockdown and the quarantine measures, the dynamics of the infectious class depends on the following ratio:
\begin{equation}\label{7.1}
R_0= \frac{\mu}{\chi+\alpha} \frac{S}{N_{Tot.}}
\end{equation}
\noindent with $N_{Tot.}$ denoting the {\it Total Population}. $R_0$ is the {\it basic reproduction number}. This parameter provides the expected number of new infections from a single infection in a population by assuming that all subjects are susceptible \cite{baley,sonia}. The epidemic only starts if $R_0$ is greater than $1$, otherwise the spread of the disease stops right from the start.

\subsection{Application of the Model and Appearance of the Second Wave of SARS-CoV-2 Infection}\label{Applications}
\noindent Let us now apply our model to the case of a small Country, Belgium, and to other two big Countries, France and Germany. Real data are provided by the various National Health agencies (Belgium-{\it Sciensano} \cite{dataBE}; France-{\it Sant{\'e} Publique France} \cite{dataFR}; Germany -{\it Robert Koch Institut. Country data from Worldbank.org} \cite{dataGermany}) and compiled, among others, by European Centre for Disease Prevention and Control (ECDC). It should be noted that this measures does not generally provide the true new cases rate but reflect the overall trend since most of the infected will not be tested \cite{ourworldindata}. It should also be specified that real data provided by ECDC refer to the {\it new cases per day}, which we denote by $\Delta I_{new}(t)$. By definition, $\Delta I_{new}(t)$ corresponds to the new infected people generated from step $I+S\xrightarrow{\mu} 2I$ solely during 1 day, and {\it not} to the compartment $I$. Hence, the ECDC data have to be confronted vs the theoretical predictions provided by the solutions for $S(t)$ and $S_{L}(t)$ of our model, according to the relation $\Delta  I_{new}(t) = -\Delta S(t) -\Delta S_{L}(t)$. The values of the parameters used to perform these comparisons are shown in Table~\ref{table}. 
\begin{table}[htp]
\caption{List of the Parameters}
\begin{center}
\begin{tabular}{|l|c|c|c|}
\hline
Parameters & Belgium & France & Germany\\
\hline
 Density [$km^{-2}$] & 377 & 119 & 240 \\
        Surface [$km^{2}$]& 30530 & 547557 & 348560 \\
        $\mu$ [$d^{-1} km^{2}$]& 0.00072 & 0.002 & 0.00093 \\
        $\mu$ after $L_{1}$ &  0.000288 & 0.00087 & 0.000387\\
        $\chi$ [$d^{-1}$]& 0.062 & 0.062 & 0.0608 \\ 
        $\alpha$ [$d^{-1}$]& 0.05 $\chi$  & 0.05 $\chi$  & 0.02 $\chi$ \\
        $k_{L} $  [$d^{-1}$]& 0.07 & 0.06 & 0.06 \\
        $k_Q$ [$d^{-1}$]& 0.02 & 0.01 & 0.01 \\
        $L_m$ [$km^{-2}$]  & 377.0 & 119 & 240 \\
        $k_1$ [$d^{-1} km^{2}$] & 0.01 & 0.01 & 0.01 \\
$k_d + k_r$ [$d^{-1}$] & 0.2 & 0.2 & 0.21 \\
        $\frac{k_d}{k_r} $ & 0.5  & 0.5 & 0.1 \\
        $t_r $ [$d$]& 7  & 7 & 7 \\
        $t_d$ [$d$]& 7 & 7 & 7\\
        $t_R$ [$d$]& 8 & 8 & 8 \\
        $t_D$ [$d$]& 8 & 8 & 8 \\
        $C$ [$km^{-2}$]  & 0.0655 &  0.0091 & 0.023 \\
        $I(60)$ [$km^{-2}$]  & 0.0023 & 0.0018 & 0.0014 \\
        Start $L_{1}$ [$d$]& 77 & 71 & 76 \\
        End $L_{1}$ [$d$]& 124 & 131 & 125 \\
        Start $L_{2}$ [$d$]& 306 & 303 & 306 \\
        \hline
\end{tabular}
\end{center}
\label{table}
\end{table}
\noindent Initial $\mu$ and $k_1$ values have been estimated (fitted) from the measurements using the short period at the start of the pandemic using simple exponential solution valid during that period. $I(60)$ is the initial value of infected from March 1, 2020 (day 60) obtains from the respective measurements. Hospital capacity is evaluated from the different Countries published capacity. Lockdown starting dates and duration are retrieved from each country Covid policies \cite{countries_measures}. Other parameters have been estimated by best fit of new cases during the first wave. We draw attention to the fact that the constants $\mu$, $L_m$, $k_1$, $C$, and $I(60)$ have been normalised with respect to the surface of the Country. As it can be seen, the values of the re-normalised constants are the same values, at least in terms of orders of magnitude, irrespective of the magnitude of the Country in question (Belgium, France, and Germany). However, we are aware that the interpretation may vary from one Country to another. Finally, numerical solutions to the time delayed ordinary differential equations have been obtained by making use of the MATLAB \texttt{dde23} module with a constant time delay. Discontinuities have been avoid for the historical values and a Runge-Kutta implicit scheme is used \cite{matlab}.

\begin{figure}[h]
\begin{center}
\centering\centering\includegraphics[scale=.30]{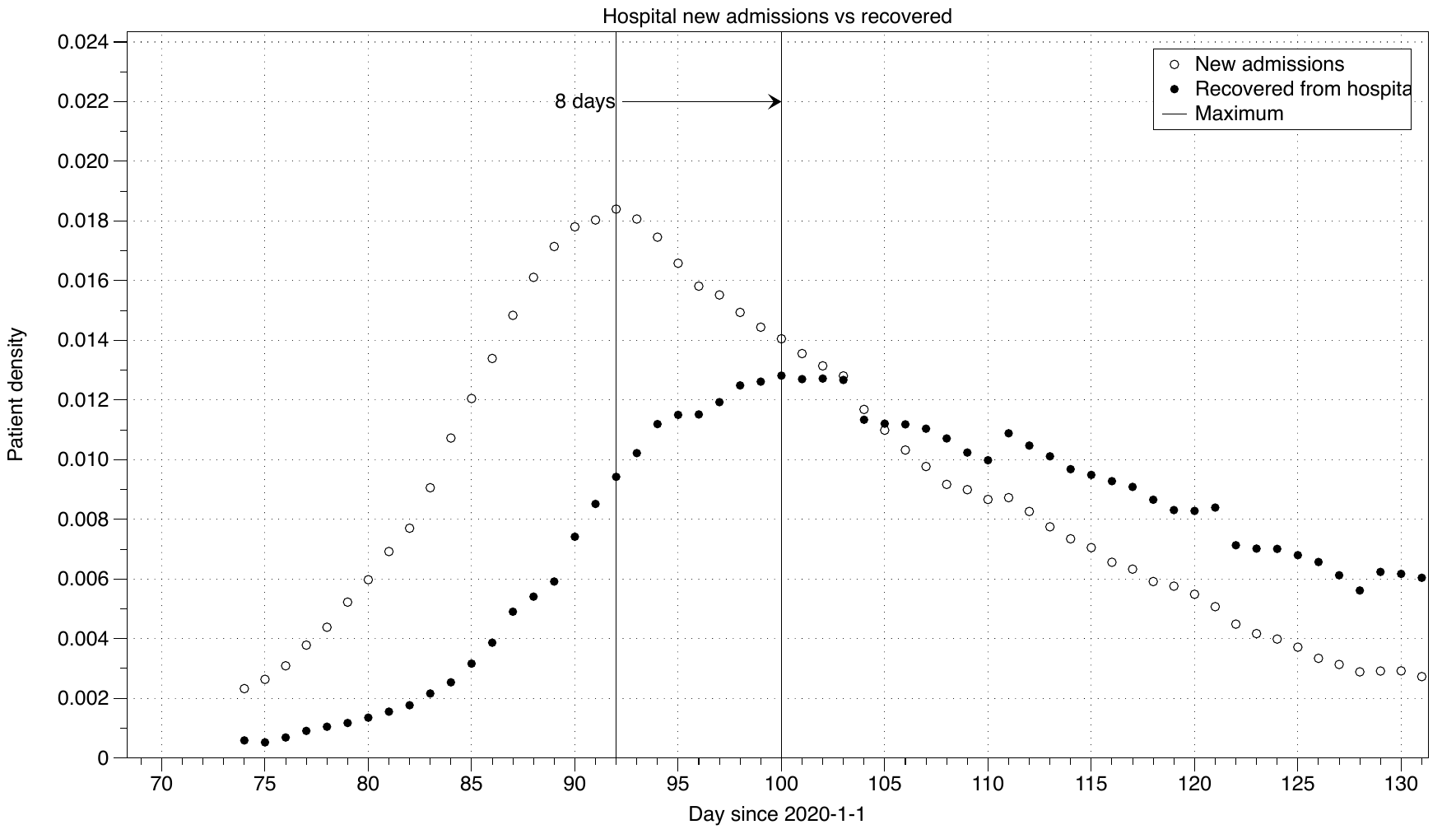}
\caption{\textit{Estimation of the time-delay. The time-delays have been estimated by considering the \textit{time-interval} between the peak of the new admission and the peak of the recovered people from hospitals. This figure corresponds to the Belgian case.}}
\label{delay}
\end{center}
\end{figure}

\noindent During the first lockdown, Countries have taken various actions to limit Coronavirus spreading (social distancing, wearing masks, reducing high density hotspots etc.). In order to include these measures in a simple way, we assumed that the net effect is to reduce the actual infection kinetic rate $\mu$ by some constant factor. This is given in the table as $\mu$ after $L_1$. Note that the transition occurs instantaneously in our model; this leads to the sharp drop in the total infected at that time shown in the figures. Other parameters are tuned to account for the actual variability of $\Delta I_{new}$ (but not its absolute value) and official number of deaths ($D_{Tot.}(t)=D(t) + d_h(t)$). The delay for recovery and death processes has been estimated from the measurements of hospitalisation recovery in a Country. For instance, Fig.~\ref{delay} shows the estimation of the recovery time-delay for Belgium: it corresponds to the \textit{time-interval} between the peak of the new admission and the peak of the recovered people from hospitals. A similar procedure has been adopted for estimating the recovery and death time-delays also for France and Germany.
%%%%%%%%%%%%%%%%%%%%%%%%%%%%%%%%%%%%%%%%
%%%%%%%%%%%%%%%%%%%%%%%%%%%%%%%%%%%%%%%%

\noindent $\bullet$ {\bf Belgian Case}.

\begin{figure}[h]
\centering\centering\includegraphics[scale=.30]{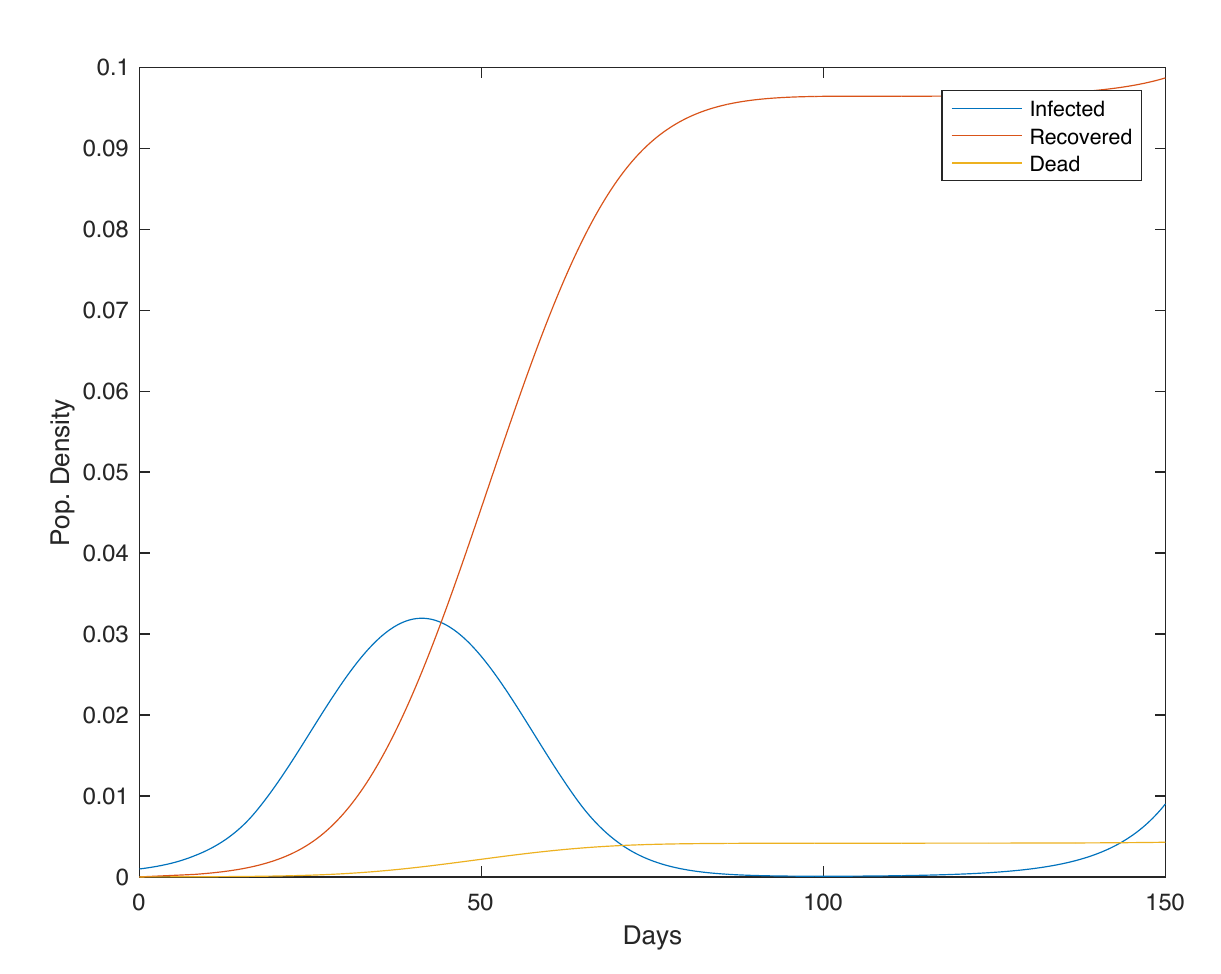}
\caption{\textit{Theoretical solutions for infectious ($I$), cumulative number of recovered  people ($R$) and deaths ($D$) for Belgium.}}
\label{BE_IRD}
\end{figure}

\begin{figure}[h]
\centering\centering\includegraphics[scale=.30]{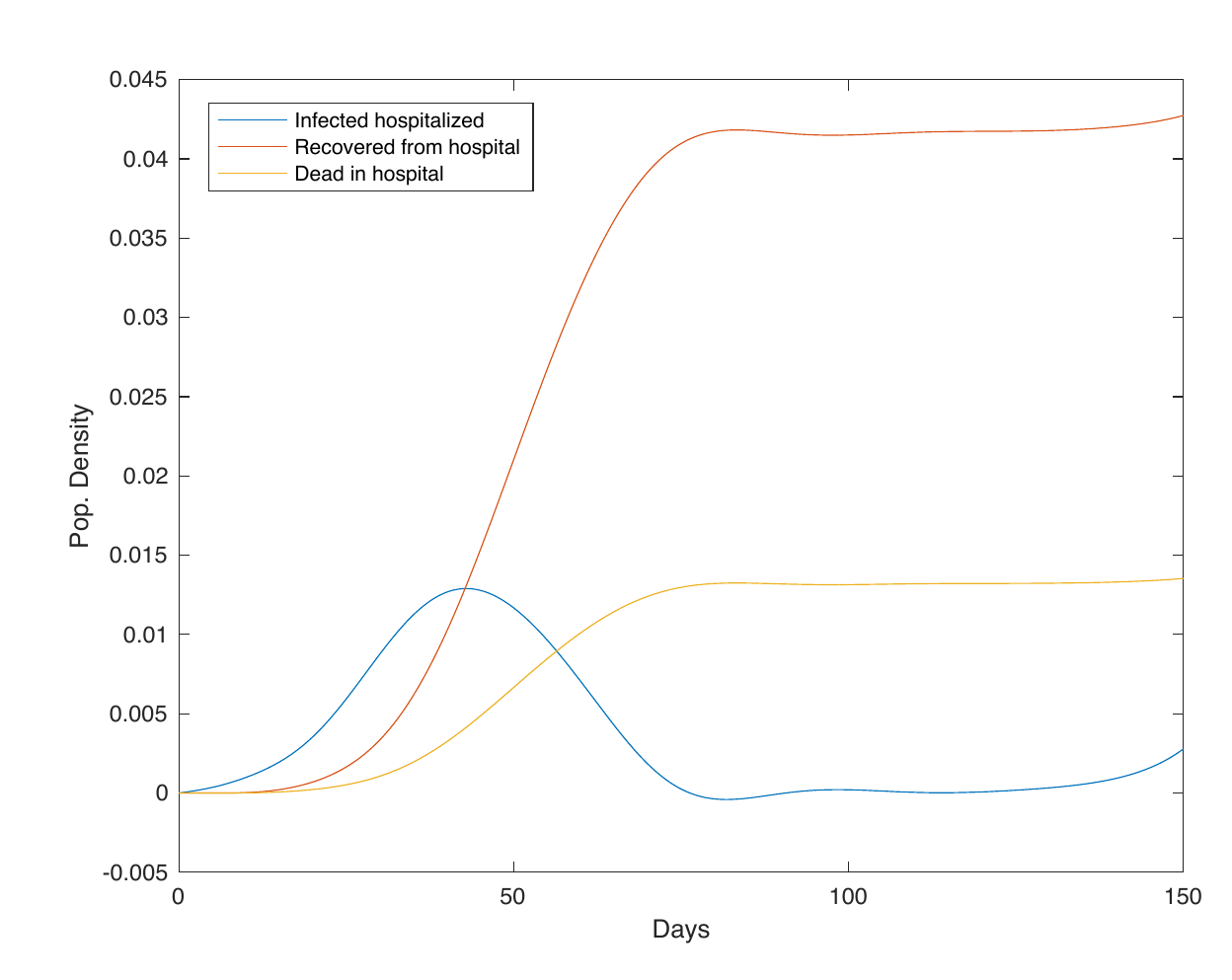}
\caption{\textit{Theoretical solutions for hospitalised infectious ($I_h$), total recovered ($r_h$) and total deceased ($d_h$) people, previously hospitalised, for Belgium.}}
\label{BE_IRD_h}
\end{figure}

\begin{figure*}[htb]
    \centering\centering\includegraphics[scale=.30]{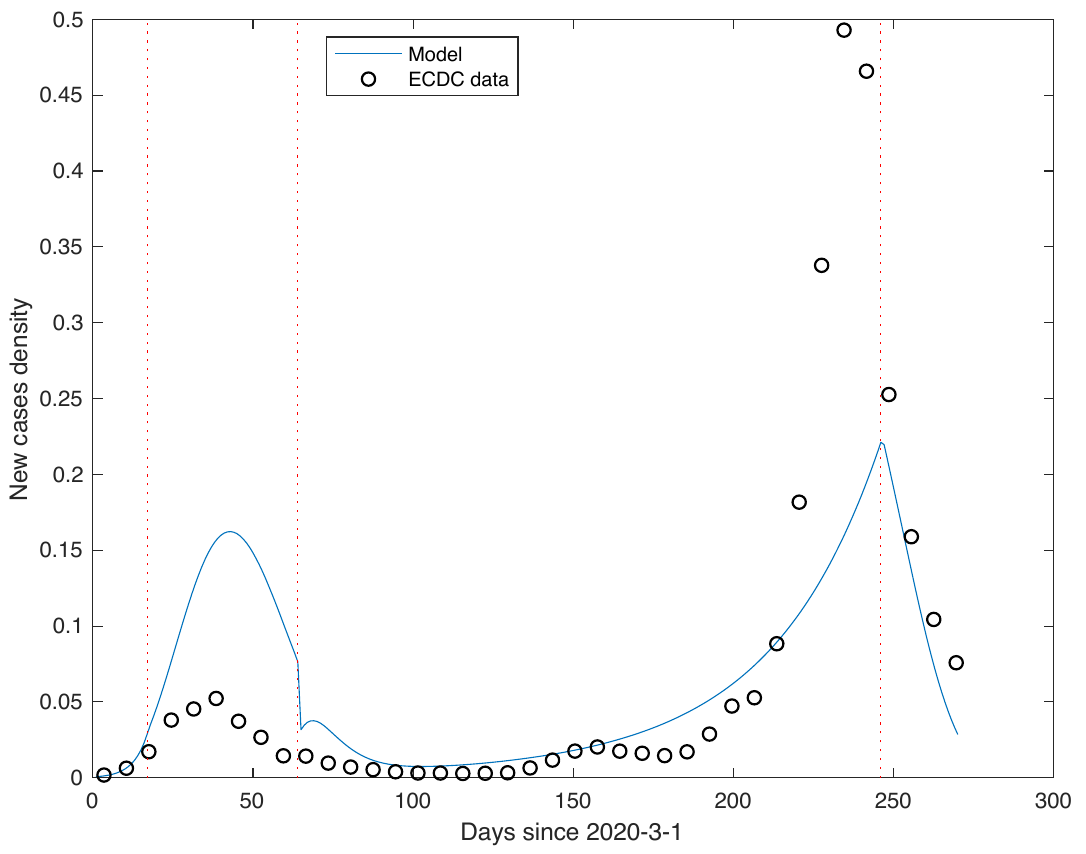}
    \caption{\textit{Comparison between the theoretical prediction for $\Delta I_{New}$ with real data provided by the data base \textit{Sciensano}, for Belgium.}}
    \label{I_new_BE}
      \centering\centering\includegraphics[scale=.30]{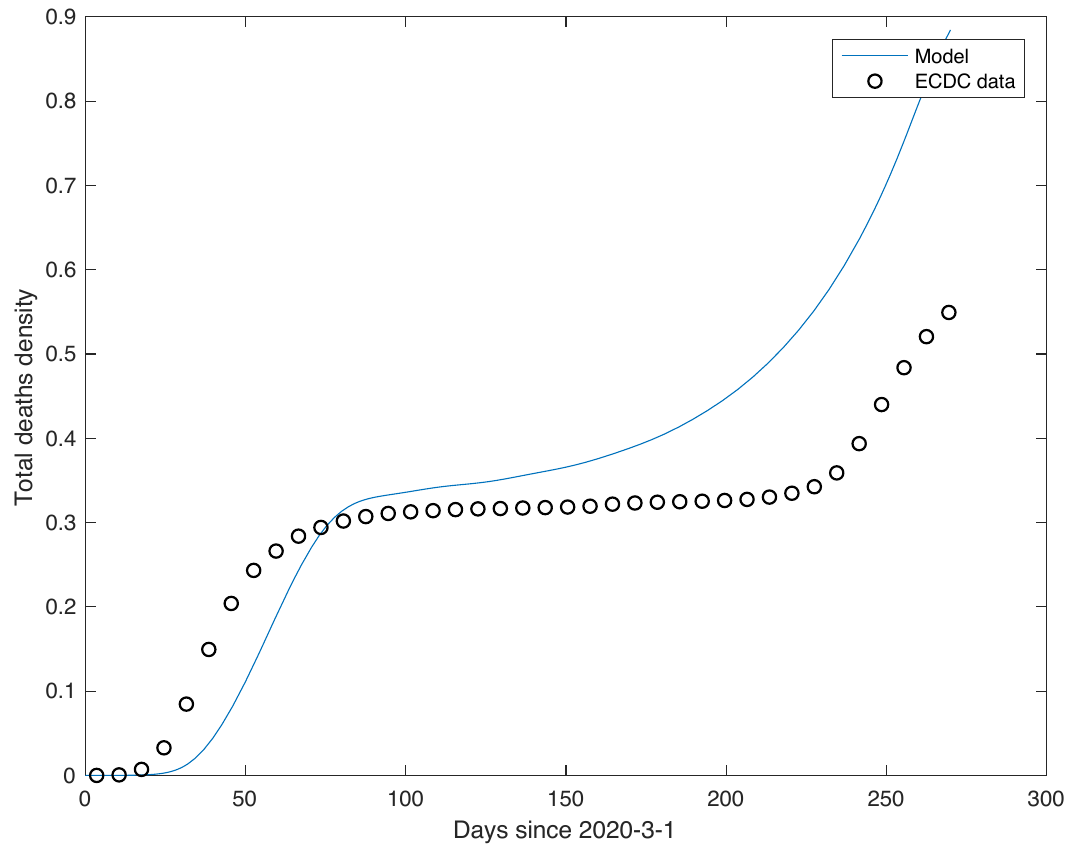}
      \caption{\textit{Comparison between the theoretical solution of our model for Deaths with real data provided by the database \textit{Sciensano}, for Belgium.}}
      \label{D_BE}
\end{figure*}

\noindent Figs~(\ref{BE_IRD}) refer to the Belgian case. In particular, Fig~(\ref{BE_IRD}) shows the solutions of our model for the infectious ($I$), total recovered ($R$) and total deceased ($D$) people. Fig.~(\ref{BE_IRD_h}) illustrates the theoretical solutions for hospitalised infectious ($I_h$), the total recovered ($r_h$) and total deceased ($d_h$) people previously hospitalised. 
%%%%%%%%%%%%%%%%%%%%%%%%%%%%%%%%%%%%%%%%
%%%%%%%%%%%%%%%%%%%%%%%%%%%%%%%%%%%%%%%%
%%%%%%%%%%%%%%%%%%%%%%%%%%%%%%%%%%%%%%%%
\noindent Figs~(\ref{I_new_BE}) and (\ref{D_BE}) shows the comparison between the theoretical predictions for $\Delta I_{new}(t)$ and deaths and real data for Belgium (according to the database \textit{Sciensano}). Notice in Fig.~\ref{I_new_BE} the prediction of the \textit{second wave of infection by SARS-CoV-2}

%%%%%%%%%%%%%%%%%%%%%%%%%%%%%%%%%%%%%%%%
%%%%%%%%%%%%%%%%%%%%%%%%%%%%%%%%%%%%%%%%
\noindent $\bullet$ {\bf French Case}.

\begin{figure*}[htb]
   \centering\centering\includegraphics[scale=.30]{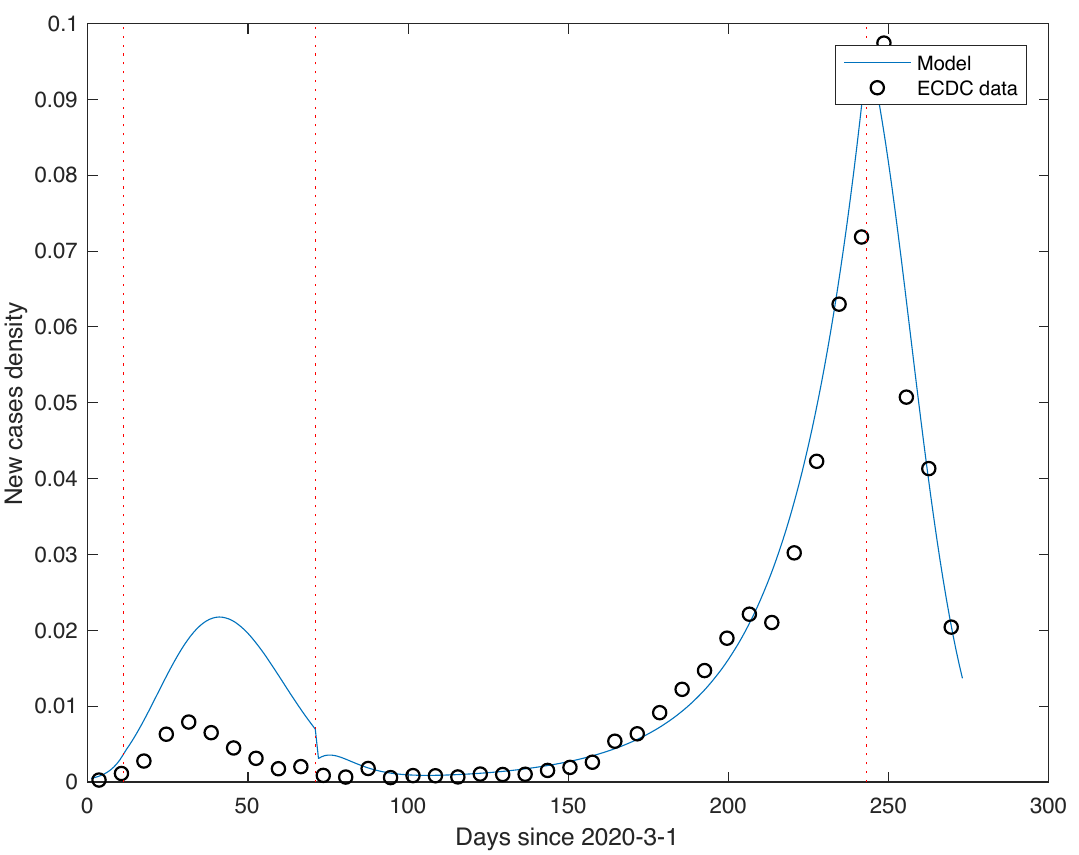}
    \caption{\textit{Comparison between the theoretical prediction for $\Delta I_{New}$ with real data provided by the data base \textit{Sant{\'e} Publique France}, for France.}}
    \label{I_new_FR}
      \centering\centering\includegraphics[scale=.30]{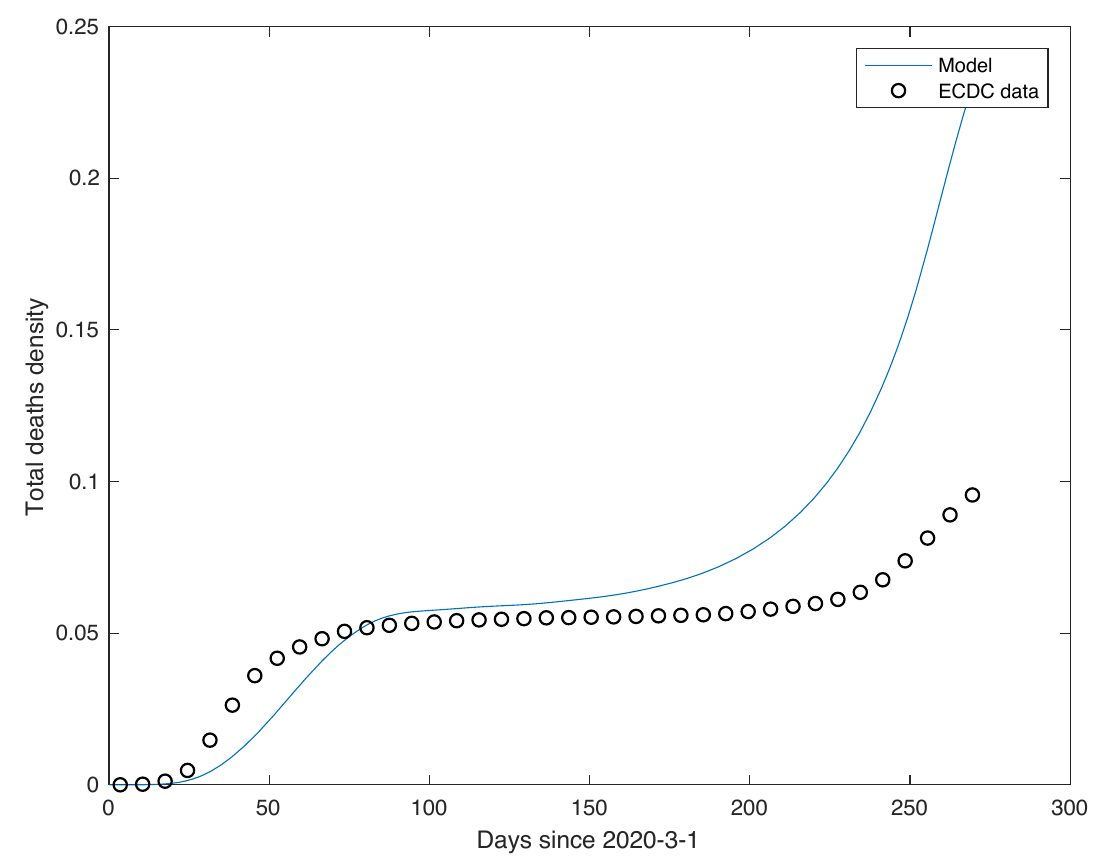}
      \caption{\textit{Comparison between the theoretical solution of our model for Deaths with real data provided by the database \textit{Sant{\'e} Publique France}, for France.}}
      \label{D_FR}
\end{figure*}

\noindent Figs~(\ref{I_new_FR}) and (\ref{D_FR}) shows the comparison between the theoretical predictions for $\Delta I_{new}(t)$ and deaths and real data for Belgium (according to the database \textit{Sant{\'e} Publique France}). Notice in Fig.~\ref{I_new_FR} the prediction of the \textit{second wave of infection by SARS-CoV-2}

%%%%%%%%%%%%%%%%%%%%%%%%%%%%%%%%%%%%%%%%
%%%%%%%%%%%%%%%%%%%%%%%%%%%%%%%%%%%%%%%%
\noindent $\bullet$ {\bf German Case}.

\begin{figure*}[htb]
    \centering\centering\includegraphics[scale=.30]{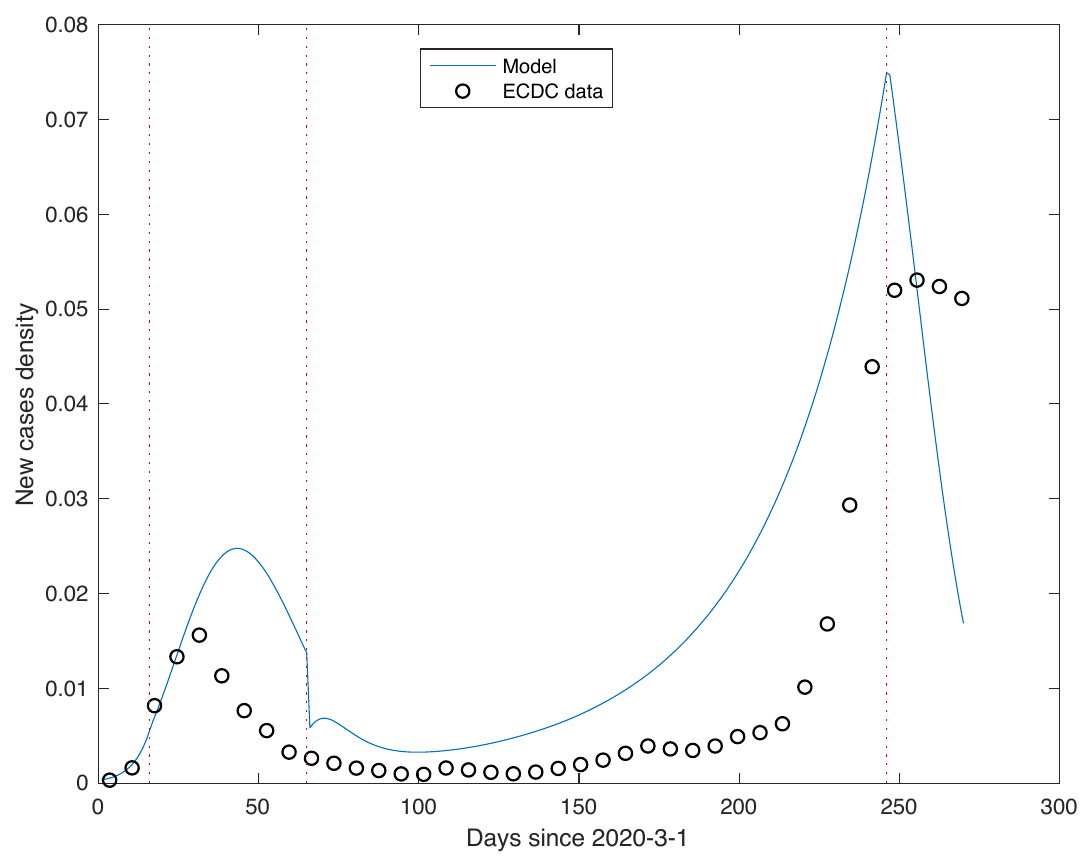}
    \caption{\textit{Comparison between the theoretical prediction for $\Delta I_{New}$ with real data provided by the data base \textit{(Robert Koch Institut. Country data from Worldbank.org}, for Germany.}}
    \label{I_new_DE}
      \centering\centering\includegraphics[scale=.30]{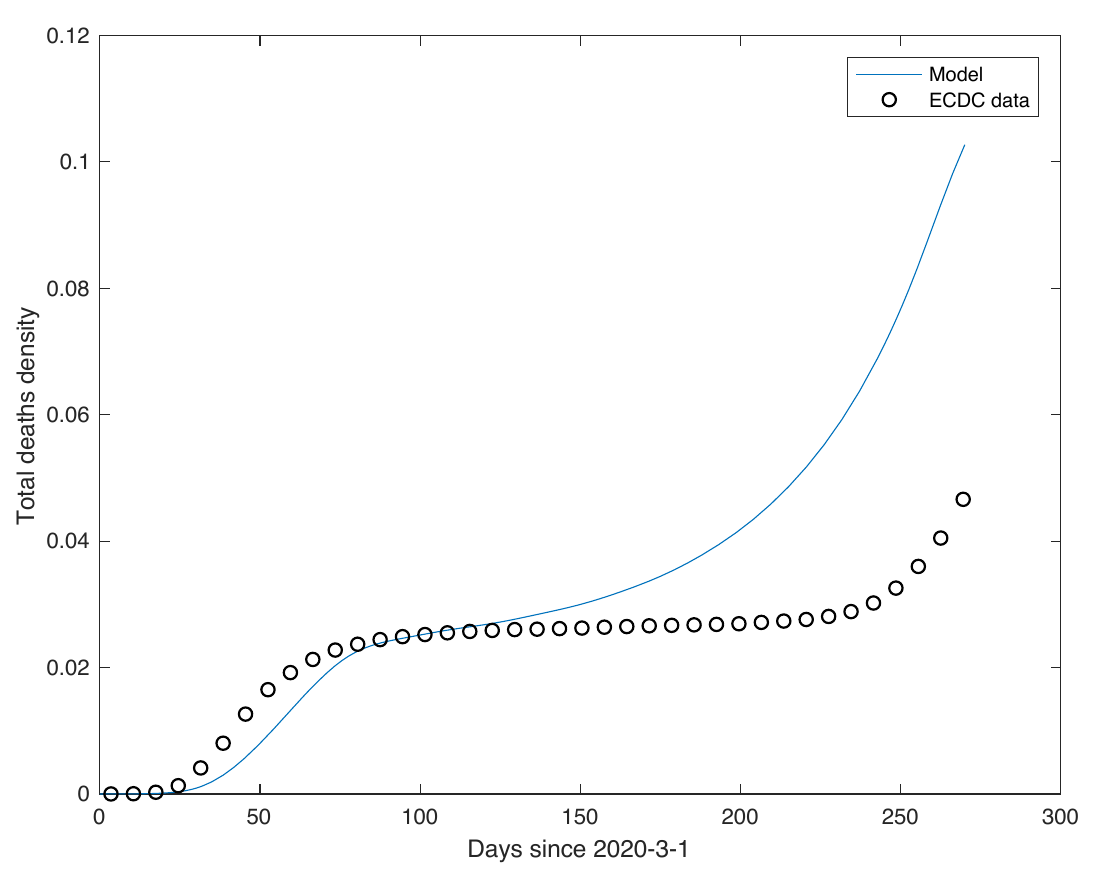}
      \caption{\textit{Comparison between the theoretical solution of our model for Deaths with real data provided by the database \textit{(Robert Koch Institut. Country data from Worldbank.org}, for Germany.}}
      \label{D_DE}
\end{figure*}

\noindent Figs~(\ref{I_new_DE}) and (\ref{D_DE}) shows the comparison between the theoretical predictions for $\Delta I_{new}(t)$ and deaths and real data for Belgium (according to the database \textit{(Robert Koch Institut). Country data from Worldbank.org}). Notice in Fig.~\ref{I_new_DE} the prediction of the \textit{second wave of infection by SARS-CoV-2}
%%%%%%%%%%%%%%%%%%%%%%%%%%%%%%%%%%%%%%%%
%%%%%%%%%%%%%%%%%%%%%%%%%%%%%%%%%%%%%%%%

\section{Modelling Descending Phase in Case of Disappearance of SARS-CoV2 Infection}\label{dp}
In Section~\ref{exp} we have modelled the descending phase on the basis of the law for the growth of a Malthusian population \cite{murray} (see Eq.~(\ref{In3}). This phase is characterised by the fact that the basic reproduction number $R_0$ in less than $1$. The full disappearance of the virus can be associated with the fact that the total number of cases $N$ (see Eq.~(\ref{6.4}) tends to reach a {\it plateau}\footnote{Notice that, in this case, $N$ is not constant since we do not assume that $N$ coincides with the total population of a country $N_{pop.}$. Hence $N_{pop.}-N$ may vary, with  $N_{pop.}=const.$}. The objective of this Section is to determine the trend of the curve of {\it positive people} during the descending phase by assuming that the total number of cases reached a maximum value, corresponding to a plateau. This task is accomplished by taking into account the appropriate equations for the recovered people and the deceased people for COVID-19 provided by System~(\ref{6.1}). Since we have assumed that $N$ reached its maximum value, during the descent phase the number of infectious people over time must satisfy a conservation equation. This allows determining the time-evolution for the positive people. 

\subsubsection{Dynamics of the recovered people and the deceased individuals}
Clearly, the {\it number of the recovered people, previously hospitalised, at the step} $n$ (i.e. $r_n$), is linked to the {\it total} number of the recovered people previously hospitalised at the step $n$ (denoted by $h_n$) by the relation
\begin{equation}\label{dp1}
r_n=h_n-h_{n-1}\quad{\rm or}\quad h_t=\sum_{n=1}^{n=t/\Delta t} r_n\quad ({\rm with}\ \  \Delta t\simeq 1\ {\rm day})
\end{equation}
\noindent where we have set $h_0=0$. Eqs~(\ref{6.1}) provide the dynamic equations for $r_n$, $R_n$ and $d_t$:
\begin{align}\label{dp2}
&{\dot r}_h=k_r{I_h}_{(t-t_r)}\\
&{\dot R}_n=\chi I_{(t-t_R)}+\chi {I_Q}_{(t-t_{R})}\nonumber\\
&{\dot d}_h=k_d {I_h}_{(t-t_d)}\nonumber
\end{align} 

\subsection{Equation for the Positive People}
\noindent Of course, during the descent phase, the number of active people $I_t$ satisfies a simple law of conservation: {\it If we are in the situation where there are no longer new cases of people tested positive for COVID-19 and if we assume that the active people cannot leave their country of origin (or else, if they do, they will be rejected by the host Country), then the number of infected people cannot but decrease either because some people are deceased or because others have been recovered}. In mathematical terms
\begin{equation}\label{dp3}
I_t=I_{0}-(h_t-h_0)-(d_t-d_0)=N_{Max}-h_t-d_t
\end{equation}
\noindent with $h_0$, $d_0$ and $I_0$ denoting the values of $h_t$, $d_t$ and $I_t$ when the infected people are evaluated at the time $t=t_{Max}$ i.e., the time that maximises the number of the total cases (for the definition of $h_t$ see the forthcoming Eq.~(\ref{dp4})). It should be noted that the conservation law~(\ref{dp3}) applies only when there are no longer new cases of people tested positive to COVID-19\footnote{So, Eq.~(\ref {dp3}) does not apply necessarily as soon as the number $n_t$ (the number of people tested positive for COVID-19) starts to decrease. Indeed, it may happen that $n_t$ decreases because, for example, the number of new cases of people tested positive is less than the number of the people who have recovered in the meantime. Conservation law~(\ref {dp3}) applies only from the moment where the number of new cases of people tested positive is strictly equal to zero.}. Here, by the {\it descending phase} we mean the phase where Eq.~(\ref {dp3}) applies. 

\subsection{Equations for the Descending Phase}
To summarise, the equations that must be satisfied for the total regression of the SARS-CoV2 infection read \cite{sonnino2}:
\begin{align}\label{dp4}
&{\dot r}_h=k_r{I_h}_{(t-t_r)}\\
&{\dot R}_n=\chi I_{(t-t_R)}+\chi {I_Q}_{(t-t_{R})}\nonumber\\
&{\dot d}_h=k_d {I_h}_{(t-t_d)}\nonumber\\
&I_t=N_{Max}-h_t-d_t\qquad\qquad\qquad\qquad\ \ {\rm with}\ \ n_{\infty}=0\nonumber\\
&h_t=\sum_{n=1}^{n=t/\Delta t} r_n\qquad\qquad\qquad\qquad\qquad\quad\ \ \!{\rm where}\ \  \Delta t\simeq 1\ {\rm day}\nonumber
\end{align}
\noindent Notice that the first three equations of system~(\ref{dp4}) are also valid during the {\it ascending-phase}. Of course, in this case, the initial conditions are $r_{t=0}=0$, $d_{t=0}=0$ and $I_{t=0}=0$. 

\subsection{Typical Trends for the Descending Phase for Italy and Belgium if $N$ Reaches a Plateau}
\begin{figure}[h]
\centering\centering\includegraphics[scale=.30]{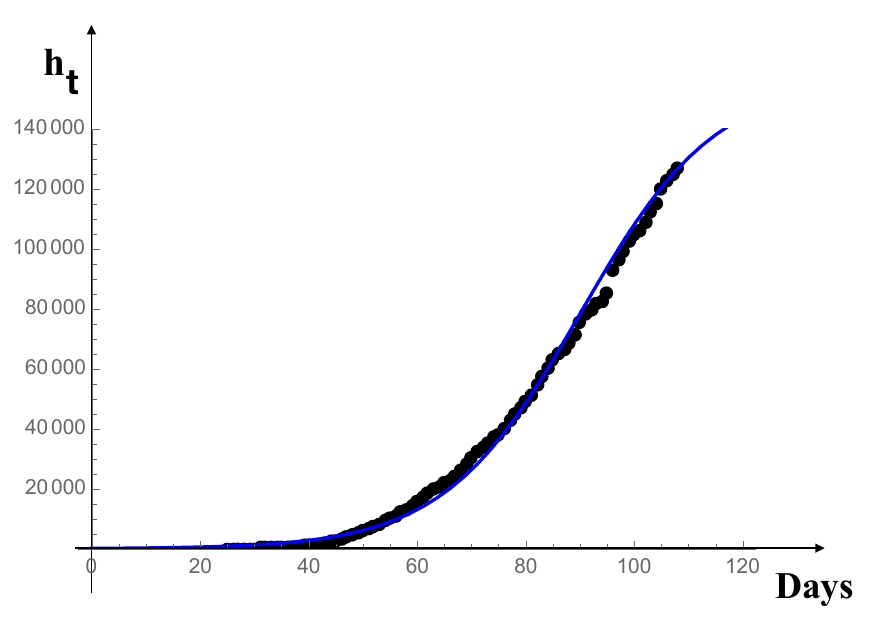}
\caption{\textit{Italy situation}. Theoretical predictions (blue line) against the experimental data (black circles) for the recovered people.}
\label{Fig.12}
\end{figure}

\begin{figure}[h]
\centering\centering\includegraphics[scale=.30]{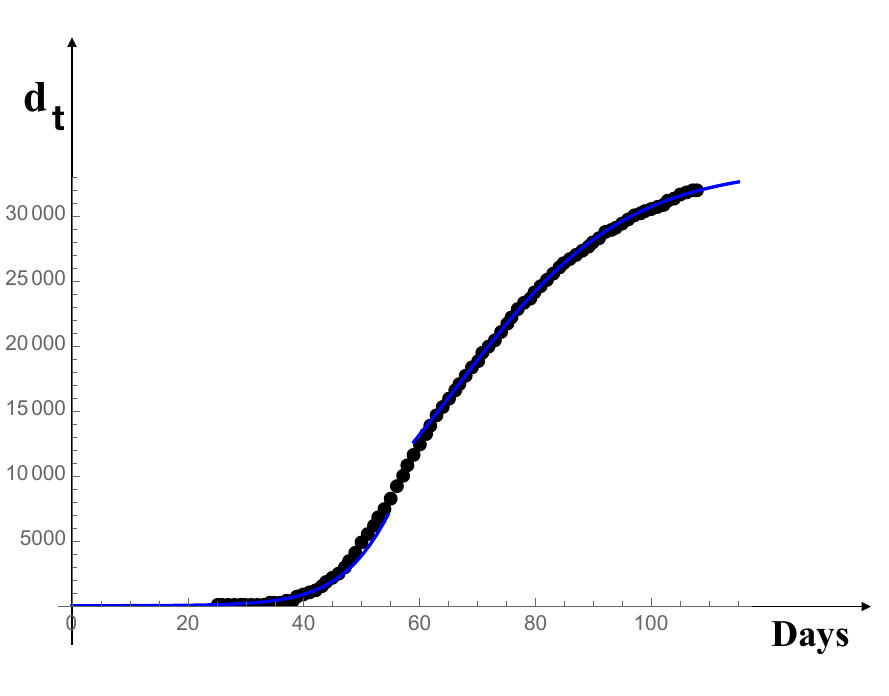}
\caption{\textit{Italy situation}. Theoretical predictions (blue line) against the experimental data (black circles) for the deceased people.}
\label{Fig.13}
\end{figure}

\begin{figure}[th!]
\centering\centering\includegraphics[scale=.30]{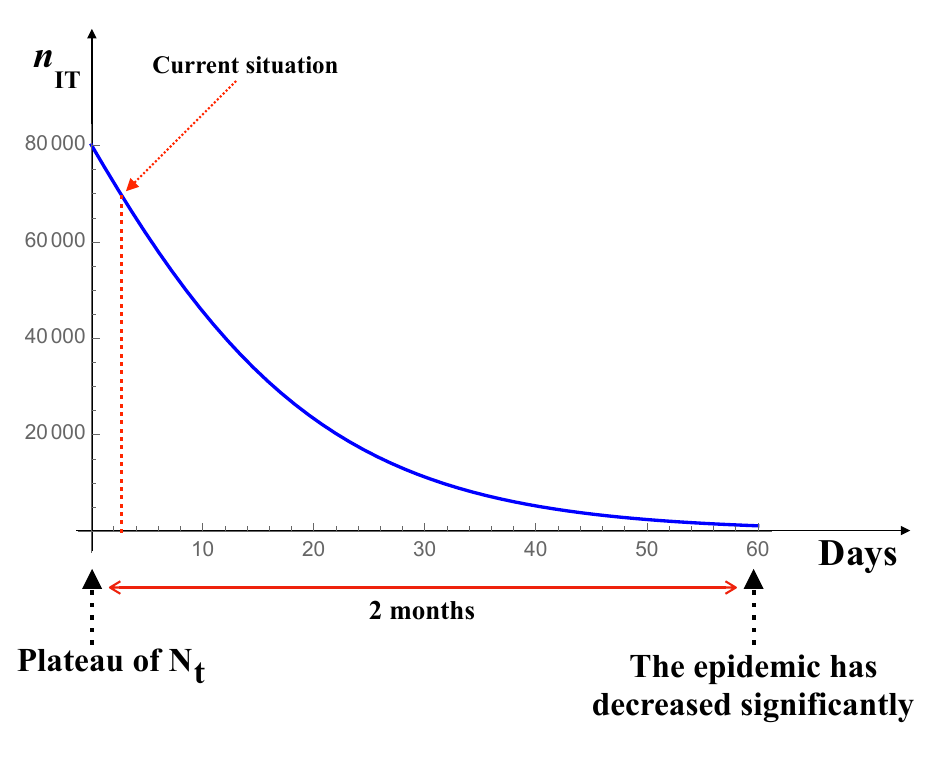}
\caption{\textit{The descending phase for Italy}. {\it If $N$ reaches a plateau,} after two months the lockdown measures may heavily be lightened and we can return to normal work.}
\label{Fig.14}
\end{figure}

In this subsection, we report the numerical solutions of Eqs~(\ref{dp4}) for Italy and Belgium. Fig.~\ref{Fig.12} and Fig.~\ref{Fig.13} concern the Italian situation. They show the numerical solution of Eqs~(\ref{dp4}) for the number of recovered people and deaths, respectively. Fig.~(\ref{Fig.14}), illustrates the descendant-phase for Italy {\it if $N$ reaches a plateau}.
%%%%%%%%%%%%%%%%%%%%%%%%%%%%%%%%%%%%%%%%
%%%%%%%%%%%%%%%%%%%%%%%%%%%%%%%%%%%%%%%%
%%%%%%%%%%%%%%%%%%%%%%%%%%%%%%%%%%%%%%%%
%%%%%%%%%%%%%%%%%%%%%%%%%%%%%%%%%%%%%%%%
%%%%%%%%%%%%%%%%%%%%%%%%%%%%%%%%%%%%%%%%%%
%%%%%%%%%%%%%%%%%%%%%%%%%%%%%%%%%%%%%%%%%%

\begin{figure}[th!]
\centering\centering\includegraphics[scale=.30]{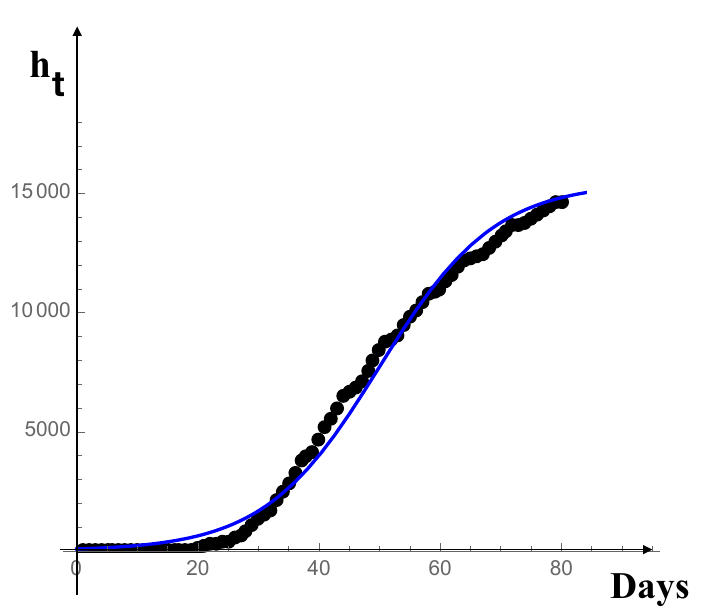}
\caption{\textit{Belgian situation}. Theoretical predictions (blue line) against the experimental data (black circles) for the recovered people.}
\label{Fig.15}
\end{figure}

\begin{figure}[th!]
\centering\centering\includegraphics[scale=.30]{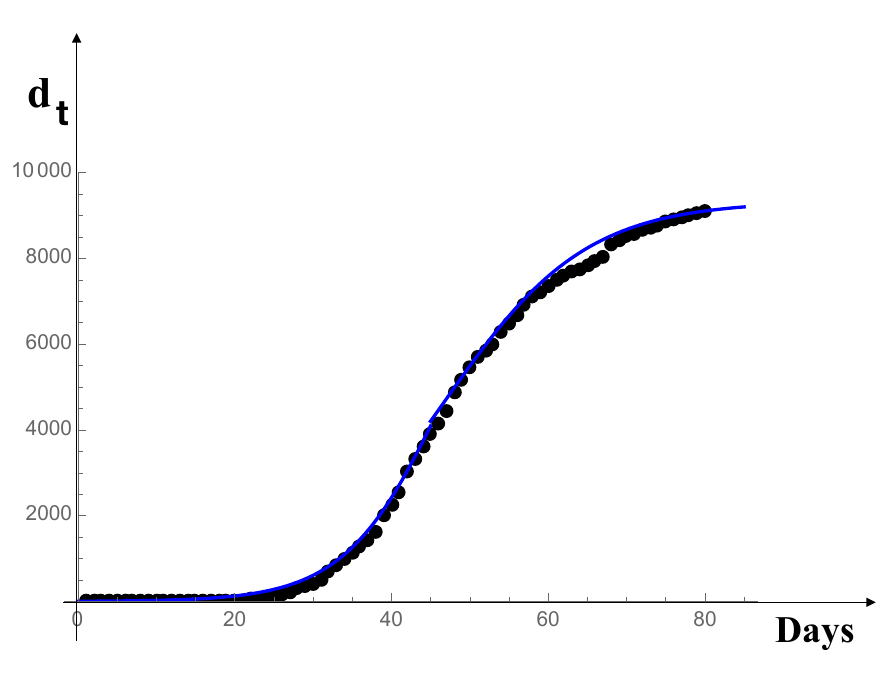}
\caption{\textit{Belgian situation}. Theoretical predictions (blue line) against the experimental data (black circles) for the deceased people.}
\label{Fig.16}
\end{figure}

\begin{figure}[th!]
\centering\includegraphics[scale=.30]{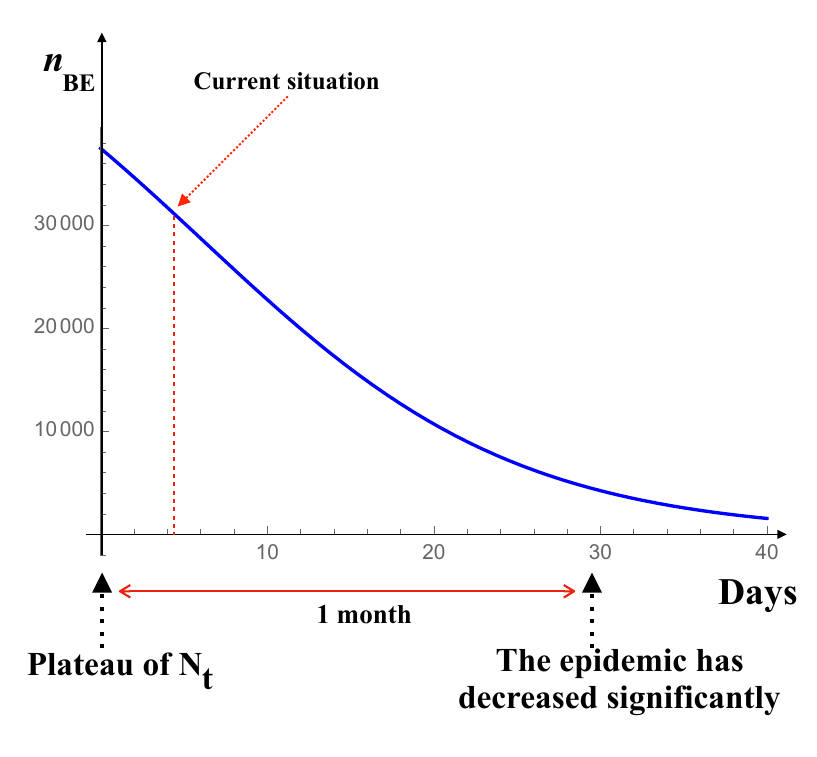}
\caption{\textit{The descending phase for Belgium}. {\it If $N$ reaches a plateau}, after one month the lockdown measures may heavily be lightened and we may return to normal work.}
\label{Fig.17}
\end{figure}

\noindent Figs~(\ref{Fig.15}) and (\ref{Fig.16}) refer to the Belgian situation. The figures illustrate the numerical solutions of Eqs~(\ref{dp4}) for the number of recovered people and deaths, respectively. Fig.~(\ref{Fig.17}) shows the descendant-phase for Belgium {\it if $N$ reaches a plateau}.
%%%%%%%%%%%%%%%%%%%%%%%%%%%%%%%%%%%%%%%%%%
%%%%%%%%%%%%%%%%%%%%%%%%%%%%%%%%%%%%%%%%%%
%%%%%%%%%%%%%%%%%%%%%%%%%%%%%%%%%%%%%%%%%%
%%%%%%%%%%%%%%%%%%%%%%%%%%%%%%%%%%%%%%%%%%

%%%%%%%%%%%%%%%%%%%%%%%%%%%%%%%%%%%%%%%%%%
%%%%%%%%%%%%%%%%%%%%%%%%%%%%%%%%%%%%%%%%%%
\noindent In Fig.~\ref{Fig.14} and \ref{Fig.17} refer to the Italian and Belgian cases during the first wave of SARS-CoV2 infection,  respectively. In these figures a red arrow appears which indicates the moment when the total compartment $N$ was close to reaching the plateau. However, the restrictive measures have been loosened considerably allowing, unfortunately, the virus to reinvigorate itself again giving rise to the so-called "second wave" of SARS-CoV2 infection. Basically, from a mathematical point of view, the total regression of the virus is obtained if the following two conditions are simultaneously satisfied:
\begin{itemize}
\item{{\it The compartment $N$ reached the plateau}};
\item{{\it Restrictive measures are maintained with severity until the value of the {\it effective reproduction number}, $R_t$\footnote{The {\it effective reproduction number} $R_t$ is defined as the mean number of secondary cases generated by a typical primary case at time $t$ in a population, calculated for the whole period over a $5$-day moving average.}, is less than $1$}.}
\end{itemize}
\noindent Of course, it is very difficult, if not harmful, to impose severe restrictive conditions for a long time due to the negative, if not catastrophic, impact at the social and economic level. So, we have to learn to coexist with the virus by remaining, however, vigilant and respecting hygiene rules, even when its rate of presence is low.

\section{Perspectives}\label{C}

\noindent It is worth noting the \textit{degree of the flexibility} of our model. For example, let us suppose that we need to set up a model able to distinguish old population (over 65 year old) from the young one (with age not exceeding 35 years), by assuming that the older population is twice as likely to get infected by Coronavirus with respect to the younger one. In this case, it is just sufficient to replace the scheme $I+S\xrightarrow{\mu} 2I$ with the scheme
\begin{align}\label{C1}
&I+S_Y \xrightarrow{\mu_y} 2I\\
&I+2S_O \xrightarrow{\mu_o} 3I\nonumber\\
&S=S_Y+S_O\nonumber
\end{align}
\noindent with $S_Y$ and $S_o$ denoting the \textit{susceptible young people} and the \textit{susceptible old people}, respectively. Another example could be the following. Let us suppose that we need to distinguish two class of infected individuals: 

\noindent {\bf 1)} infected people (denoted by $I_1$) able to transmit the Coronavirus to susceptible according to the (standard) scheme $I_1+S\rightarrow 2I$;

\noindent {\bf 2)} Infected people (denoted by $I_2$) having the capacity to transmit the virus, say, 7 times higher with respect to the category {\bf 1)}. In this case, the corresponding scheme reads:
\begin{align}\label{C2}
&I_1+S \xrightarrow{\mu_1} 2I\\
&I_2+7S \xrightarrow{\mu_2} 8I\nonumber\\
&I=I_1+I_2\nonumber
\end{align}
\noindent It is then easy to write the ordinary differential equations associated to schemes (\ref {C1}) and (\ref {C2}). The above example draws attention to the great flexibility offered by the Kinetic-type approach. For example, in the introduction we mentioned the new sub-variants of Omicron, BA.4 and BA.5. It is commonly accepted that BA.4, BA.5, and BA.2.12.1 are more contagious than past versions of Omicron, which is allowing them to spread even faster \cite{ladyzhets}. The kinetic scheme (\ref{C2}) can be easily readjusted to treat the evolution of variants B.4 and B.5 once the degree of contagion $n$ of these two sub-variants is known, establishing the {\it kinetic reaction}
\begin{equation}\label{C2a}
I+S \xrightarrow{\mu} n I
\end{equation}
\noindent Let us now consider another aspect of the model. In the Subsection~(\ref{LQM}), we have introduced scheme~(\ref{LQM1}) that models the lockdown measures. As mentioned, such measures are imposed by national governments to all susceptible population. However, we can also take into consideration the hypothesis that these measures are not rigorously respected by the population and this for various reasons: neglect of the problem, depression due to prolonged isolation, lack of confidence in the measures adopted by the Government, desire to attend parties with friends and relatives, refusal to wear masks in crowded environments, etc. These actions invalidate the effectiveness of lockdown measures significantly. Scheme~(\ref{LQM1}) still adapts to describe these kind of situations with the trick of replacing Fig.~\ref{LEP} with another one that models the {\it emotional behaviour} of susceptible people (or with an analytic expression that may be obtained by using the {\it mathematical basis} introduced in \cite{sonnino2,sonnino,sonnino3,sonnino4}. The O.D.E.s read
\begin{align}\label{C3}
&{\dot S}=-\mu SI - k_E S(E_{Max}-S_E)+(1-k_E)(E_{Max}-E)\\
&{\dot S}_E=k_E SE-k^{-1}_ES_E\nonumber
\end{align}
\noindent where $E$ stands for {\it Emotional}. 

\noindent This paper, together with Sonnino {\it et al.} \cite{sonnino3}, are the first contributions to the overall objectives aiming to obtain the correct space-time stochastic differential equations able to describe realistic situations of spread of SARS-CoV2 infection in large countries. As  mentioned in the Introduction, our task will be accomplished if are able to
\begin{enumerate}
\item model the distribution of hospitals in a country;
\item model the distribution of the poles of attraction of susceptible people (e.g., shopping centres workplaces, etc.);
\item identify a mechanism that allows to establish when a pole of attraction becomes "saturated" with infected people by proposing alternative poles of attraction;
\item  model correctly the Lockdown and the Quarantine measures adopted by the Government of the Country;
\item determine the nature of the intrinsic (ie spontaneous) fluctuations to which a macroscopic system is subjected, determining the correlation function by statistical mechanics.
\end{enumerate}

\noindent At first glance, such a work program would appear to be too ambitious and as said, to our knowledge, the state-of-the-art of the current alternative techniques are unable to resolve the issues listed above. The approach, "kinetic-type reactions" (KTR) proposed by us is very promising and allows to achieve this goal in a relatively simple way. With the axioms enunciated in the Introduction, the "kinetic-type reactions" approach
\begin{itemize}
\item models each actor by a dedicated “chemical species” that can only be created or destroyed as the result of one, or several, elementary steps, 
\item allows to determine the dynamics of the system starting from this set of elementary steps;
\item  and due to its flexibility, allows to analyse complex situations where several variables are involved, such as $R$, $Q$, $R_h$, $I_h$ etc;
\end{itemize}

\section{Conclusions}\label{D}
We showed that our models are able to produce predictions not only on the first but also on the successive waves of SARS-CoV2 infections, Omicron and its sub-variants. The theoretical predictions are in agreement with the official number of cases with minimal parameter fitting. We discussed the strengths and limitations of the proposed models regarding the long-term predictions and, above all, the duration of how long the lockdown and the quarantine measures should be taken in force in order to limit as much as possible the intensities of subsequent SARS-CoV-2 infection waves. This task has been carried out by taking into account the theoretical results recently appeared in literature \cite{sonnino1,sonnino2,sonnino,sonnino3,sonnino4} and without neglecting the delay in the reactions steps. Our models (the $(SISI_h)_L$-model and the {\it kinetic-type reactions model}) emphasise, and demonstrate, the crucial role played by the Hospitals. More specifically, we showed that the Health Care Institutions directly enter into the dynamics of the infectious individuals by influencing the outcome of the outbreak significantly, limiting, and even dampening, the spread of the Coronavirus. We applied our results in two (very) different situations: the spreading of the Coronavirus in a small European Country (Belgium) and in big Countries (Italy, France, Germany and USA). We have also incorporated real data into a stochastic model. The goal of this series of works is to obtain a comparative analysis against the deterministic one, in order to use the new theoretical results to predict the number of new cases of infected people and to propose possible changes to the measures of isolation.

\noindent  {\bf Acknowledgements}

\noindent We are grateful to Dr Fernando Mora of the Institut de Physique de Nice, Universit{\'e} C$\hat{\rm o}$te d’Azur, for his collaboration in the study of the stochastic modeling of COVID 19.

%%%%%%%%%%%%%%%% t

%\bigskip
\end{document}